\documentclass{article}

\usepackage[preprint,nonatbib]{neurips_2023}

\usepackage{xcolor}
\usepackage{graphicx}
\usepackage{amsmath}
\newtheorem{theorem}{Theorem}
\newtheorem{definition}{Definition}
\newtheorem{lemma}{Lemma}

\newtheorem{corollary}{Corollary}
\newtheorem{remark}{Remark}
\usepackage{bbm}
\usepackage[retainorgcmds]{IEEEtrantools}
\usepackage{subcaption}
\usepackage{wrapfig}
\usepackage{adjustbox}
\usepackage{makecell}

\usepackage[font=small,labelfont=bf]{caption}

\usepackage{footnote}
\makesavenoteenv{tabular}

\usepackage[utf8]{inputenc} % allow utf-8 input
\usepackage[T1]{fontenc}    % use 8-bit T1 fonts
\usepackage{hyperref}       % hyperlinks
\usepackage{url}            % simple URL typesetting
\usepackage{booktabs}       % professional-quality tables
\usepackage{amsfonts}       % blackboard math symbols
\usepackage{nicefrac}       % compact symbols for 1/2, etc.
\usepackage{microtype}      % microtypography
\usepackage{tabularx}
\usepackage{pifont}% http://ctan.org/pkg/pifont
\newcommand{\xmark}{\ding{55}}%
\usepackage{enumitem}
\usepackage{cite}

\usepackage[utf8]{inputenc} % allow utf-8 input
\usepackage[T1]{fontenc}    % use 8-bit T1 fonts
\usepackage{hyperref}       % hyperlinks
\usepackage{url}            % simple URL typesetting
\usepackage{booktabs}       % professional-quality tables
\usepackage{amsfonts}       % blackboard math symbols
\usepackage{nicefrac}       % compact symbols for 1/2, etc.
\usepackage{microtype}      % microtypography
\usepackage{xcolor}         % colors
\usepackage{rotating}

\title{On the Choice of Perception Loss Function  for Learned Video Compression}

\author{%
  Sadaf Salehkalaibar\thanks{Equal Contribution} \\
  ECE Department\\
  University of Toronto\\
  \texttt{sadafs@ece.utoronto.ca} \\
   \And
   Buu Phan\textsuperscript{*}\\
   ECE Department\\
  University of Toronto\\
   \texttt{truong.phan@mail.utoronto.ca} \\
   \And
   Jun Chen \\
 ECE Department\\
  McMaster University\\
   \texttt{chenjun@mcmaster.ca} \\
   \And
   Wei Yu \\
ECE Department\\
  University of Toronto\\
   \texttt{weiyu@ece.utoronto.ca} \\
   \And
   Ashish Khisti \\
ECE Department\\
  University of Toronto\\
   \texttt{akhisti@ece.utoronto.ca} \\
}
 
\begin{document}

\maketitle

\begin{abstract}

We study causal, low-latency, sequential video compression when the output is subjected to both a mean squared-error (MSE) distortion loss as well as a perception loss to target realism. Motivated by prior approaches, we consider two different perception loss functions (PLFs). The first, PLF-JD,  considers the joint distribution (JD) of all the video frames up to the current one, while the second metric, PLF-FMD,  considers the framewise marginal distributions (FMD) between the source and reconstruction. Using information theoretic analysis and deep-learning based experiments, we demonstrate that the choice of PLF can have a significant effect on the reconstruction, especially at low-bit rates. In particular, while the reconstruction based on PLF-JD can better preserve the temporal correlation across frames, it also imposes a significant penalty in distortion  compared to PLF-FMD and further makes it more difficult to recover from errors  made in the earlier output frames. Although the choice of PLF decisively affects  reconstruction quality, we also demonstrate that it may not be essential to commit to a particular PLF during encoding and the choice of PLF can be delegated to the decoder. In particular, encoded representations generated by training a system to minimize the MSE (without requiring either PLF) can be  {\em near universal}  and can generate close to optimal reconstructions for either choice of PLF at the decoder.  We validate our results using (one-shot) information-theoretic analysis, detailed study of the rate-distortion-perception tradeoff of the Gauss-Markov source model as well as deep-learning based experiments on moving MNIST, KTH and UVG datasets. 
\iffalse
 On the theoretical side, we establish the rate-distortion-perception tradeoff for first-order Markov sources and study the case of first-order Gauss-Markov sources in detail. For general sources, we establish that for any given encoder, the decoder can switch from a representation that minimizes MSE to a representation that achieves zero PLF with at-most a factor of $2$ increase in distortion. We also validate our results with deep-learning based experiments on moving MNIST, KTH and UVG datasets. 
\fi
\end{abstract}

\section{Introduction}

There is an increasing demand for video compression algorithms that are able to generate visually pleasing videos at low bitrates.  Most of the current video codecs use distortion measures such as PSNR \cite{PSNR1, PSNR2, PSNR3, PSNR4}, MSE and MS-SSIM \cite{SSIM, PSNR3, PSNR4} to generate reconstructions which tend to be blurry at extremely low bitrates. In recent years, there has been a  growing interest (see e.g.,~\cite{zhang2021dvc,video1,yang2021perceptual,video-joint, GAN3}) in using deep generative models to make the reconstructions look more realistic. Such techniques introduce an additional perception loss function that measures a distance between distributions of the source and reconstruction, with {\em perfect} perception corresponding requiring that the two distributions be identical.

In compression systems, improving realism comes at the price of increasing distortion. The  work of Blau and Michaeli~\cite{blau2019rethinking} establishes the theoretical rate-distortion-perception (RDP) tradeoff which has also been validated in~\cite{image-comp1, image-comp2, image-comp3, image-comp4}. Furthermore \emph{universal} encoded representations were proposed in \cite{Jun-Ashish2021} where the representation is fixed at the encoder and the decoder is adapted to achieve a performance near the optimal RDP tradeoff curve. The extension of these works to video compression involves many challenges. First, the compression system must not only account for spatial redundancy as in image compression, but also exploit the temporal redundancy across video frames, making the system design more complex. Secondly, unlike the case of image compression, there may be no clear choice of the perception loss function (PLF). Indeed, some prior works~\cite{video1}  consider PLF that preserves framewise marginal distribution (PLF-FMD) between the source and reconstruction, while other works consider joint distribution (PLF-JD) across multiple frames~\cite{video-joint}.

\begin{figure}[t]
%\centering
\begin{subfigure}[b]{0.4\textwidth}
  \includegraphics[width=\textwidth]{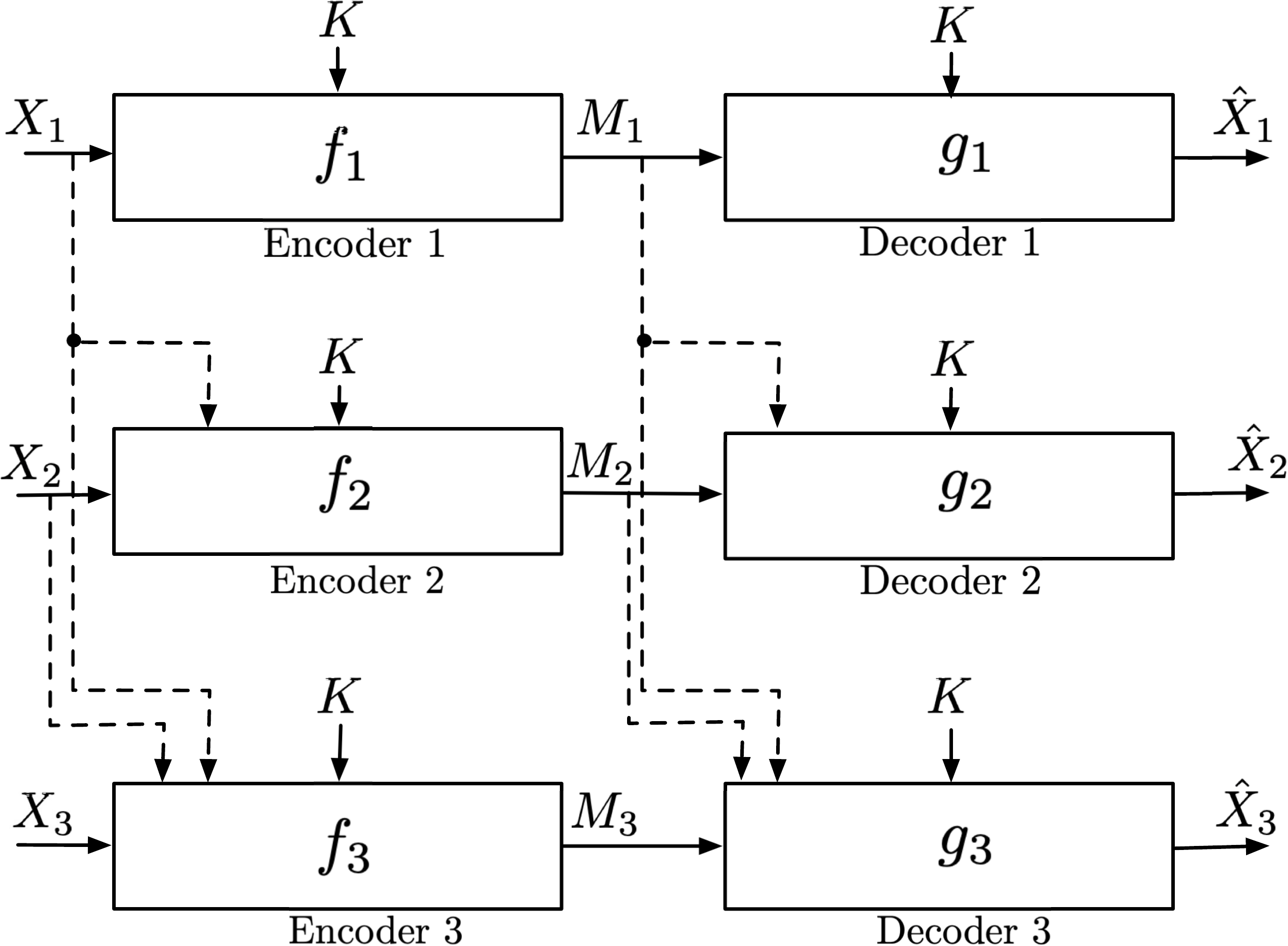}
\caption{Encoders and decoders for $T{=}3$ frames.}
\label{enc-dec}
\end{subfigure}
\centering
\hspace{0.5cm}\begin{subfigure}[b]{0.55\textwidth}
   \includegraphics[width=\textwidth]{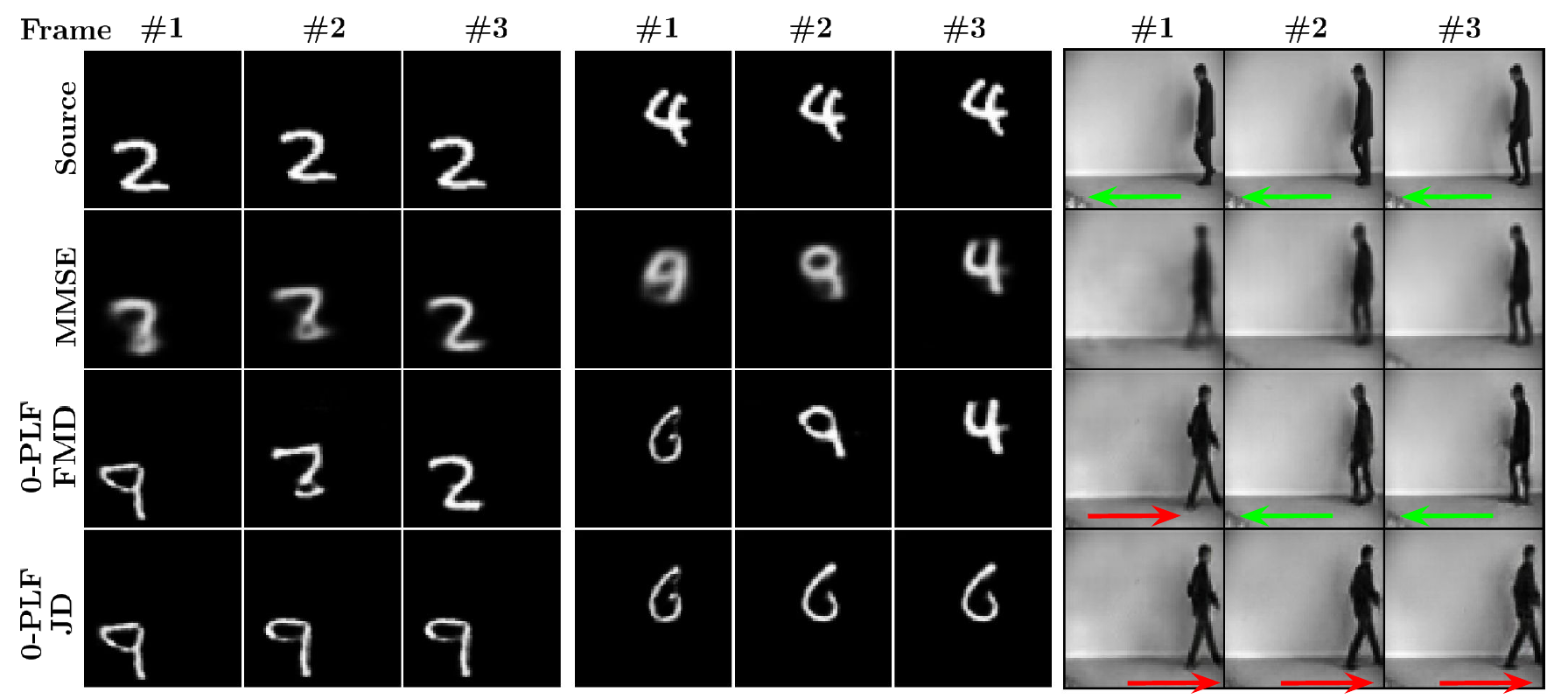}
    \caption{Effects of different PLFs on reconstructions for Moving MNIST and KTH datasets (best view in the monitor). }
   \label{figb:error permanence2}
\end{subfigure}
\label{fig:KTH error permanence}

\begin{subfigure}[b]{1.0\textwidth}
   \includegraphics[width=\textwidth]{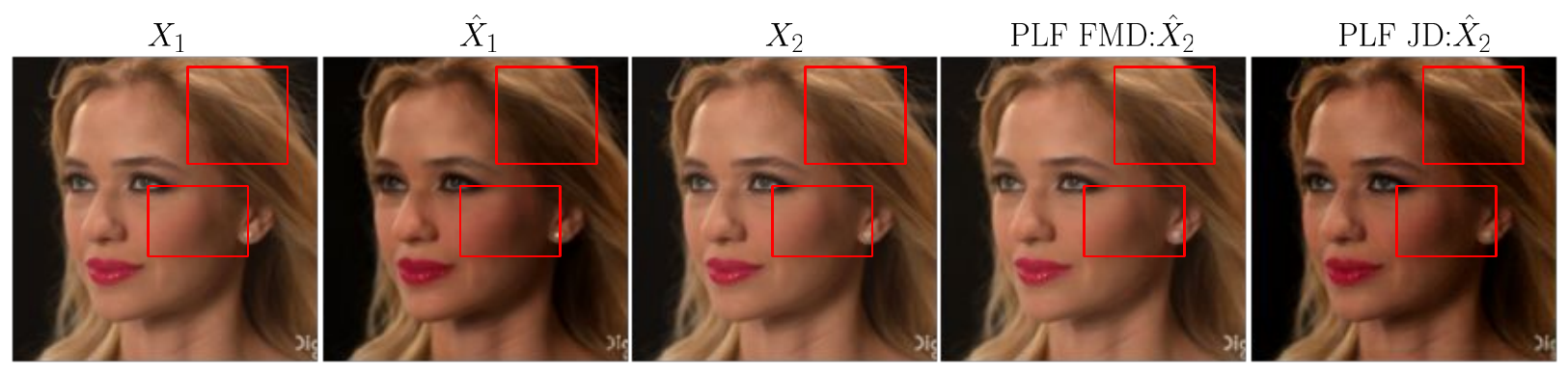}
    \caption{Error permanence on the UVG dataset. The PLF-JD reconstructions propagate the flaws in the color tone from
the previous I-frame reconstruction while the decoder based on PLF-FMD is able to fix these flaws. }
   \label{figb:error permanence uvg}
\end{subfigure}
\caption{(a) Proposed System Model (b,c)  Error permanence phenomenon under different PLFs. High fidelity but incorrect  I-frame reconstruction propagates the error to subsequent P-frames in PLF-JD reconstructions. The MMSE and 0-PLF-FMD reconstructions do not suffer from this issue.}
\vspace{0cm}
\end{figure}

As illustrated in Fig.~\ref{enc-dec}, we study causal, low-latency, sequential video compression when the output is subjected to both a mean squared-error (MSE) distortion loss  and either a  PLF-JD or PLF-FMD metric for perception loss. Our main results are as follows:

\vspace{-0.5em}

\begin{itemize}[leftmargin=0.2in]
\item \emph{Differences in reconstruction quality based on the choice of PLF}: We demonstrate that the choice of PLF can decisively affect the reconstruction quality especially in the low bit-rate regime. We  approximately characterize the operational RDP region on a per-frame basis for a first-order Markov source model and analyze the special case of Gauss-Markov sources in detail. We show that there is a significant penalty in distortion when using PLF-JD in the low-rate regime.  On the experimental side, we demonstrate that while  PLF-JD preserves better temporal consistency across video frames,  it suffers from the \emph{permanence of error} phenomenon in which the mistakes in reconstructions propogate to future frames. On the other hand, the PLF-FMD metric shows more capability  in correcting mistakes across frames (see Fig.~\ref{figb:error permanence2} for visualizations involving three-frame videos and  Fig.~\ref{figb:error permanence uvg} for the effects on the high resolution UVG dataset).   On the other hand, if the first frame is transmitted at high bit-rate, we demonstrate  that PLF-JD performs better than  PLF-FMD. 

\item \emph{Universality of minimum mean square error (MMSE) reconstructions}: We demonstrate that encoded representations generated from an encoder trained to minimize MSE reconstruction (without considering any PLF)  suffice to produce close-to-optimal reconstructions for either choice of PLF.  For general sources,  we show that when using PLF-FMD, the MMSE reconstruction can be transformed to a reconstruction satisfying perfect perceptual quality by increasing the distortion at most by a \emph{factor of two}. While a similar result does not hold for PLF-JD in general, it is satisfied for a special  class of encoders which operate  in the low-rate regime. For the Gauss-Markov source model we demonstrate exact universality i.e.,  encoded representation for the MMSE reconstruction can be adapted to achieve any other reconstruction in the RDP region. We  also use deep learning based experiments to provide experimental evidence of these results.

We note that the above notion of universal encoded representations based on MSE reconstruction can have significant advantages in practice. First, although the  reconstructions associated with different choices of PLFs can be  visually very different, universal representations delegate the choice of PLF to the decoder rather than requiring the encoder to commit to a specific PLF. Secondly, the motion estimation step during compression  requires an MSE based reconstruction to generate the best set of motion vectors. A universal representation seamlessly enables this even when the system is designed to achieve a different operating point in the RDP region.
\end{itemize}

\subsection*{Related Work}

\noindent {\em{Perceptual Lossy Video Compression.}} Distribution preserving framework using Generative Adversarial Networks (GAN) has been widely adopted as a surrogate metric for perceptual quality in image\cite{gao2022flexible,image-comp1,GAN,agustsson2019generative} and, recently, video compression \cite{zhang2021dvc,video1,yang2021perceptual,video-joint}. Unlike image compression, where the choice of PLF is straightforward, there is currently no agreed-upon objective  for lossy video compression. For instance, DVC-P\cite{zhang2021dvc} employs PLF-FMD to improve the visual quality per frame, ignoring the temporal coherence. Similarly, Mentzer et al.\cite{video1} utilize the per-frame metric with a conditional GAN model, and found no significant differences when using a GAN objective with multiple frames. Other works target temporal consistency by incorporating  multiple frames in their GAN objective. This includes the work by Yang et al. \cite{yang2021perceptual}, where every two consecutive frames are included, and by Veerabadran et al.\cite{video-joint}, where they employ the PLF-JD metric in a non-causal setting. Unlike previous works, we study the impact of two different perception objectives, i.e. PLF-JD and PLF-FMD, on reconstructions in the causal setting, presenting theoretical properties that are verified by deep learning experiments.  For the PLF-JD metric, we demonstrate the {\em error permanence phenomenon}, which, unlike the error propagation issue\cite{GAN,lu2020content}, cannot be resolved by increasing the code rate assigned to the $P$ frames.

\noindent {\em{RDP Tradeoff and Principle of Universality.}} Targeting the distribution preserving framework in lossy image compression, several theoretical works have shown the presence of RDP tradeoff\cite{blau2018perception,blau2019rethinking,freirich2021theory,saldi2013randomized}, where perfect perception comes at a cost of increasing distortion by at most a factor of 2. Furthermore, an encoder that generates universal representations exists \cite{Jun-Ashish2021,yan2021perceptual,agustsson2022multi}, which enables the decoder to freely choose the level of distortion-perception tradeoff it desires. Our work explores these avenues in the context of causal video compression.   As discussed previously we show that the MMSE representation can be used as a universal representation for both perception metrics which has several advantages in the context of video compression..

\section{System Model}
Let $(X_1,\ldots,X_T)\in\mathcal{X}_1\times\ldots\times \mathcal{X}_T$ be $T$ frames in a video  (with each $\mathcal{X}_i \subseteq {\mathbb R}^d$) distributed according to $P_{X_1\ldots X_T}$ . The frames are available for encoding sequentially;  $X_1$ is available first, then $X_2$ arrives, followed by $X_3$ and so on. There is a shared randomness $K\in\mathcal{K}$ which is available at all encoders and decoders.
The following (possibly stochastic) mappings define the encoding and decoding functions:
\begin{IEEEeqnarray}{rCl}
f_{j}&\colon& \mathcal{X}_1\times \ldots \times \mathcal{X}_j\times \mathcal{K}\to \mathcal{M}_{j},\qquad\qquad\qquad j=1,\ldots,T,\\
g_{j} &\colon& \mathcal{M}_{1}\times \mathcal{M}_{2}\times \ldots \times \mathcal{M}_{j}\times \mathcal{K}\to \hat{\mathcal{X}}_j,\;\;\label{decoding-function}
\end{IEEEeqnarray}
where $\mathcal{M}_{j} \in \{0,1\}^\star$ denotes the set of  (variable-length) messages assigned by the $j$th encoder and $\hat{\mathcal{X}}_j \subseteq {\mathbb R}^d$ is the $j$-th reconstruction alphabet (see Fig.~\ref{enc-dec}). Let $P_{\hat{X}_1\ldots \hat{X}_T|X_1\ldots X_T}$ be the conditional distribution of the reconstructed video given the original video which is basically determined by the mappings $\{f_{j}\}_{j=1}^T$ and $\{g_{j}\}_{j=1}^T$.  The above setting is a  \emph{one-shot} setup as only a single source sample is compressed at a time. For each frame $j$,  a distortion metric is imposed on the output, which we assume throughout is the mean squared-error (MSE) function i.e. $d(x_j, \hat{x}_j) = ||x-\hat{x}_j||^2$, which is commonly used in many applications. From a perceptual point of view, for given probability distributions $P_{X_{1}\ldots X_{j}}$ and $P_{\hat{X}_{1}\ldots\hat{X}_{j}}$ on the original and reconstructed frame $j$, let $\phi_j(P_{X_{1}\ldots X_{j}},P_{\hat{X}_{1}\ldots\hat{X}_{j}})$ be the perception function capturing the difference between them. Note that the function $\phi_j$ is defined based on the joint distribution of all first $j$ frames. We call this metric as \emph{perception loss function based on joint distribution (PLF-JD)}. 
Note that when $\phi_j(P_{X_1\ldots X_j},P_{\hat{X}_1\ldots \hat{X}_j})=0$, we have: \begin{IEEEeqnarray}{rCl}
P_{X_1\ldots X_j}=P_{\hat{X}_1\ldots\hat{X}_j},\qquad j=1,\ldots,T.\label{percep-joint}
\end{IEEEeqnarray}

We refer to this case as \emph{zero-perception loss function based on joint distribution ($0$-PLF-JD)}. Alternatively, the \emph{perception loss function based on framewise marginal distribution (PLF-FMD)} is denoted by $\xi_j(P_{X_j},P_{\hat{X}_j})$ and is based on only the marginal distribution of the $j$-th frame.
In particular, note that $0$-PLF-FMD implies that $P_{X_j}=P_{\hat{X}_j}$ for each $j$.   In most of the paper, for simplicity of presentation, we provide some of our results for $T=3$ frames.   In that case, we use the shorthand notation $\mathsf{X}$ to denote the tuple $(X_1,X_2,X_3)$, e.g., $\mathsf{M}:=(M_1,M_2,M_3)$, $\mathsf{D}:=(D_1,D_2,D_3)$, $\mathsf{f}:=(f_1,f_2,f_3)$.

%Moreover, for given distributions $P_{X_j}$ and $P_{\hat{X}_j}$, $W_2^2(P_{X_j},P_{\hat{X}_j})$ denotes the ``Wasserstein-2 distance'' defined as $\inf\mathbbm{E}[\|X_j-\hat{X}_j\|^2]$ where the infimum is over all joint distributions of $(X_j,\hat{X}_j)$ with marginals $P_{X_j}$ and $P_{\hat{X}_j}$.}

%{\color{red}We note the following distribution which is induced by the encoding and decoding functions: 
%\begin{IEEEeqnarray}{rCl}
%&&P_{\mathsf{M}\mathsf{X}\hat{\mathsf{X}}|K}(\mathsf{m},\mathsf{x},\hat{\mathsf{x}}|k):=P_{\mathsf{X}}(\mathsf{x})\cdot\mathbbm{1}\{\mathsf{m}=\mathsf{f}(\mathsf{x},k)\}\cdot\mathbbm{1}\{\hat{\mathsf{x}}=\mathsf{g}(\mathsf{m},k)\},\nonumber\\
%&&\hspace{2.5cm}\forall\; \mathsf{m}\in \mathcal{M}_1\times  \mathcal{M}_2\times  \mathcal{M}_3, \mathsf{x}\in \mathcal{X}_1\times \mathcal{X}_2\times \mathcal{X}_3, \hat{\mathsf{x}}\in \hat{\mathcal{X}}_1\times \hat{\mathcal{X}}_2 \times \hat{\mathcal{X}}_3, k\in\mathcal{K}.\label{P-general-distribution}
%\end{IEEEeqnarray}}

\section{Distortion Analysis for a Fixed Encoder and Zero-perception Loss}
\label{sec:fixed}
In this section, we assume that the encoding functions ${\mathsf f}$  are fixed,  but the decoding functions ${\mathsf g}$ can be optimized to generate different reconstructions. Equivalently, the distribution $P_{\mathsf{M}|\mathsf{X}K}{:=}\mathbbm{1}\{\mathsf{M}=\mathsf{f}(\mathsf{X},K)\}$  is fixed, while by varying the reconstruction distribution $P_{\hat{\mathsf{X}}|\mathsf{M}K}{:=} \mathbbm{1}\{\hat{\mathsf{X}}=\mathsf{g}(\mathsf{M},K)\},$ one attains different reconstructions 
$\hat{\mathsf{X}}$, where $\mathbbm{1}\{.\}$ denotes the indicator function. Furthermore  defining $D_j{:=}\mathbbm{E}_P[\|X_{j}-\hat{X}_{j} \|^2]$, we denote  $\mathsf{D}$ as the achievable distortion tuple associated with  $P_{\hat{\mathsf{X}}|\mathsf{M}K}$.

\iffalse
{\color{blue}\begin{definition}[Feasible Reconstruction for a Given Encoder]
\label{def:dec_fun}
For an encoder $P_{\mathsf{M}|\mathsf{X}K}$, a feasible reconstruction  is  the conditional distribution $P_{\hat{\mathsf{X}}|\mathsf{M}K}$ defined in~\eqref{P-general-distribution}.
Furthermore with $D_j = \mathbbm{E}_P[\|X_{j}-\hat{X}_{j} \|^2]$, let $\mathsf{D}$ be the achievable distortion tuple associated with the reconstruction function $P_{\hat{\mathsf{X}}|\mathsf{M}K}$.
\end{definition}}
\fi

One natural choice of  reconstructions is the minimum mean squared error (MMSE) reconstruction function. At step $j$, the reconstruction, which we denote in this case by $\tilde{X}_j$, is obtained by taking the conditional expectation of $X_j$ given all information at the decoder up to time $j$ i.e.,  $\tilde{X}_j{:=}\mathbbm{E}_P[X_j|M_{1} \ldots M_{j},K]$ for each $j=1,2,3$. It is well known that the MMSE reconstruction functions minimize the reconstruction distortion i.e., if we define the set
\begin{IEEEeqnarray}{rCl}
&&\hspace{-0.5cm}\Phi_{\mathsf{D}^{\min}}(P_{\mathsf{M}|\mathsf{X}K})=\{\mathsf{D}:D_j \geq \mathbbm{E}_P[\|X_j-\tilde{X}_j\|^2],  \qquad j=1,2,3
\}\label{distortion-min}
\end{IEEEeqnarray}
then the distortion tuple ${\mathsf{D}}$ associated with any reconstruction  $P_{\hat{\mathsf{X}}|\mathsf{M}K}$ satisfies  
${\mathsf{D}} \in \Phi_{\mathsf{D}^{\min}}(P_{\mathsf{M}|\mathsf{X}K})$.

The main result of this section is that assuming fixed encoder, the achievable distortions under $0$-PLF-FMD is at most twice of that under the MMSE distortion loss alone. The same conclusion also holds for 0-PLF-JD for a class of encoders operating at low rate. We first consider the case of $0$-PLF-FMD. 

%\subsection{Distortion Analysis for the $0$-PLF FMD}
\begin{definition}[$0$-PLF-FMD Distortion]\label{def-zero-per}  For an encoder $P_{\mathsf{M}|\mathsf{X}K}$, the set $\Phi_{\mathsf{D}^0}(P_{\mathsf{M}|\mathsf{X}K})$ denotes the set of all distortion tuples $\mathsf{D}$ for which there exists a reconstruction $P_{\hat{\mathsf{X}}|\mathsf{M}K}$ satisfying  $P_{ X_{j}}=P_{\hat{X}_{j}}$ for each $j \in \{1,2,3\}$.
\end{definition}

\begin{theorem}\label{thm-universal-one-shot} The set $\Phi_{\mathsf{D}^0}(P_{\mathsf{M}|\mathsf{X}K})$ is characterized as follows:
\begin{IEEEeqnarray}{rCl}
&&\hspace{-0.2cm}\Phi_{\mathsf{D}^0}(P_{\mathsf{M}|\mathsf{X}K})=\{\mathsf{D}:D_j \geq \mathbbm{E}_P[\|X_j-\tilde{X}_j\|^2]+W_2^2(P_{\tilde{X}_j},P_{X_j}),\; j=1,2,3
\},\label{distortion-0-percep}
\end{IEEEeqnarray}
 where $W_2^2(P_{X_j},P_{\hat{X}_j})$ denotes the Wasserstein-2 distance between the two distributions \cite{Wasserstein-distance}.  Furthermore, we also have that:
\begin{IEEEeqnarray}{rCl}&&\hspace{-0.5cm}\Phi_{\mathsf{D}^0}(P_{\mathsf{M}|\mathsf{X}K})\supseteq \{\mathsf{D}:  D_j \geq 2\mathbbm{E}_P[\|X_j-\tilde{X}_j\|^2],\;\;\; j=1,2,3
\},
\end{IEEEeqnarray}
i.e., minimum achievable distortion with $0$-PLF-FMD is at most twice the MMSE distortion.
\end{theorem}
\begin{IEEEproof} See Appendix~\ref{PLF-FMD-factor-two}.
\end{IEEEproof}
\iffalse
\begin{IEEEeqnarray}{rCl}
&&\mathbbm{E}[\|X_j-\tilde{X}_j\|^2]+W_2^2(P_{\tilde{X}_j},P_{X_j})\leq 2\mathbbm{E}[\|X_j-\tilde{X}_j\|^2], \;\;j=1,2,3,
\end{IEEEeqnarray}
\fi

 We remark that the proof of Theorem~\ref{thm-universal-one-shot}, operationally demonstrates that the MMSE reconstruction can be converted to another reconstruction satisfying $0$-PLF-FMD with at-most a factor of $2$ increase in distortion,  generalizing the result in \cite{Jun-Ashish2021} for the single frame scenario (see also \cite{blau2018perception}).

%\subsection{Distortion Analysis for the $0$-PLF JD}
We next consider the case when zero perception loss is satisfied under the PLF-JD metric. Analogous to $\Phi_{\mathsf{D}^0}(P_{\mathsf{M}|\mathsf{X}K})$ in Definition~\ref{def-zero-per}, one can define $\Phi^{\text{joint}}_{\mathsf{D}^0}(P_{\mathsf{M}|\mathsf{X}K})$ to be the set of distortions associated with reconstruction functions that satisfy~\eqref{percep-joint}. The analysis of  $\Phi^{\text{joint}}_{\mathsf{D}^0}(P_{\mathsf{M}|\mathsf{X}K})$ is discussed in Appendix~\ref{joint-PLF-app} as it is more involved. In general, the {\em factor of two bound} as in Theorem~\ref{thm-universal-one-shot} cannot be realized in this case as demonstrated by a counter-example in Appendix~\ref{joint-PLF-app}.
Nevertheless, for a special  family of encoders we can obtain a counterpart of Theorem~\ref{thm-universal-one-shot}. In this family of encoders, the source $X_j$ at time $j$ is nearly independent of the encoder outputs up to and including time $j$, i.e., we can express: 
\begin{IEEEeqnarray}{rCl}
 P^{\text{noisy}}_{X_j|M_1\ldots M_jK}=(1-\mu)P_{X_j}+\mu Q^{\text{noisy}}_{X_j|M_1\ldots M_jK},\qquad j=1,2,3.\label{noisy-representation}
 \end{IEEEeqnarray}
where $\mu$ is a sufficiently small constant and the distribution $Q^{\text{noisy}}(\cdot)$ could be arbitrary conditional distribution with same marginal as $P_{X_j}$. We note that such encoders are studied in a variety of problems in information theory (see e.g.,~\cite{noisy}) that correspond to the low rate operating regime. The following result states that the factor-two bound holds approximately for such encoders.  \begin{theorem}\label{factor-two-thm}  For the class of encoders given by~\eqref{noisy-representation}, we have
\begin{IEEEeqnarray}{rCl}&&\hspace{-0.5cm}\Phi^{\text{joint}}_{\mathsf{D}^0}(P^{\text{noisy}}_{\mathsf{M}|\mathsf{X}K})\supseteq \{\mathsf{D}:  D_j \geq 2\mathbbm{E}_{P^{\text{noisy}}}[\|X_j-\tilde{X}_j\|^2]+O(\mu),\;\;\; j=1,2,3
\}.
\end{IEEEeqnarray}
\end{theorem}
\begin{IEEEproof} See Appendix~\ref{section-low-rate-app}.
\end{IEEEproof}

We note that the low-rate operating regime is practically important, as at higher rates MMSE based reconstructions can suffice  and the use of PLF metrics may be less relevant.

\section{Rate-Distortion-Perception Region}
\label{sec:RDP}

In this section, we assume that both the encoder $P_{\mathsf{M}|\mathsf{X}K}$ as well as the reconstruction $P_{\hat{\mathsf{X}}|\mathsf{M}K}$ can be optimized and study the associated rate-distortion-perception (RDP) tradeoff. We remind the reader that for PLF-JD and PLF-FMD,  the PLFs are denoted by $\phi_j(P_{X_1\ldots X_{j}},P_{\hat{X}_1\ldots\hat{X}_j})$ and $\xi_j(P_{X_j},P_{\hat{X}_j})$, respectively. In this case, the operational RDP region in the one-shot setting is defined as follows.
\begin{definition}[ Operational RDP region]\label{definition-information-function} For a given $P_{\mathsf{X}}$, an RDP tuple $(\mathsf{R},\mathsf{D},\mathsf{P})$ is said to be achievable for the one-shot setting if there exist an encoder $P_{\mathsf{M}|\mathsf{X}K}$ and a reconstruction $P_{\hat{\mathsf{X}}|\mathsf{M}K}$ satisfying:
\begin{IEEEeqnarray}{rCl}
 \mathbbm{E}[\ell(M_{j})] \leq & R_j, \;\;\;
\mathbbm{E}[\|X_j-\hat{X}_j\|^2] \leq & D_j, \;\;\;
\phi_j(P_{X_1\ldots X_j}, P_{\hat{X}_1\ldots \hat{X}_j}) \leq  P_j,\;\; j=1,2,3,\label{perception-condition}
\end{IEEEeqnarray}
where $\ell(M_{j})$ denotes the length of the message $M_{j}$.  The closure of the set of all achievable tuples, denoted by $\mathcal{C}^o_{\mathsf{RDP}}$, is the  operational RDP region. Moreover, for a given $(\mathsf{D},\mathsf{P})$, the operational DP rate region, denoted by $\mathcal{R}^o(\mathsf{D},\mathsf{P})$, is the closure of the set of all tuples $\mathsf{R}$ such that $(\mathsf{R},\mathsf{D},\mathsf{P})\in \mathcal{C}^o_{\mathsf{RDP}}$.
\end{definition}
The region $\mathcal{C}_{\mathsf{RDP}}^o$
cannot be directly computed as it involves all possible one-shot encoders/decoders. But for first-order Markov source, it has a tractable approximation in terms of mutual information.
%\vspace{-0.1cm}In this section we assume that {\both} the encoder $P_{\mathsf{M}|\mathsf{X}K}$ and  reconstruction $P_{\hat{\mathsf{X}}|\mathsf{M}K}$ can be varied and study the associated rate-distortion-perception (RDP) tradeoff. Again, for PLF JD and PLF FMD,  the perception functions are  denoted by $\phi_j(P_{X_1\ldots X_{j}},P_{\hat{X}_1\ldots\hat{X}_j})$ and $\xi_j(P_{X_j},P_{\hat{X}_j})$, respectvely.  We will focus on the class of first-order Markov sources, which lead to insightful analysis.

\subsection{RDP Region of First-Order Markov Sources}

 We first define the first-order Markov sources and then introduce an iRDP region which is computable.

\begin{definition}\label{1st-order-Markov} We call $\mathsf{X}$ as a first-order Markov source if the Markov chain $X_1{\to} X_2{\to} X_3$ holds. 
\end{definition}
 
\begin{definition}[Information RDP region]\label{iRDP-region}  For first-order Markov sources, the information RDP (iRDP) region, denoted by $\mathcal{C}_{\mathsf{RDP}}$, is the set of all tuples $(\mathsf{R},\mathsf{D},\mathsf{P})$ which satisfy the following
\begin{IEEEeqnarray}{rCl}
&R_1 \geq  I(X_1;X_{r,1}), \qquad R_2 \geq  I(X_2;X_{r,2}|X_{r,1}),\qquad  R_3 \geq  I(X_3;X_{r,3}|X_{r,1},X_{r,2}) \label{rate1-3}\\
&D_j \geq  \mathbbm{E}[\|X_j-\hat{X}_j\|^2], \qquad P_j \geq  \phi_j(P_{X_1\ldots X_j},P_{\hat{X}_1\ldots\hat{X}_j}), \qquad j=1,2,3,\label{perception}
\end{IEEEeqnarray}
for auxiliary random variables $(X_{r,1},X_{r,2},X_{r,3})$ and $(\hat{X}_1,\hat{X}_2,\hat{X}_3)$ satisfying the following
\begin{IEEEeqnarray}{rCl}
\hat{X}_1&=& \eta_1(X_{r,1}), \;\;\hat{X}_2=\eta_2(X_{r,1},X_{r,2}),\;\; \hat{X}_3=X_{3,r},\label{function-condition}\\
X_{r,1}&\to& X_1\to (X_2,X_3),\;\; X_{r,2}\to (X_2,X_{r,1})\to (X_1,X_3),\label{Markov-conditions1}\\X_{r,3}&\to& (X_3,X_{r,1},X_{r,2})\to (X_1,X_2),\label{Markov-conditions3}
\end{IEEEeqnarray}
for some deterministic functions $\eta_1(.)$ and $\eta_2(.,.)$.  Moreover, for a given $(\mathsf{D},\mathsf{P})$, the information DP (iDP) rate region, denoted by $\mathcal{R}(\mathsf{D},\mathsf{P})$, is the closure of the set of all tuples $\mathsf{R}$ that $(\mathsf{R},\mathsf{D},\mathsf{P})\in \mathcal{C}_{\mathsf{RDP}}$.   
\end{definition}

The expression for the iRDP region involves a search over auxiliary random variables $\mathsf{X}_r$ and $\hat{\mathsf{X}}$ that satisfy ~\eqref{function-condition}-\eqref{Markov-conditions3} subject to the constraints in~\eqref{rate1-3}--\eqref{perception}.  For first-order Markov sources, the following theorem states that the operational DP rate region can be approximated by the iDP rate region.

\begin{theorem}\label{thm-one-shot-information-function} For first-order Markov sources, a given $(\mathsf{D},\mathsf{P})$ and $\mathsf{R}\in \mathcal{R}(\mathsf{D},\mathsf{P})$, we have
\begin{IEEEeqnarray}{rCl}
\mathsf{R}+\log(\mathsf{R}+1)+5\in \mathcal{R}^{o}(\mathsf{D},\mathsf{P}) \subseteq \mathcal{R}(\mathsf{D},\mathsf{P}).\label{1-shot-thm-inner}
\end{IEEEeqnarray}
\end{theorem} 
\vspace{-0.2cm}\begin{IEEEproof} See Appendix~\ref{one-shot-app}.
\end{IEEEproof}

From Theorem~\ref{thm-one-shot-information-function}, it follows that  $\mathcal{R}(\mathsf{D},\mathsf{P})$ with overhead $\log(\mathsf{R}+1)+5$ is an inner bound to $\mathcal{R}^{o}(\mathsf{D},\mathsf{P})$. On the other hand, $\mathcal{R}(\mathsf{D},\mathsf{P})$ provides an outer bound to $\mathcal{R}^o(\mathsf{D},\mathsf{P})$. The two bounds match with each other in high rates. It can be shown that the overhead also vanishes in the  large-blocklength setting where multiple symbols are encoded at a time. In the remainder of this paper, we will approximate $\mathcal{R}^{o}(\mathsf{D},\mathsf{P})$ with $\mathcal{R}(\mathsf{D},\mathsf{P})$ and use the latter region for our analysis.

\begin{remark}(Encoded Representations): {The proof of the inner bound in Theorem~\ref{thm-one-shot-information-function}  in Appendix~\ref{one-shot-app} provides an operational interpretation to the auxiliary random variables $\mathsf{X}_r=(X_{r,1},X_{r,2},X_{r,3})$ defined in iRDP region in Definition~\ref{iRDP-region}. In particular, $X_{r,j}$ is a lossy version of the source sample $X_j$ generated by the encoder in step $j$. It is compressed and transmitted to the decoder at rate $R_j$ in~\eqref{rate1-3}. We refer to $\mathsf{X}_{r}$ as the {\em encoded representation} of the source $\mathsf{X}$. The Markov chains~\eqref{Markov-conditions1}--\eqref{Markov-conditions3} indicate that without loss of optimality, an encoded representation $X_{r,j}$ can be computed from the source $X_j$ and past reconstructions $X_{r,1},\ldots, X_{r,j-1}$ without using past source samples $X_1,\ldots,X_{j-1}$. }
\end{remark}

\begin{remark}(Deterministic Reconstructions):
 Note that the reconstruction functions generating $\hat{\mathsf{X}}$ in Definition~\ref{iRDP-region} are deterministic functions of the encoded representations (c.f.~\eqref{function-condition}). In particular, the shared randomness $K$ is not required in the reconstruction functions. However, as the proof of the inner bound of Theorem~\ref{thm-one-shot-information-function} illustrates, the shared randomness is required in the compression and construction of $X_{r,j}$.  Moreover, by following the arguments  in \cite{MaIshwar},  one can set the reconstruction function of the last frame to be identity.  Thus, in Definition~\ref{iRDP-region}, we have set $\hat{X}_3=X_{r,3}$ in~\eqref{function-condition} where $T=3$. In the sequel, for $T$ frames we will set $\hat{X}_T=X_{r,T}$. 

 \end{remark}

\begin{remark}
The result in Theorem~\ref{thm-one-shot-information-function} also holds for the PLF-FMD.  That is, one can replace the PLF in \eqref{perception-condition} and \eqref{perception} by $\xi_j(P_{X_j},P_{\hat{X}_j})$ and get a similar result (see Appendix~\ref{one-shot-app} for the justification).
\end{remark}

\subsection{Gauss-Markov Source Model: RDP Region}
\label{sec:GM}
In this section, we obtain  practical insights through the analysis of the special case of first-order Gauss-Markov sources.  We assume that $X_1 \sim {\mathcal N}(0, \sigma_1^2)$,
\begin{IEEEeqnarray}{rCl}
X_2 = \rho_1\frac{\sigma_2}{\sigma_1}X_1+N_1, \qquad
X_3 = \rho_2\frac{\sigma_3}{\sigma_2}X_2+N_2,\label{Gaus-def2}
\end{IEEEeqnarray}
where $N_j$ is independent of $X_j$ with mean zero and variance $(1-\rho_{j}^2)\sigma_{j+1}^2$ for $j=1,2$.    Note that the model extends naturally to the case of $T$ time-steps.
 The perception metric is assumed to be the Wasserstein-2 distance, i.e., $\phi_j(P_{X_1\ldots X_j},P_{\hat{X}_1\ldots \hat{X}_j}):=W_2^2(P_{X_1\ldots X_j},P_{\hat{X}_1\ldots\hat{X}_j})$. For the PLF-FMD, the perception metric is given by 
$\xi_j(P_{X_j},P_{\hat{X}_j}):=W_2^2(P_{X_j},P_{\hat{X}_j})$. %We remark that the Wasserstein-2 distance can also be replaced by the KL-divergence in most of our analysis. The common properties between these two measures are convexity and the fact that they both depend on only second-order statistics when restricted to Gaussian source model.

 The following result states that for Gaussian sources, jointly Gaussian reconstructions are optimal. Thus, for a given tuple $(\mathsf{D},\mathsf{P})$, the characterization of $\mathcal{R}(\mathsf{D},\mathsf{P})$ becomes computable.
\begin{theorem}\label{thm-Gaussian-optimality} For the Gauss-Markov source model, any  tuple $(\mathsf{R},\mathsf{D},\mathsf{P})\in \mathcal{C}_{\mathsf{RDP}}$ can be attained by a jointly Gaussian distribution over $\mathsf{X}_r$ and identity mappings for $\eta_j(\cdot)$ in Definition~\ref{iRDP-region}. 
\end{theorem}
\vspace{-0.2cm}\begin{IEEEproof} See Appendix~\ref{Gaussian-optimality-proof}.
\end{IEEEproof}
Generally, the optimized distribution in the above theorem may not admit a simple form. In the special case of $T=2$ frames, the optimal reconstructions are given in Appendix~\ref{Gaussian-optimality-proof}. To obtain practical insights, we consider various asymptotic operating regimes  and provide a detailed analysis in Appendix~\ref{extreme-rates-app} for the case of $T=2$ frames  and with $\sigma_1^2=\sigma_2^2$. A summary of these results is provided in Table~\ref{table-ach-recons} in the same Appendix.  We briefly summarize some of these results next.

\subsection{Gauss-Markov Source Model:  Extremal Rates}
\label{sec:GM-ext}
One of the key observations of this paper is that the choice of PLF has implication on the rate allocation across different frames. Specifically, first consider the case when both $R_1 {=} R_2$ are small i.e., $R_1{=}R_2{=}\epsilon$ (for small enough $\epsilon$). We discuss how each PLF  affects the reconstruction in the second step. In the first step, we note that reconstruction in both cases must be identical and of the form $\hat{X}_1^G{=}\sqrt{2\epsilon\ln 2}X_1{+}Z_1$ where $Z_1{\sim} \mathcal{N}(0,(1{-}2\epsilon\ln 2)\sigma^2)$ is independent of $X_1$; the resulting distortion is given by $D_1{=}2(1{-}\sqrt{2\epsilon\ln 2})\sigma^2$. However, the reconstructions in the second steps will be different for the two measures. For simplicity, we consider the extreme case when $\rho{=}1$ (i.e., when $X_2{=}X_1$). Here, the PLF-JD metric is required to preserve perfect correlation and thus has to set $\hat{X}_2^G{=}\hat{X}_1^G$ and results in $D_2{=}D_1$.  In other words, the decoder in the second step is unable to use any information transmitted in the second step as 0-PLF-JD enforces the stringent constraint $\hat{X}_2^G{=}\hat{X}_1^G$. In contrast, for the PLF-FMD metric, it can be shown that the reconstruction in the second step for $\rho{=}1$ reduces to $\hat{X}_2^G{=}\sqrt{2}\sqrt{2\epsilon\ln 2}X_1{+}Z_{\text{FMD}}$ and the associated distortion is given by $D_2{=}2(1{-}\sqrt{4\epsilon\ln 2})\sigma^2$, which is lower than PLF-JD. Extending this example to $T$ steps (with $\rho{=}1$), we note that PLF-JD will always be forced to output $\hat{X}_1$, while the reconstruction using PLF-FMD will successively improve. The following theorem formalizes this observation.

\begin{theorem}\label{error-permanence-thm} For sufficiently small $\epsilon$, let $R_j=\epsilon$ and suppose that  $\rho_j=\rho$ and $\sigma_j=\sigma$, for $j=1,\ldots, T$. The achievable distortions
$D_{\text{FMD},j}$  (for $0$-PLF-FMD) and $D_{\text{JD},j}$ (for $0$-PLF-JD) are: 
\begin{IEEEeqnarray}{rCl}
%D_{\text{FMD},1}&=& 2(1-\sqrt{2\epsilon\ln 2})\sigma^2,\\
D_{\text{FMD},j}=2(1-\Delta_{\text{FMD},j}\sqrt{2\epsilon\ln 2})\sigma^2,\quad  D_{\text{JD},j}=2(1-\Delta_{\text{JD},j}\sqrt{2\epsilon\ln 2})\sigma^2,
\end{IEEEeqnarray}
where  $\Delta_{\text{FMD},j}:=\sqrt{1+\rho^2\frac{(2\rho^2)^{j-1}-1}{2\rho^2-1}}$ and $\Delta_{\text{JD},j}:=\rho^{2(j-1)}+\mathbbm{1}\{j\geq 2\}\cdot\sqrt{1-\rho^2}(\sum_{i=0}^{j-2}\rho^{2i})$.
\end{theorem}
\vspace{-0.2cm}\begin{IEEEproof} See Appendix~\ref{comparison-app}.
\end{IEEEproof}
In particular, specializing to $\rho=1$,  $\Delta_{\text{FMD},j}=2^{\frac{j-1}{2}}$ and $\Delta_{\text{JD},j}=1$. This shows that the decrease in $D_{\text{FMD},j}$ is exponential at each step which implies the ability of decoder based on $0$-PLF-FMD in correcting mistakes and not propagating them in future reconstructions. However, as discussed previously the decoder which uses $0$-PLF-JD is stuck at $\hat{X}_j = \hat{X}_1$ when $\rho=1$ and results in no subsequent improvement in the distortion. We call this behaviour as \emph{permanence of error}.  This phenomenon is  magnified in the case when $R_1 \rightarrow 0$ and $R_2 \rightarrow \infty$, treated in Table~\ref{table-ach-recons} (Appendix~\ref{extreme-rates-app}) as the PLF-JD severely constrains the decoder to copy the previous noisy reconstruction while the flexibility provided by PLF-FMD reduces the distortion in the second step.

The case when $R_1 \rightarrow \infty$ and $R_2 = \epsilon$ treated in Table~\ref{table-ach-recons} in Appendix~\ref{extreme-rates-app} corresponds to the case when $X_1$  is sent at a sufficiently high rate (as is the case with some I-frames)  while $X_2$  is sent at a low rate. Naturally, we have $\hat{X}_1^G\approx X_1$ for both PLFs. On the other hand, we once again see a qualitatively different behaviour in the reconstruction of $X_2$. For the case of  $0$-PLF-FMD, we have $\hat{X}_2^G \approx (1- O(\epsilon))\hat{X}_1^G + O(\epsilon) X_2 $, i.e., the decoder essentially copies the previous frame with little contribution from the second step. In contrast, for the case of $0$-PLF-JD, it can be shown that $\hat{X}_2^G\approx \rho_1\hat{X}_1^G+ O(\sqrt{\epsilon}) X_2 +Z_{\text{JD}}$, where $Z_{\text{JD}}$ is independent Gaussian noise with variance close to $1-\rho^2$. We note that the PLF-JD metric prevents the decoder from simply ``copying'' the previous frame, but instead forces the decoder to generate a more diverse representation consistent with the joint distribution between the two frames.

\subsection{Universal Representations for Gauss-Markov Source Model}\label{universal-section}

In this section, we show that the Gauss-Markov source model admits universal encoded representations. Such representations can be transformed through appropriate reconstruction functions to achieve the entire DP rate region.  This is the counterpart of the result for general sources in Theorem~\ref{thm-universal-one-shot} where it is shown that the MMSE reconstructions can be transformed to some target reconstructions satisfying the $0$-PLF-FMD with at most a factor-2 increase in distortion. In contrast, we demonstrate that the Gauss-Markov model admits {\em exact universality} i.e., target reconstructions proposed in this section achieve all points in the iDP rate region. Interestingly, the transformation is linear with possibly some additive noise. First, we formalize the notion of universal representations.

\begin{definition}[iDP-Tradeoff] For a given rate tuple $\mathsf{R}$, the optimal iDP-tradeoff is the closure of the set of all tuples $(\mathsf{D},\mathsf{P})$ such that $(\mathsf{R},\mathsf{D},\mathsf{P})\in \mathcal{C}_{\mathsf{RDP}}$ and is denoted by $\mathcal{DP}(\mathsf{R})$.
\end{definition}
%Now, for a given rate, we introduce universal representations which can achieve each $(\mathsf{D},\mathsf{P})$ in DP-tradeoff $\mathcal{DP}(\mathsf{R})$. 
\begin{definition}[Universal Representation] A given encoded representation $\mathsf{X}_r$ is called universal with respect to rate tuple $\mathsf{R}$ if it satisfies the rate constraints \eqref{rate1-3} and the Markov chains in \eqref{Markov-conditions1}--\eqref{Markov-conditions3}
and for each $(\mathsf{D},\mathsf{P})\in \mathcal{DP}(\mathsf{R})$, there exists a reconstruction $\hat{\mathsf{X}}$ generated from  $P_{\hat{\mathsf{X}}|\mathsf{X}_r}$ achieving it.
\end{definition}

For the Gauss-Markov source model, we show that the MMSE reconstruction admits a universal representation. We consider the reconstruction $\hat{\mathsf{X}}_r$ that achieves minimum distortion in the $\mathcal{DP}(\mathsf{R})$ region. This point is explicitly characterized in Appendix~\ref{MMSE-representation-appendix}. Furthermore,  following Theorem~\ref{thm-Gaussian-optimality}, since the reconstruction functions $\eta_j(\cdot)$ are identity, the MMSE reconstruction is equivalent to MMSE representation i.e.,  $\hat{\mathsf{X}}_r=\mathsf{X}_r^{\text{RD}}$.The following theorem establishes that any point in $\mathcal{DP}(\mathsf{R})$ can be achieved from $\hat{\mathsf{X}}_r$.

\iffalse
\begin{definition}[MMSE Representations] For a given rate tuple $\mathsf{R}$, let $\mathsf{D}^\mathrm{min}$ denote the tuple of minimum achievable distortions which are explicitly characterized in Appendix~\ref{MMSE-representation-appendix}. Let $\mathsf{X}_r^{\text{RD}}$ be the associated representation such that the reconstruction $\hat{\mathsf{X}}= \mathsf{X}_r^{\text{RD}}$ achieves $\mathsf{D}^\mathrm{min}$ (see Appendix~\ref{MMSE-representation-appendix}) Then, $\mathsf{X}_r^{\text{RD}}$ is defined as the MMSE representation associated with rate tuple $\mathsf{R}$.
\end{definition}

We next establish that the MMSE representations are also universal representations.
 \fi
\begin{theorem}\label{universal-thm} For the Gauss-Markov source model and a given rate tuple $\mathsf{R}$ with strictly positive components, let the MMSE representation  be denoted as $\mathsf{X}_r^{\text{RD}}= (X^{\text{RD}}_{r,1},X^{\text{RD}}_{r,2},X^{\text{RD}}_{r,3})$. Let $(\mathsf{D},\mathsf{P})\in\mathcal{DP}(\mathsf{R})$ and let $\hat{\mathsf{X}}=(\hat{X}_1,\hat{X}_2,\hat{X}_3)$ be the corresponding reconstruction achieving it. Then  there exist $\kappa_1$, $\theta_1$, $\theta_{2}$, $\psi_{1}$, $\psi_{2}$ and $\psi_{3}$ and noise variables $(Z_1$, $Z_2$, $Z_3)$ independent of $(X^{\text{RD}}_{r,1},X^{\text{RD}}_{r,2},X^{\text{RD}}_{r,3})$,  which satisfy the following
\begin{IEEEeqnarray}{rCl}
\hat{X}_{1} = \kappa_{1}X^{\text{RD}}_{r,1}+Z_1,\quad
\hat{X}_{2} = \theta_{1}X^{\text{RD}}_{r,1}+\theta_2X^{\text{RD}}_{r,2}+Z_2,\quad
\hat{X}_{3} = \psi_{1}X^{\text{RD}}_{r,1}+\psi_{2}X^{\text{RD}}_{r,2}+\psi_{3}\hat{X}^{\text{RD}}_{r,3}+Z_3. \notag 
\end{IEEEeqnarray}
\end{theorem}
\begin{IEEEproof} See Appendix~\ref{Gaussian-universality-app}.
\end{IEEEproof}
The above theorem indicates that the MMSE representation can be linearly transformed to achieve any point in  $\mathcal{DP}(\mathsf{R})$. In general the MMSE representation may have to be degraded through additional noise terms. In the proof of Theorem~\ref{universal-thm} we identify conditions when such degradation is not needed.

As discussed in Appendix~\ref{Gaussian-universality-app}, Theorem~\ref{universal-thm} holds for both PLFs. This suggests the idea that one can train an encoder to get MMSE representations which are oblivious to the choice of PLF. Then, the decoder can generate a reconstruction which satisfies either of PLFs by simply applying a linear transformation to the MMSE representation. Thus, the task of choosing the right PLF can be assigned to the decoder based on distortion and perception requirements.  We conclude by noting that Appendix~\ref{Gaussian-example-app} provides an example where the coefficients in Theorem~\ref{universal-thm} can be computed explicitly.

\iffalse
We conclude this section with a simple example where the coefficients in Theorem~\ref{universal-thm} can be explicitly characterized. We assume that $\sigma_j^2=1$ and $\rho_j=\rho$ for $j=1,2,3$. Also suppose the rate tuple $\mathsf{R}$ is such that the minimum distortion tuple satisfies $D_j=D$ for $j=1,2,3$ and let $\mathsf{X}_r^\text{RD}$ be the associated representation.  Such an $\mathsf{R}$ is explicitly characterized in Appendix~\ref{Gaussian-example-app}.
Then a simple scaling of the form $\hat{X}_j = \frac{1}{\sqrt{1-D}}X_{r,j}^\text{RD}$ achieves the minimum distortion with zero perception loss for both the joint and marginal measures.
\fi

\section{Experimental Results}

%\begin{table}[t]
%  \centering
%  \begin{subtable}[b]{0.5\textwidth}
%    \centering
%    \begin{tabular}{|c|c|c|c|}
%    \hline
%     $R_2$& MMSE ($\times10^{-2}$) & 0-PLF FMD & 0-PLF JD \\ 
%    \hline
%    2 bits & 0.0109 \pm 0.0002 & 0.0171 \checkmark &  0.0203 \checkmark\\ 
%    \hline
%    4 bits & 0.0088 & 0.0135 \checkmark & 0.0144 \checkmark \\
%    \hline
%    6 bits & 0.0053 &  0.0075 \checkmark & 0.0078 \checkmark \\
%    \hline
%    \end{tabular}
%    \caption{Case 1: $R_1{=}\infty$ bits}
%    \label{tab:factora_factor2}
%  \end{subtable}
%  \quad
%  \begin{subtable}[b]{0.4\textwidth}
%    \centering
%    \begin{tabular}{|c|c|c|c|}
%    \hline
%     $R_2$& MMSE & 0-PLF FMD & 0-PLF JD \\ 
%    \hline
%    4 bits & 0.0121 &  0.0219 \checkmark&  0.0234 \checkmark\\ 
%    \hline
%    8 bits & 0.0102 &  0.0176 \checkmark& 0.0228 \xmark \\
%    \hline
%    12 bits & 0.0087 & 0.0143 \checkmark& 0.0226 \xmark\\
%    \hline
%    $\infty$ bits & 0.0 & 0.0 \checkmark&  0.0216 \xmark \\
%    \hline
%    \end{tabular}
%    \caption{Case 2{:} $R_1{=}12$ bits (case $R_1{=}\epsilon$).}
%    \label{tab:factorb_factor2}
%  \end{subtable}
%  \caption{Distortion of optimal reconstructions at each criteria for a) $R_1=\infty$ bits and b) $R_1=12$bits (\checkmark means factor of 2 holds and \xmark means otherwise.)}
%  \label{tab:factor_2}
%\end{table}

%\textcolor{red}{rewrite this to follow the structure of the theoretical analysis part}
%48-37
\begin{figure}[t]
\centering
\vspace{-0cm}
\begin{subfigure}[b]{0.48\textwidth}
  \includegraphics[width=\textwidth]{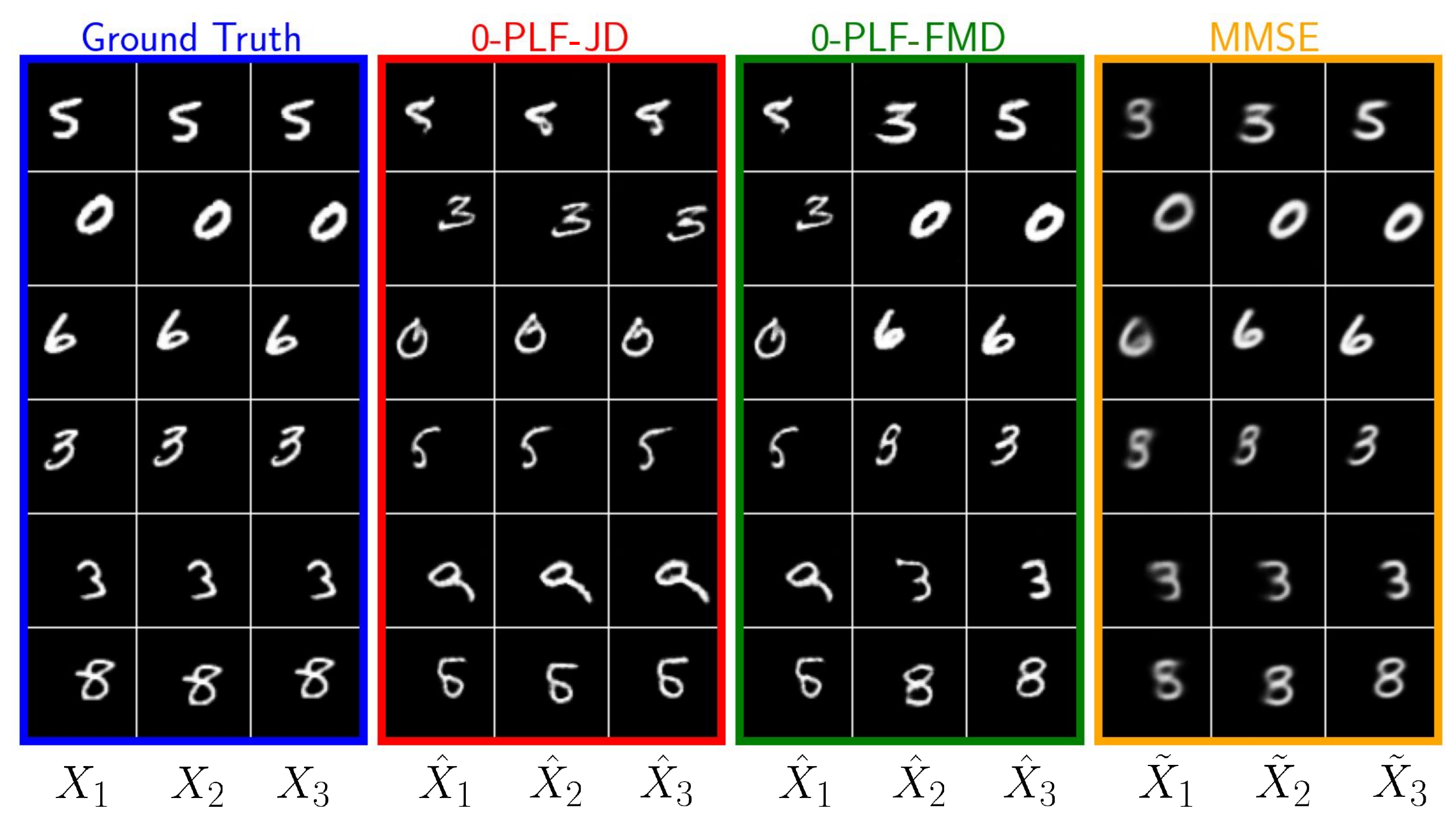}
  \caption{Ground-truth GOP and their optimal reconstructions with different PLFs for $R_1{=}R_2{=}R_3{=}12$ bits.}
  \label{figa:error permanence}
\end{subfigure}
\hspace{0.5cm}\begin{subfigure}[b]{0.37\textwidth}
   \includegraphics[width=\textwidth]{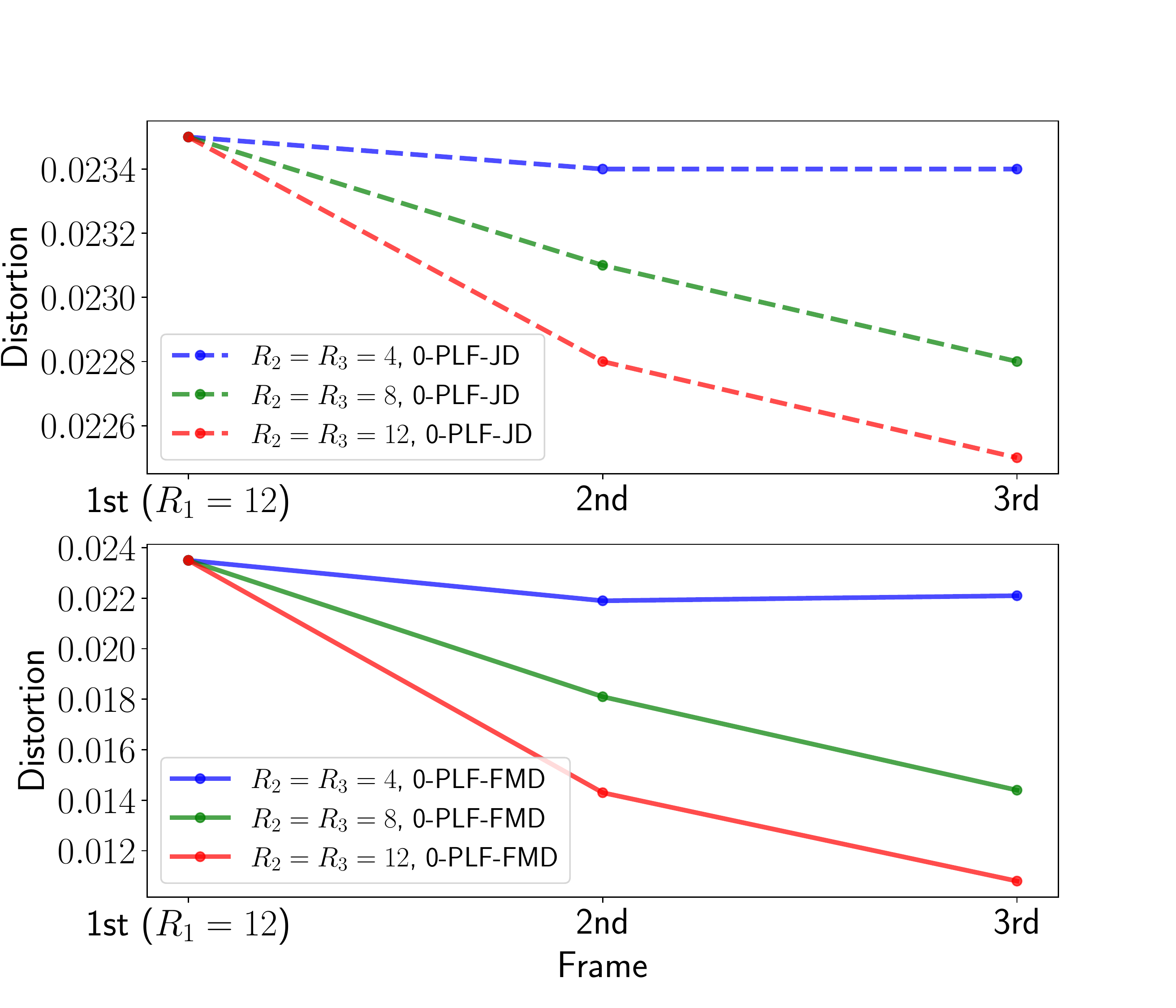}\vspace{-0.4cm}
    \caption{Distortion per frame $(X_i-\hat{X}_i)^2$ for different rates with $i=1,2,3$. }
    \label{figb:error permanence}
\end{subfigure}%\vspace{-0.3cm}
\caption{Permanence of Error Phenomenon. In (a), we visually compare the reconstructions. Note that $\hat{X}_1$ is the same for both 0-PLF-JD and 0-PLF-PMD. In (b), we show the framewise distortion for different $(R_2,R_3)$. }
\vspace{-0.4cm}
\label{fig:error permanence}
\end{figure}

 \vspace{-0.1cm}We conduct experiments on the MovingMNIST dataset \cite{srivastava2015unsupervised} (with 1 digit) using  Wasserstein GAN\cite{gulrajani2017improved}, to verify the implications of our theoretical claims to perceptual video compression. Additional results on the KTH \cite{schuldt2004recognizing} and  UVG datasets  are available in Appendices \ref{extra-experiments} and \ref{appendix-UVG-dataset}, respectively. Our compression network is built on the scale-space flow model \cite{agustsson2020scale} and conditional module \cite{li2021deep}. For a given rate and PLF, we obtain different distortion-perception tradeoff points by optimizing the weighted sum between distortion and perception losses. Details about the architecture and training procedure are available in the Appendix~\ref{exp-setup}.  The experimental setup is focused on validating our theory, rather than proposing state-of-the-art neural network architectures. Accordingly, we begin by (1) validating Theorems~\ref{thm-universal-one-shot} and~\ref{factor-two-thm}, which characterize the factor-of-two bounds on the distortion of 0-PLF reconstructions (2) empirically demonstrating the {\em error permanence} phenomenon of the PLF-JD in Section~\ref{sec:GM-ext} and (3) computing the distortion-perception tradeoff function experimentally as well as confirming that the MMSE reconstruction provide near universal representations, as motivated by the results in Section~\ref{universal-section}.

\begin{wraptable}{r}{0.55\textwidth}
\vspace{0cm}
\caption{Distortions of optimal reconstructions at different regime (\checkmark means factor of 2 holds and \xmark means otherwise). Distortion is scaled by $10^{-2}$.}\vspace{-0.2cm}
\begin{subtable}[l]{0.3\textwidth}
\centering
\small
\begin{tabular}{|c|c|c|c|}
    \hline
     $R_2$& MMSE  & 0-PLF-FMD & 0-PLF-JD \\ 
    \hline
    1  & $1.08 \pm 0.01$ & $1.74 \pm 0.02$ \checkmark &  $2.05 \pm 0.03$ \checkmark\\ 
    \hline
    2  & $0.88 \pm 0.01$ &  $1.39 \pm 0.03$\checkmark & $1.46 \pm 0.02$ \checkmark \\
    \hline
    3.17  & $0.53 \pm 0.01$ &  $0.76 \pm 0.01$ \checkmark & $0.79\pm 0.01$ \checkmark \\
    \hline
    \end{tabular}
\caption{Case 1: $R_1{=}\infty$ bits}\vspace{0cm}
    \label{tab:factora_factor2}
\end{subtable}%

\begin{subtable}[l]{0.3\textwidth}
\vspace{0cm}
\centering
\small
\begin{tabular}{|c|c|c|c|}
    \hline
     $R_2$& MMSE & 0-PLF-FMD & 0-PLF-JD \\ 
    \hline
    4 & $1.23\pm0.01$ &  $2.21 \pm0.04$\checkmark&  $2.36 \pm 0.04$ \checkmark\\ 
    \hline
    8  & $1.04 \pm 0.01$ &  $1.78\pm0.03$ \checkmark& $2.28  \pm 0.03$ $\text{\xmark}$ \\
    \hline
    12  & $0.89 \pm 0.02$ & $1.43\pm0.02$ \checkmark& $2.26 \pm 0.03$ $\text{\xmark}$ \\
    \hline
    $\infty$  & 0.0 & 0.0 \checkmark &  $2.18  \pm 0.02$ $\text{\xmark}$  \\
    \hline
    \end{tabular}
\caption{Case 2{:} $R_1{=}12$ bits$(\epsilon )$.}
    \label{tab:factorb_factor2}
\end{subtable}\vspace{-0.5cm}
  \label{tab:factor_2}
\end{wraptable}
\paragraph{}\vspace{-0.4cm}

As our first  experimental result in Table~\ref{tab:factor_2}, we validate the {\em factor of two bounds} in Theorems~\ref{thm-universal-one-shot} and~\ref{factor-two-thm}. We consider the compression of two frames $X_1$ and $X_2$ at rates $R_1$ and $R_2$ respectively. The compression of $X_1$ is performed without any prior reference and corresponds to the compression of the ``I-frame'', while the compression of $X_2$ corresponds to the ``P-frame'', using $X_1$ as the reference. We consider the cases when either $R_1 {=} \infty$ or $R_1{=}12$ bits, where the former corresponds to lossless compression of $X_1$ and the latter corresponds to the low rate regime (see Appendix~\ref{sec:just_mnist} for a justification).  The average distortion for the first frame when $R_1=12$  is $0.0124$ for the MMSE reconstruction and 
$0.0235$ for the $0$-PLF reconstruction, thus satisfying the factor of two bound.  In compression of $X_2$, we systematically vary the value of the rate $R_2{\in}\{4, 8, 12, \infty\}$.  Following Table \ref{tab:factorb_factor2}, for 0-PLF-JD reconstruction, only $R_2{=}4$ bits (low rate) satisfies the factor of two bounds as expected. Intuitively, even as more bits are acquired,  the 0-PLF-JD criteria actively restricts improving the reconstructions, resulting in persistently higher distortion. Even in the case when $R_2 = \infty$, the distortion remains non-zero as  the decoder is forced to maintain temporal consistency with $\hat{X}_1$. In contrast, for FMD, the factor of $2$ bound holds at all rates,  consistent with Theorem~\ref{thm-universal-one-shot}.

%\subsection{Permanence of Error}\label{Error Permanence}

\begin{figure}[h]
  \includegraphics[width=\textwidth,trim={0cm 0.5cm 0cm 0.05cm},clip]{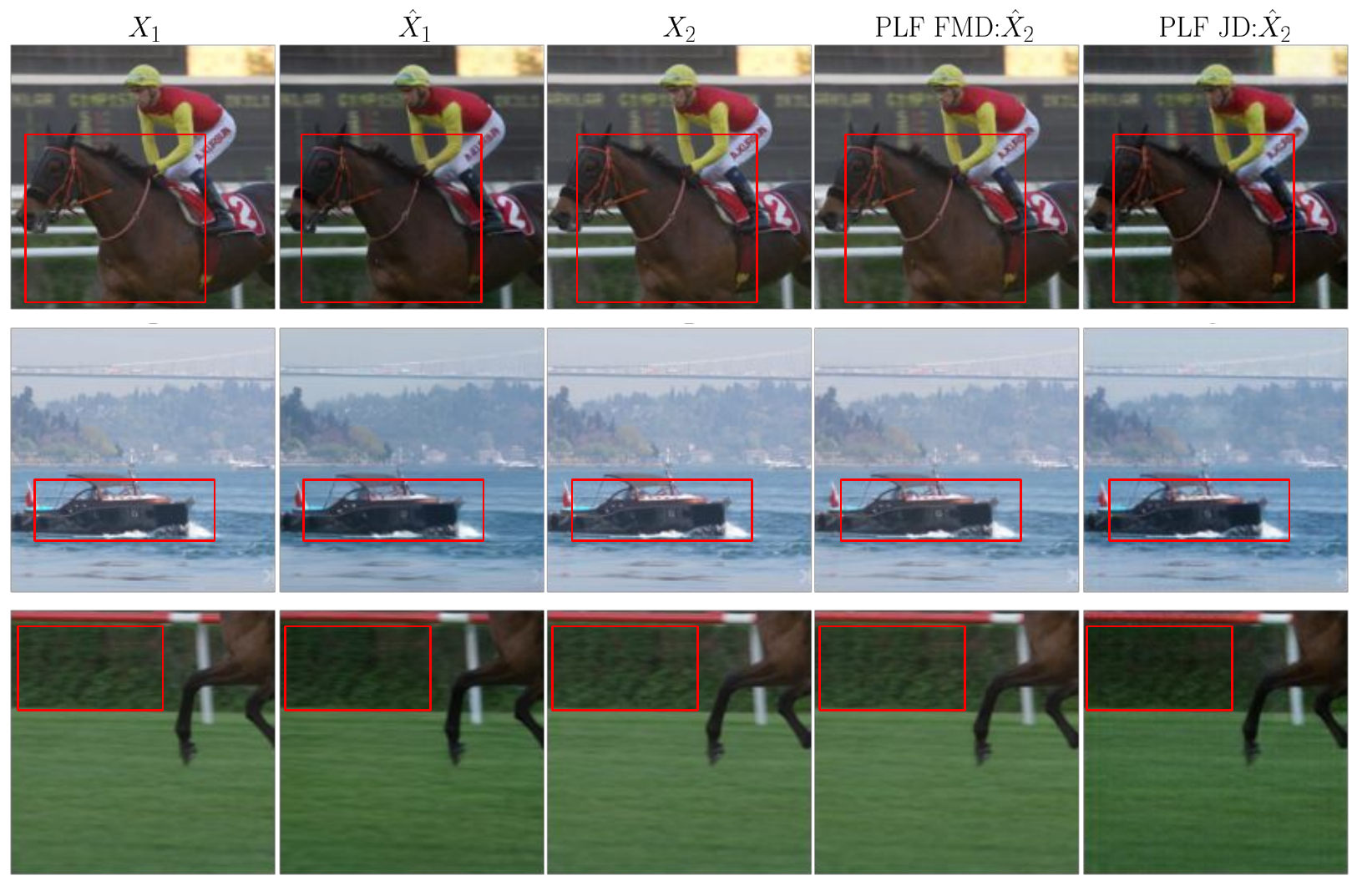}
  %\vspace{-0.95cm}
  \caption{Visualization of the error permanence phenomenon on the UVG dataset. The PLF-JD reconstructions propagate the flaws in the color tone from
the previous I-frame reconstruction while the PLF-FMD is able to fix these flaws. Compression rate for I-frame and P-frame are ~0.144bpp (low rate) and 4.632bpp (high rate) respectively.}
  \label{fig:UVG_main_papers}
\end{figure}

%\begin{figure}[t]
%\centering
%\vspace{-1cm}
%\begin{subfigure}[b]{0.45\textwidth}
%  \includegraphics[width=\textwidth]{experiment_figures/error_examples_v3.pdf}
%  \caption{Ground-truth GOP and their optimal reconstruction with different PLFs for $R_1{=}R_2{=}R_3{=}12$ bits.}
%  \label{figa:error permanence}
%\end{subfigure}
%\hspace{0.5cm}\begin{subfigure}[b]{0.32\textwidth}
%   \includegraphics[width=\textwidth]{experiment_figures/Frame-wise-distortion-3frames-v3.pdf}\vspace{-0.3cm}
%    \caption{Distortion per frame $(X_i-\hat{X}_i)^2$ for different rate with $i=1,2,3$. }
%    \label{figb:error permanence}
%\end{subfigure}\vspace{-0.2cm}
%\caption{Permanence of Error Phenomenon. In (a), we visually compare the reconstructions. Note that $\hat{X}_1$ is the same for both 0-PLF JD and 0-PLF PMD. In (b), we show the framewise distortion for different $(R_2,R_3)$. }
%\vspace{-0.5cm}
%\label{fig:error permanence}
%\end{figure}
In Fig.~\ref{fig:error permanence}, we present our experimental results  with a group of pictures (GOP) of size 3 (i.e. one I-frame followed by two P-frames). In Fig.~\ref{figa:error permanence}, we visualize sample reconstructions for MSE, 0-PLF-FMD and 0-PLF-JD cases when operating in the low-rate regime with $R_j = 12$ bits for $j=1,2,3$.  Note that given an incorrect digit reconstruction in $\hat{X}_1$, the decoder with 0-PLF-JD consistently produces incorrect digits (or content) while the 0-PLF-FMD gradually ``corrects'' it, which confirms the {\em error permanence phenomenon} discussed in the theoretical analysis in Section~\ref{sec:GM-ext} and Table~\ref{table-ach-recons}  in Appendix~\ref{extreme-rates-app}. We also plot the framewise distortion in Fig.~\ref{figb:error permanence} to show the difference in achievable distortion two perception metrics across different values for $R_2$ and $R_3$ as a function of the frame index. Consistent with Theorem~\ref{error-permanence-thm}, we note that the achievable distortion decreases much faster for 0-PLF-FMD than 0-PLF-JD for all selection of rates.  We also observe the same phenomenon on the large scale UVG dataset, as shown in Fig.~\ref{fig:UVG_main_papers}, when $R_1$ is small while $R_2$ is large.

\begin{figure}
\centering
\vspace{0cm}
\hspace{0cm}
\begin{subfigure}[t]{0.445\textwidth}
  \includegraphics[width=\textwidth,trim={0cm 0.2cm 0cm 0cm},clip]{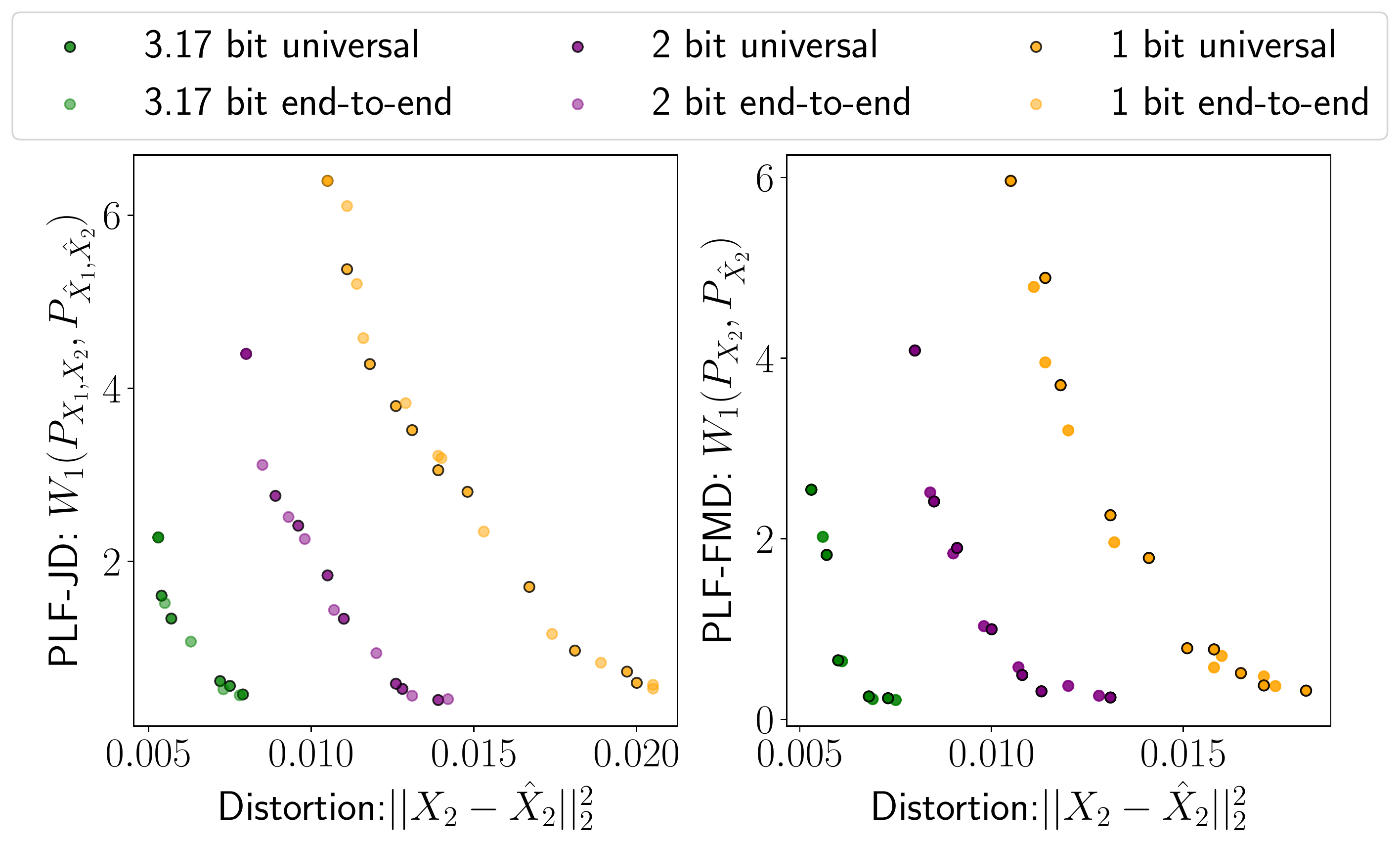}
  \caption{Case 1: $R_1{=}\infty$, $R_2{=}\{1, 2, 3.17\}$ bits}
  \label{fig:RDP_tradeoffa}
\end{subfigure}\hspace{0.2cm}
\begin{subfigure}[t]{0.50\textwidth}
  \includegraphics[width=\textwidth,trim={0cm 0.2cm 0cm 0cm},clip]{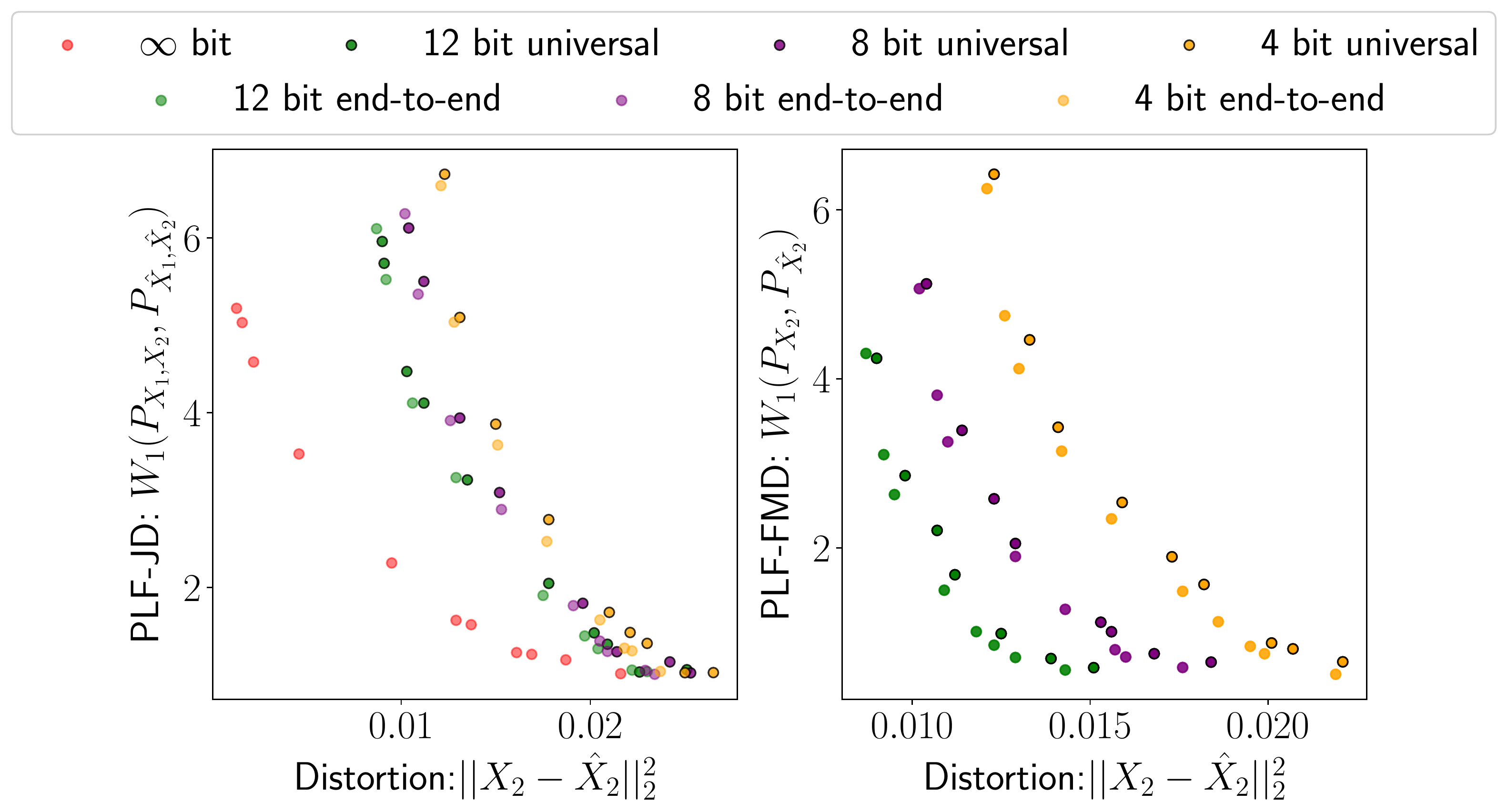}
  \caption{Case 2: $\!R_1{=}12$, $R_2{=}\{4, 8, 12, \infty\}$ bits}
  \label{fig:RDP_tradeoffb}
\end{subfigure}%\vspace{-0.15cm}
\caption{RDP tradeoff curves for end-to-end and universal models. We plot the tradeoff for the two regimes: $R_1{=}\infty$ and $R_1{=}\epsilon$ in (a) and (b) respectively. The universal and optimal curves are close to each other.}
\vspace{-0.4cm}
\label{fig:RDP_tradeoff}
\end{figure}

%-Theorem 5: 1 shot close to optimal for general sources.
%Theorem 8: General source: approximately the same for low rate (for both)
%Theorem 9: Gaussian source- optimal and universal the same
%\subsection{RDP Tradeoff and  Universal Representations}\label{universal_exp}
%As a reminder of our setup, we fix the encoder for MMSE reconstruction and train different decoders for different tradeoff pairs for universal representation. When $R_1{=}\infty$, we  send the respected ground truth frame $X_1$ losslessly to the decoder, which is now both a universal and optimal representation. We input to the P-frame decoder $\hat{X}_1$ that satisfies 0-PLF FMD in order to achieve optimal joint Wasserstein-1 distance. 
% Finally, when sending $X_1$ lossily, we reconstruct optimally $\hat{X}_1$ with 0-PLF FMD and use it for optimizing the joint Wasserstein-1 distance, since any imperfect perception $\hat{X}_1$ would yield a suboptimal solution for $P_{\hat{X}_1, \hat{X}_2}$. As a result, without the loss of generality, we use two frames to demonstrate the RDP tradeoff results since any suboptimal (joint distance) reconstruction in previous frames would also yield suboptimal tradeoffs.
In Fig.~\ref{fig:RDP_tradeoff}, we plot the tradeoff curves between distortion and perception  for the second reconstruction $\hat{X}_2$  for both optimal (end-to-end) and universal representations for two cases:  when $R_1{=}\infty$ and $R_1{=}12$ bits and for a range of values for $R_2$. In general, the curves for both universal and optimal representations are relatively close to each other at all rate regimes. The general shape of every curve is relatively similar with the exception of the PLF-JD metric in Fig.~\ref{fig:RDP_tradeoffb}, where the curves for different rates seemingly converge since increasing the rate does not significantly improve the distortion in this case as noted previously.  Finally, as the universal encoders are derived from MMSE solutions, these results imply that one can simply send the MMSE representation to the decoder and the user can flexibly change the DP tradeoff up to their requirements. We further note that even when the end-to-end model targets an operating point different from the MMSE reconstruction, the latter is still required to estimate the motion flow vectors best. The universal representation provides a natural way to reconstruct the MMSE reconstruction from the encoder output. In the plots of Fig.~\ref{fig:RDP_tradeoff}, we leverage on established universality results for I-frame compression in prior works ~\cite{Jun-Ashish2021} to construct the MMSE representation for motion compensation as we have a GOP of size $2$.

%We further note that even when the end-to-end model targets an operating point different from the MMSE reconstruction, the latter is required to provide the best estimates the motion flow vectors. %In the plots of Fig.~\ref{fig:RDP_tradeoff}, we leverage on established universality results for I-frame compression in prior works ~\cite{Jun-Ashish2021} to construct the MMSE representation for motion compensation as we have a GOP of size $2$. The case of longer GOPs is discussed in Appendix~\ref{sec:extra-experiment}.

%0.475 0.53

Finally in Appendix~\ref{sec:diversity}, we  consider the ability of the decoder to generate diverse reconstructions when operating under either PLF-JD or PLF-FMD. We focus on the case when  $X_1$ is transmitted losslessly and when $X_2$ is compressed at low rates. Consistent with the theoretical analysis in Section~\ref{sec:GM} and Table~\ref{table-ach-recons}  in Appendix~\ref{extreme-rates-app}, the decoder optimized for PLF-JD is capable of producing  diverse reconstructions by mimicking the actual motion between the frames. The PLF-FMD leads to reconstructions that are highly correlated and less desirable.

\vspace{-1em}

\section{Conclusions}

\vspace{-0.1cm}This work examines different perception loss functions for causal video coding, establishing its key theoretical properties such as the operational RDP region and universality principle. Our analysis highlights that while 0-PLF-JD reconstruction preserves temporal correlation, it is susceptible to the error permanence phenomenon. Moreover, our investigation of universality reveals that the encoder can transform the MMSE representation to other points on the DP tradeoffs, irrespective of the PLF. We suggest future research directions such as exploring region-based perceptual metrics \cite{pergament2022pim}, incorporating image-aware bits allocation, and leveraging conditional perception metric \cite{GAN}.

%\textcolor{red}{Future works:}
%\begin{itemize}
%    \item Region-based Perceptual Metric for Video Compression \cite{pergament2022pim}
%    \item Image aware bits allocation for joint perceptual metric.
%    \item Distribution-free metric.
%    \item Context adaptive bitrate and universality.
%\end{itemize}
\bibliographystyle{IEEEtran}
\bibliography{ICML_references}

% Generated by IEEEtran.bst, version: 1.14 (2015/08/26)
\begin{thebibliography}{10}
\providecommand{\url}[1]{#1}
\csname url@samestyle\endcsname
\providecommand{\newblock}{\relax}
\providecommand{\bibinfo}[2]{#2}
\providecommand{\BIBentrySTDinterwordspacing}{\spaceskip=0pt\relax}
\providecommand{\BIBentryALTinterwordstretchfactor}{4}
\providecommand{\BIBentryALTinterwordspacing}{\spaceskip=\fontdimen2\font plus
\BIBentryALTinterwordstretchfactor\fontdimen3\font minus
  \fontdimen4\font\relax}
\providecommand{\BIBforeignlanguage}[2]{{%
\expandafter\ifx\csname l@#1\endcsname\relax
\typeout{** WARNING: IEEEtran.bst: No hyphenation pattern has been}%
\typeout{** loaded for the language `#1'. Using the pattern for}%
\typeout{** the default language instead.}%
\else
\language=\csname l@#1\endcsname
\fi
#2}}
\providecommand{\BIBdecl}{\relax}
\BIBdecl

\bibitem{PSNR1}
E.~Agustsson, D.~Minnen, N.~Johnston, J.~Ball\'e, S.~J. Hwang, and G.~Toderici,
  ``Scale-space flow for end-to-end optimized video compression,'' in
  \emph{Proceedings of the IEEE Conference on Computer Vision and Pattern
  Recognition}, 2020, pp. 8503--8512.

\bibitem{PSNR2}
\BIBentryALTinterwordspacing
R.~Yang, Y.~Yang, J.~Marino, and S.~Mandt, ``Hierarchical autoregressive
  modeling for neural video compression,'' 2020. [Online]. Available:
  \url{https://arxiv.org/pdf/2010.10258.pdf}
\BIBentrySTDinterwordspacing

\bibitem{PSNR3}
\BIBentryALTinterwordspacing
O.~Rippel, A.~G. Anderson, K.~Tatwawadi, S.~Nair, C.~Lytle, and L.~Bourdev,
  ``Elf-vc: Efficient learned flexible-rate video coding,'' 2021. [Online].
  Available: \url{https://arxiv.org/abs/2104.14335}
\BIBentrySTDinterwordspacing

\bibitem{PSNR4}
J.~Li, B.~Li, and Y.~Lu, ``Deep contextual video compression,'' in
  \emph{Advances in Neural Information Processing Systems}, 2021, pp.
  18\,114--18\,125.

\bibitem{SSIM}
A.~Golinski, R.~Pourreza, Y.~Yang, G.~Sautiere, and T.~S. Cohen, ``Feedback
  recurrent autoencoder for video compression,'' in \emph{Proceedings of the
  Asian Conference on Computer Vision}, 2020.

\bibitem{zhang2021dvc}
S.~Zhang, M.~Mrak, L.~Herranz, M.~G. Blanch, S.~Wan, and F.~Yang, ``Dvc-p: Deep
  video compression with perceptual optimizations,'' in \emph{2021
  International Conference on Visual Communications and Image Processing
  (VCIP)}.\hskip 1em plus 0.5em minus 0.4em\relax IEEE, 2021, pp. 1--5.

\bibitem{video1}
F.~Mentzer, E.~Agustsson, J.~Ball\'e, D.~Minnen, N.~Johnston, and G.~Toderici,
  ``Neural video compression using gans for detail synthesis and propagation,''
  in \emph{European Conference on Computer Vision}, 2022.

\bibitem{yang2021perceptual}
R.~Yang, L.~Van~Gool, and R.~Timofte, ``Perceptual learned video compression
  with recurrent conditional gan,'' \emph{arXiv preprint arXiv:2109.03082},
  vol.~1, 2021.

\bibitem{video-joint}
\BIBentryALTinterwordspacing
V.~Veerabadran, R.~Pourreza, A.~Habibian, and T.~Cohen, ``Adversarial
  distortion for learned video compression,'' 2021. [Online]. Available:
  \url{https://arxiv.org/pdf/2004.09508.pdf}
\BIBentrySTDinterwordspacing

\bibitem{GAN3}
Y.~Wang, P.~Bilinski, F.~Bremond, and A.~Dantcheva, ``G$^3$an: Disentangling
  appearance and motion for video generation,'' in \emph{The IEEE Conference on
  Computer Vision and Pattern Recognition (CVPR)}, 2020.

\bibitem{blau2019rethinking}
Y.~Blau and T.~Michaeli, ``Rethinking lossy compression: The
  rate-distortion-perception tradeoff,'' in \emph{International Conference on
  Machine Learning}.\hskip 1em plus 0.5em minus 0.4em\relax PMLR, 2019, pp.
  675--685.

\bibitem{image-comp1}
E.~Agustsson, M.~Tschannen, F.~Mentzer, R.~Timofte, and L.~Van~Gool,
  ``Generative adversarial networks for extreme learned image compression,'' in
  \emph{Proceedings of the IEEE International Conference on Computer Vision},
  2019, pp. 221--231.

\bibitem{image-comp2}
J.~Ball\'e, V.~Laparra, and E.~P. Simoncelli, ``End-to-end optimized image
  compression,'' in \emph{5th International Conference on Learning
  Representations}, 2017.

\bibitem{image-comp3}
L.~Theis, W.~Shi, A.~Cunningham, and F.~Husz\'ar, ``Lossy image compression
  with compressive autoencoders,'' in \emph{5th International Conference on
  Learning Representations}, 2017.

\bibitem{image-comp4}
F.~Mentzer, E.~Agustsson, M.~Tschannen, R.~Timofte, and L.~V. Gool,
  ``Conditional probability models for deep image compression,'' in \emph{The
  IEEE Conference on Computer Vision and Pattern Recognition (CVPR)}, 2018.

\bibitem{Jun-Ashish2021}
G.~Zhang, J.~Qian, J.~Chen, and A.~Khisti, ``Universal
  rate-distortion-perception representations for lossy compression,'' in
  \emph{Advances in Neural Information Processing Systems}, 2021, pp.
  11\,517--11\,529.

\bibitem{gao2022flexible}
C.~Gao, T.~Xu, D.~He, Y.~Wang, and H.~Qin, ``Flexible neural image compression
  via code editing,'' \emph{Advances in Neural Information Processing Systems},
  vol.~35, pp. 12\,184--12\,196, 2022.

\bibitem{GAN}
F.~Mentzer, G.~Toderici, M.~Tschannen, and E.~Agustsson, ``High-fidelity
  generative image compression,'' in \emph{Advances in Neural Information
  Processing Systems}, 2020.

\bibitem{agustsson2019generative}
E.~Agustsson, M.~Tschannen, F.~Mentzer, R.~Timofte, and L.~V. Gool,
  ``Generative adversarial networks for extreme learned image compression,'' in
  \emph{Proceedings of the IEEE/CVF International Conference on Computer
  Vision}, 2019, pp. 221--231.

\bibitem{lu2020content}
G.~Lu, C.~Cai, X.~Zhang, L.~Chen, W.~Ouyang, D.~Xu, and Z.~Gao, ``Content
  adaptive and error propagation aware deep video compression,'' in
  \emph{Computer Vision--ECCV 2020: 16th European Conference, Glasgow, UK,
  August 23--28, 2020, Proceedings, Part II 16}.\hskip 1em plus 0.5em minus
  0.4em\relax Springer, 2020, pp. 456--472.

\bibitem{blau2018perception}
Y.~Blau and T.~Michaeli, ``The perception-distortion tradeoff,'' in
  \emph{Proceedings of the IEEE Conference on Computer Vision and Pattern
  Recognition}, 2018, pp. 6228--6237.

\bibitem{freirich2021theory}
D.~Freirich, T.~Michaeli, and R.~Meir, ``A theory of the distortion-perception
  tradeoff in wasserstein space,'' \emph{Advances in Neural Information
  Processing Systems}, vol.~34, pp. 25\,661--25\,672, 2021.

\bibitem{saldi2013randomized}
N.~Saldi, T.~Linder, and S.~Y{\"u}ksel, ``Randomized quantization and optimal
  design with a marginal constraint,'' in \emph{2013 IEEE International
  Symposium on Information Theory}.\hskip 1em plus 0.5em minus 0.4em\relax
  IEEE, 2013, pp. 2349--2353.

\bibitem{yan2021perceptual}
Z.~Yan, F.~Wen, R.~Ying, C.~Ma, and P.~Liu, ``On perceptual lossy compression:
  The cost of perceptual reconstruction and an optimal training framework,'' in
  \emph{International Conference on Machine Learning}.\hskip 1em plus 0.5em
  minus 0.4em\relax PMLR, 2021, pp. 11\,682--11\,692.

\bibitem{agustsson2022multi}
E.~Agustsson, D.~Minnen, G.~Toderici, and F.~Mentzer, ``Multi-realism image
  compression with a conditional generator,'' \emph{arXiv preprint
  arXiv:2212.13824}, 2022.

\bibitem{Wasserstein-distance}
V.~M. Panaretos and Y.~Zemel, \emph{An invitation to statistics in Wasserstein
  space}.\hskip 1em plus 0.5em minus 0.4em\relax Springer, 2020.

\bibitem{noisy}
A.~Makur, \emph{Information contraction and decomposition}.\hskip 1em plus
  0.5em minus 0.4em\relax PhD Thesis, MIT, 2019.

\bibitem{MaIshwar}
N.~Ma and P.~Ishwar, ``On delayed sequential coding of correlated sources,''
  \emph{IEEE Trans.~on Info.~Theory}, vol.~57, no.~6, pp. 3763--3782, 2011.

\bibitem{srivastava2015unsupervised}
N.~Srivastava, E.~Mansimov, and R.~Salakhudinov, ``Unsupervised learning of
  video representations using lstms,'' in \emph{International conference on
  machine learning}.\hskip 1em plus 0.5em minus 0.4em\relax PMLR, 2015, pp.
  843--852.

\bibitem{gulrajani2017improved}
I.~Gulrajani, F.~Ahmed, M.~Arjovsky, V.~Dumoulin, and A.~C. Courville,
  ``Improved training of wasserstein gans,'' \emph{Advances in neural
  information processing systems}, vol.~30, 2017.

\bibitem{schuldt2004recognizing}
C.~Schuldt, I.~Laptev, and B.~Caputo, ``Recognizing human actions: a local svm
  approach,'' in \emph{Proceedings of the 17th International Conference on
  Pattern Recognition, 2004. ICPR 2004.}, vol.~3.\hskip 1em plus 0.5em minus
  0.4em\relax IEEE, 2004, pp. 32--36.

\bibitem{agustsson2020scale}
E.~Agustsson, D.~Minnen, N.~Johnston, J.~Balle, S.~J. Hwang, and G.~Toderici,
  ``Scale-space flow for end-to-end optimized video compression,'' in
  \emph{Proceedings of the IEEE/CVF Conference on Computer Vision and Pattern
  Recognition}, 2020, pp. 8503--8512.

\bibitem{li2021deep}
J.~Li, B.~Li, and Y.~Lu, ``Deep contextual video compression,'' \emph{Advances
  in Neural Information Processing Systems}, vol.~34, pp. 18\,114--18\,125,
  2021.

\bibitem{pergament2022pim}
E.~Pergament, P.~Tandon, O.~Rippel, L.~Bourdev, A.~G. Anderson, B.~Olshausen,
  T.~Weissman, S.~Katti, and K.~Tatwawadi, ``Pim: Video coding using perceptual
  importance maps,'' \emph{arXiv preprint arXiv:2212.10674}, 2022.

\bibitem{LiElGamal}
C.~T. Li and A.~El~Gamal, ``Strong functional representation lemma and
  applications to coding theorems,'' \emph{IEEE Trans.~on Info.~Theory},
  vol.~64, no.~11, pp. 6967--6978, 2018.

\bibitem{Skoglund}
P.~Stavrou, M.~Skoglund, and T.~Tanaka, ``Sequential source coding for
  stochastic systems subject to finite rate constraints,'' \emph{IEEE Trans.~on
  Automatic Control}, vol.~67, no.~8, pp. 3822--3835, 2022.

\bibitem{Ashishproof}
A.~Khina, V.~Kostina, A.~Khisti, and B.~Hassibi, ``Tracking and control of
  gauss-markov processes over packet-drop channels with acknowledgments,''
  \emph{IEEE Trans.~on Cont. of Net. Systems}, vol.~6, no.~2, pp. 549--560,
  2019.

\bibitem{KimElGamal}
A.~El~Gamal and Y.~H. Kim, \emph{Network Information Theory}.\hskip 1em plus
  0.5em minus 0.4em\relax Cambridge University Press, 2011.

\bibitem{SVG}
E.~Denton and R.~Fergus, ``Stochastic video generation with a learned prior,''
  in \emph{International conference on machine learning}.\hskip 1em plus 0.5em
  minus 0.4em\relax PMLR, 2018, pp. 1174--1183.

\bibitem{kwon2019predicting}
Y.-H. Kwon and M.-G. Park, ``Predicting future frames using retrospective cycle
  gan,'' in \emph{Proceedings of the IEEE/CVF Conference on Computer Vision and
  Pattern Recognition}, 2019, pp. 1811--1820.

\bibitem{hong2019diversity}
S.~Hong, D.~Yang, Y.~Jang, T.~Zhao, and H.~Lee, ``Diversity-sensitive
  conditional generative adversarial networks,'' in \emph{7th International
  Conference on Learning Representations, ICLR 2019}.\hskip 1em plus 0.5em
  minus 0.4em\relax International Conference on Learning Representations, ICLR,
  2019.

\end{thebibliography}
\newpage
\appendix
\setcounter{theorem}{0}
\setcounter{definition}{0}

\section{Distortion Analysis for $0$-PLF-FMD}\label{PLF-FMD-factor-two}
Recall the definition of Wasserstein-2 distance \cite{Wasserstein-distance} as follows. For given distributions $P_{X_j}$ and $P_{\tilde{X}_j}$, let
\begin{IEEEeqnarray}{rCl}
W_2^2(P_{\tilde{X}_j},P_{X_j}):=\inf \mathbbm{E}[\|X_j-\tilde{X}_j\|^2],
\end{IEEEeqnarray}
where the infimum is over all joint distributions of $(X_j,\tilde{X}_j)$ with marginals $P_{X_j}$ and $P_{\tilde{X}_j}$.

\begin{theorem} The set $\Phi_{\mathsf{D}^0}(P_{\mathsf{M}|\mathsf{X}K})$ is characterized as follows:
\begin{IEEEeqnarray}{rCl}
&&\hspace{-0.2cm}\Phi_{\mathsf{D}^0}(P_{\mathsf{M}|\mathsf{X}K})=\{\mathsf{D}:D_j \geq \mathbbm{E}_P[\|X_j-\tilde{X}_j\|^2]+W_2^2(P_{\tilde{X}_j},P_{X_j}),\; j=1,2,3
\},
\end{IEEEeqnarray}
Furthermore, we also have that:
\begin{IEEEeqnarray}{rCl}&&\hspace{-0.5cm}\Phi_{\mathsf{D}^0}(P_{\mathsf{M}|\mathsf{X}K})\supseteq \{\mathsf{D}:  D_j \geq 2\mathbbm{E}_P[\|X_j-\tilde{X}_j\|^2],\;\;\; j=1,2,3
\},
\end{IEEEeqnarray}
i.e., minimum achievable distortion with $0$-PLF-FMD is at most twice the MMSE distortion.
\end{theorem}
\begin{IEEEproof} Define
\begin{IEEEeqnarray}{rCl} 
&&\hspace{-0.5cm}\mathcal{D}^0:= \{\mathsf{D}: D_j \geq \mathbbm{E}[\|X_j-\tilde{X}_j\|^2]+W_2^2(P_{\tilde{X}_j},P_{X_j}),\;\; j=1,2,3
\}.
\end{IEEEeqnarray}

First, we show that $\Phi_{\mathsf{D}^0}(P_{\mathsf{M}|\mathsf{X}K})\subseteq \mathcal{D}^0$. For any $\mathsf{D}\in \Phi_{\mathsf{D}^0}(P_{\mathsf{M}|\mathsf{X}K})$, there exists $\hat{\mathsf{X}}_{\mathsf{D}^0}=(\hat{X}_{D_1^0},\hat{X}_{D_2^0},\hat{X}_{D_3^0})$ jointly distributed with $(\mathsf{M},\mathsf{X},K)$ such that 
\begin{IEEEeqnarray}{rCl}
\mathbbm{E}[\|X_j-\hat{X}_{D_j^0} \|^2]&\leq& D_j,\qquad\;\; j=1,2,3, \\
P_{ X_j}&=&P_{\hat{X}_{D_j^0}}.
\end{IEEEeqnarray}
Then, for example,  the analysis for the second frame is as follows
\begin{IEEEeqnarray}{rCl}
D_2 &\geq & \mathbbm{E}[\|X_2-\hat{X}_{D_2^0}\|^2]\label{orthog-1st-step}\\
&=&\mathbbm{E}[\|(X_2-\tilde{X}_2)-(\hat{X}_{D_2^0}-\tilde{X}_2)\|^2]\\
&=& \mathbbm{E}[\|X_2-\tilde{X}_2\|^2]+\mathbbm{E}[\|\tilde{X}_2-\hat{X}_{D_2^0}\|^2]\label{orthog-step-next}\\
&\geq & \mathbbm{E}[\|X_2-\tilde{X}_2\|^2]+ W_2^2(P_{\tilde{X}_2},P_{\hat{X}_{D_2^0}})\\
&=& \mathbbm{E}[\|X_2-\tilde{X}_2\|^2]+ W_2^2(P_{\tilde{X}_2},P_{X_{2}}),\label{zero-perception-step}
\end{IEEEeqnarray}
 where \eqref{orthog-step-next} holds because  both $\tilde{X}_2$ and $\hat{X}_{D_2^0}$ are functions of $(M_{1},M_{2},K)$ and thus the MMSE $(X_2-\tilde{X}_2)$ is uncorrelated with $(\hat{X}_{D_2^0}-\tilde{X}_2)$; \eqref{zero-perception-step} follows because the $0$-PLF-FMD implies that $P_{\hat{X}_{D_2^0}}$ $=P_{X_{2}}$. Following similar steps for other frames, we get $\Phi_{\mathsf{D}^0}(P_{\mathsf{M}|\mathsf{X}K})\subseteq \mathcal{D}^0$.

Next, we show that $\mathcal{D}^0\subseteq \Phi_{\mathsf{D}^0}(P_{\mathsf{M}|\mathsf{X}K})$. Assume that $\mathsf{D}\in \mathcal{D}^0$.  Let $\hat{X}^*_1$ be an auxiliary random variable jointly distributed with $(M_{1}, K)$ such that it satisfies the following conditions
\begin{IEEEeqnarray}{rCl}
P_{\hat{X}^*_1}=P_{X_1},\label{1st-step-condition1}
\end{IEEEeqnarray} 
and  
\begin{IEEEeqnarray}{rCl}
 P_{\tilde{X}_1\hat{X}^*_1}= \arg \inf_{\substack{\bar{P}_{\tilde{X}_1\hat{X}^*_1}:\\\bar{P}_{\tilde{X}_1}=P_{\tilde{X}_1}\\\bar{P}_{\hat{X}^*_1}=P_{\hat{X}^*_1}}} \mathbbm{E}_{\bar{P}}[\|\tilde{X}_1-\hat{X}^*_1\|^2].\label{X1hat-ind10}
\end{IEEEeqnarray}
Moreover, let $\hat{X}^*_2$ be an auxiliary random variable jointly distributed with $(M_1,M_2,K)$ such that the following two conditions are satisfied
\begin{IEEEeqnarray}{rCl}
P_{\hat{X}^*_2}=P_{X_2},
\end{IEEEeqnarray} 
and  
\begin{IEEEeqnarray}{rCl}
 P_{\tilde{X}_2\hat{X}^*_2}= \arg \inf_{\substack{\bar{P}_{\tilde{X}_2\hat{X}^*_2}:\\\bar{P}_{\tilde{X}_2}=P_{\tilde{X}_2}\\\bar{P}_{\hat{X}^*_2}=P_{\hat{X}^*_2}}} \mathbbm{E}_{\bar{P}}[\|\tilde{X}_2-\hat{X}^*_2\|^2].\label{X2hat-ind}
\end{IEEEeqnarray}
Similarly, we define $\hat{X}_3^*$.
Now, notice that since $\mathsf{D}\in \mathcal{D}^0$, we have:
\begin{IEEEeqnarray}{rCl}
D_2 \geq \mathbbm{E}[\|X_2-\tilde{X}_2\|^2]+W_2^2(P_{\tilde{X}_2},P_{X_2}).
\end{IEEEeqnarray}
It then directly follows that
\begin{IEEEeqnarray}{rCl}
\mathbbm{E}[\|X_2-\hat{X}^*_2\|^2]
&=&\mathbbm{E}[\|X_2-\tilde{X}_2\|^2]+\mathbbm{E}[\|\tilde{X}_2-\hat{X}^*_2\|^2]\label{orthog-just-a}\\
&= & \mathbbm{E}[\|X_2-\tilde{X}_2\|^2]+W_2^2(P_{\tilde{X}_2},P_{\hat{X}^*_2})\label{cont-just-a}\\
&= & \mathbbm{E}[\|X_2-\tilde{X}_2\|^2]+W_2^2(P_{\tilde{X}_2},P_{X_2})\label{W2-final-a}\\
&\leq & D_2,
\end{IEEEeqnarray}
where
\begin{itemize}
\item \eqref{orthog-just-a} follows because $\tilde{X}_2$ and $\hat{X}_{2}^*$ are functions of $(M_1,M_{2},K)$ and thus the MMSE $(X_2-\tilde{X}_2)$ is uncorrelated with $(\hat{X}^*_{2}-\tilde{X}_2)$;
\item \eqref{cont-just-a} follows from \eqref{X2hat-ind};
\item \eqref{W2-final-a} follows because $P_{\hat{X}^*_2}=P_{X_2}$.
\end{itemize}
Following similar steps for other frames, we get $\mathsf{D}\in \Phi_{\mathsf{D}^0}(P_{\mathsf{X}_{\text{r}}|\mathsf{X}})$.

Now, notice that $W_2^2(P_{\tilde{X}_2},P_{X_2})\leq \mathbbm{E}[\|X_2-\tilde{X}_2\|^2]$ since the Wasserstein-2 distance takes the infimum over all possible joint distributions $(X_2,\tilde{X}_2)$, but the expectation in $\mathbbm{E}[\|X_2-\tilde{X}_2\|^2]$ is taken over the given $P_{X_2\tilde{X}_2}$. Thus, we get 
\begin{IEEEeqnarray}{rCl}
\mathbbm{E}[\|X_2-\tilde{X}_2\|^2]+W_2^2(P_{\tilde{X}_2},P_{X_2})\leq 2\mathbbm{E}[\|X_2-\tilde{X}_2\|^2].
\end{IEEEeqnarray}

This concludes the proof. 
\end{IEEEproof}

\section{Distortion Analysis for $0$-PLF-JD}\label{joint-PLF-app}
Let $\hat{X}_1^*$ be defined as in \eqref{1st-step-condition1}--\eqref{X1hat-ind10}. Moreover, let $\hat{X}^*_2$ be an auxiliary random variable jointly distributed with $(M_1,M_2,K)$ such that the following conditions are satisfied
\begin{IEEEeqnarray}{rCl}
P_{\hat{X}^*_2|\hat{X}^*_1=x_1}=P_{X_2|X_1=x_1},\qquad \forall x_1\in\mathcal{X}_1,\label{X2hatx1-cond1}
\end{IEEEeqnarray} 
and  
\begin{IEEEeqnarray}{rCl}
&&\hspace{-0.5cm} P_{\tilde{X}_2\hat{X}^*_2|\hat{X}^*_1=x_1}= 
\arg \hspace{-0.5cm}\inf_{\substack{\bar{P}_{\tilde{X}_2\hat{X}^*_2|\hat{X}^*_1=x_1}:\\\bar{P}_{\tilde{X}_2|\hat{X}^*_1=x_1}=P_{\tilde{X}_2|\hat{X}^*_1=x_1}\\\bar{P}_{\hat{X}^*_2|\hat{X}^*_1=x_1}=P_{\hat{X}^*_2|\hat{X}^*_1=x_1}}}\hspace{-0.5cm} \mathbbm{E}_{\bar{P}}[\|\tilde{X}_2-\hat{X}^*_2\|^2|\hat{X}^*_1=x_1],\qquad \forall x_1\in \mathcal{X}_1.\label{step81}
\end{IEEEeqnarray}
Then, the following result holds.
\begin{theorem}\label{Distortion-analysis-thm-JD} We have
\begin{IEEEeqnarray}{rCl}&&\Phi^{\text{joint}}_{\mathsf{D}^0}(P_{\mathsf{M}|\mathsf{X}K})\supseteq\{\mathsf{D}: D_1 \geq \mathbbm{E}[\|X_1-\tilde{X}_1\|^2]+W_2^2(P_{\tilde{X}_1},P_{X_1}),\nonumber\\
&&\hspace{0.7cm} D_2 \geq \mathbbm{E}[\|X_2-\tilde{X}_2\|^2]+\sum_{x_1}P_{X_1}(x_1)W_2^2(P_{\tilde{X}_2|\hat{X}^*_{1}=x_1},P_{X_2|X_1=x_1}),\nonumber\\&&
\hspace{0.7cm}D_3 \geq \mathbbm{E}[\|X_3-\tilde{X}_3\|^2]+\sum_{x_1,x_2}P_{X_1X_2}(x_1,x_2)W_2^2(P_{\tilde{X}_3|\hat{X}^*_1=x_1,\hat{X}^*_2=x_2},P_{X_3|X_1=x_1,X_2=x_2})
\}.\nonumber\\
\end{IEEEeqnarray}
\end{theorem}
\begin{IEEEproof} Define
\begin{IEEEeqnarray}{rCl}
&&\mathcal{D}^0_{\text{joint}}:=\{\mathsf{D}: D_1 \geq \mathbbm{E}[\|X_1-\tilde{X}_1\|^2]+W_2^2(P_{\tilde{X}_1},P_{X_1}),\nonumber\\
&&\hspace{0.7cm} D_2 \geq \mathbbm{E}[\|X_2-\tilde{X}_2\|^2]+\sum_{x_1}P_{X_1}(x_1)W_2^2(P_{\tilde{X}_2|\hat{X}^*_{1}=x_1},P_{X_2|X_1=x_1}),\nonumber\\&&
\hspace{0.8cm}D_3 \geq \mathbbm{E}[\|X_3-\tilde{X}_3\|^2]+\sum_{x_1,x_2}P_{X_1X_2}(x_1,x_2)W_2^2(P_{\tilde{X}_3|\hat{X}^*_1=x_1,\hat{X}^*_2=x_2},P_{X_3|X_1=x_1,X_2=x_2})
\}.\nonumber\\
\end{IEEEeqnarray}
Now, assume that $\mathsf{D}\in \mathcal{D}^0_{\text{joint}}$. For the first frame,  recall that $\hat{X}^*_1$ is an auxiliary random variable jointly distributed with $(M_1,K)$ such that it satisfies \eqref{1st-step-condition1}--\eqref{X1hat-ind10}.
From similar steps to \eqref{orthog-just-a}--\eqref{W2-final-a}, it then follows that
\begin{IEEEeqnarray}{rCl}
\mathbbm{E}[\|X_1-\hat{X}^*_1\|^2]&=& \mathbbm{E}[\|X_1-\tilde{X}_1\|^2]+W_2^2(P_{\tilde{X}_1},P_{X_1})\\
&\leq & D_1.
\end{IEEEeqnarray}

For the second frame, since $\mathsf{D}\in \mathcal{D}^0_{\text{joint}}$, we have:
\begin{IEEEeqnarray}{rCl}
D_2 &\geq& \mathbbm{E}[\|X_2-\tilde{X}_2\|^2]+\sum_{x_1} P_{X_1}(x_1)W_2^2(P_{\tilde{X}_2|X_1=x_1},P_{X_2|X_1=x_1}).
\end{IEEEeqnarray}
Recall that $\hat{X}^*_2$ is an auxiliary random variable jointly distributed with $(M_1,M_2,K)$ such that \eqref{X2hatx1-cond1}--\eqref{step81} hold. It then directly follows that
\begin{IEEEeqnarray}{rCl}
\mathbbm{E}[\|X_2-\hat{X}^*_2\|^2]
&=& \mathbbm{E}[\|X_2-\tilde{X}_2\|^2]+\mathbbm{E}[\|\tilde{X}_2-\hat{X}^*_2\|^2]\label{orthog-just}\\
&=&\mathbbm{E}[\|X_2-\tilde{X}_2\|^2]+\sum_{x_1}P_{\hat{X}^*_1}(x_1)\mathbbm{E}[\|\tilde{X}_2-\hat{X}^*_2\|^2|\hat{X}^*_1=x_1]\\
&= & \mathbbm{E}[\|X_2-\tilde{X}_2\|^2]+\sum_{x_1}P_{\hat{X}^*_1}(x_1)W_2^2(P_{\tilde{X}_2|\hat{X}^*_{1}=x_1},P_{\hat{X}^*_2|\hat{X}^*_{1}=x_1})\label{cont-just}\\
&= & \mathbbm{E}[\|X_2-\tilde{X}_2\|^2]+\sum_{x_1}P_{X_1}(x_1)W_2^2(P_{\tilde{X}_2|\hat{X}^*_{1}=x_1},P_{X_2|X_1=x_1}),\label{W2-final}
\end{IEEEeqnarray}
where
\begin{itemize}
\item \eqref{orthog-just} follows because $\tilde{X}_2$ and $\hat{X}_{2}^*$ are functions of $(M_1,M_2,K)$ and thus the MMSE $(X_2-\tilde{X}_2)$ is uncorrelated with $(\hat{X}^*_{2}-\tilde{X}_2)$,
\item \eqref{cont-just} follows from \eqref{step81},
\item \eqref{W2-final} follows because $P_{\hat{X}^*_1\hat{X}^*_2}=P_{X_1X_2}$.
\end{itemize}
Following similar steps for the third frame, we get $\mathsf{D}\in \Phi_{\mathsf{D}^0}(P_{\mathsf{M}|\mathsf{X}K})$. This concludes the proof. 
\end{IEEEproof}

\subsection{A Counterexample for Factor-Two Bound in Case of $0$-PLF-JD}\label{counterexample}
Assume that we have only two frames, i.e., $D_3\to \infty$. Let $M_1$ be independent of $X_1$  and $M_2=X_2$. Then, we have $\tilde{X}_1=\emptyset$ and $\tilde{X}_2=X_2$. Consider the achievable distortion region of Theorem~\ref{Distortion-analysis-thm-JD}. The distortion of the first step is given by the following
\begin{IEEEeqnarray}{rCl}
\mathbbm{E}[\|X_1-\tilde{X}_1\|^2]+W_2^2(P_{\tilde{X}_1},P_{X_1})=2\mathbbm{E}[X_1^2].\label{D1-joint}
\end{IEEEeqnarray}
For the second frame, we have
\begin{IEEEeqnarray}{rCl}
&&\hspace{-1cm}\mathbbm{E}[\|X_2-\tilde{X}_2\|^2]+\sum_{x_1}P_{X_1}(x_1)W_2^2(P_{\tilde{X}_2|\hat{X}^*_{1}=x_1},P_{X_2|X_1=x_1})\nonumber\\&=&\sum_{x_1}P_{X_1}(x_1)W_2^2(P_{X_2|\hat{X}^*_{1}=x_1},P_{X_2|X_1=x_1})\label{example-app1}\\
&=& \sum_{x_1}P_{X_1}(x_1)W_2^2(P_{X_2},P_{X_2|X_1=x_1}),\label{example-app2}
\end{IEEEeqnarray}
where \eqref{example-app1} follows because $\tilde{X}_2=X_2$ and \eqref{example-app2} follows because $X_2$ is independent of $\hat{X}_1^*$ ($M_1$ is independent of $X_1$, then $\hat{X}_1^*$, which is a function of $(M_1,K)$, would be independent of $X_1$ and hence independent of $X_2$).

Now, notice that the MMSE distortion of the second step is zero since $\tilde{X}_2=X_2$. However, the achievable distortion of the second step for the reconstruction satisfying $0$-PLF JD is given in \eqref{example-app2} which clearly does not satisfy the factor-two bound.

\section{Fixed Encoders Operating at  Low rate regime}\label{section-low-rate-app}
We consider the class of noisy encoders where the encoder distribution can be written as follows
\begin{IEEEeqnarray}{rCl}
 P^{\text{noisy}}_{X_j|M_1\ldots M_jK}=(1-\mu)P_{X_j}+\mu Q^{\text{noisy}}_{X_j|M_1\ldots M_jK},\qquad j=1,2,3.\label{noisy-representation2}
 \end{IEEEeqnarray}
where $\mu$ is a sufficiently small constant and the distribution $Q^{\text{noisy}}(\cdot)$ could be arbitrary conditional distribution with same marginal as $P_{X_j}$.

\begin{theorem}\label{low-rate-app}  For the class of encoders given by~\eqref{noisy-representation2}, we have
\begin{IEEEeqnarray}{rCl}&&\hspace{-0.5cm}\Phi^{\text{joint}}_{\mathsf{D}^0}(P^{\text{noisy}}_{\mathsf{M}|\mathsf{X}K})\supseteq \{\mathsf{D}:  D_j \geq 2\mathbbm{E}_{P^{\text{noisy}}}[\|X_j-\tilde{X}_j\|^2]+O(\mu),\;\;\; j=2,\ldots,3
\}.
\end{IEEEeqnarray}
\end{theorem}
\begin{IEEEproof} We analyze the distortion for the second frame. A similar argument holds for other frames.

Denote the reconstruction of the second step by $\hat{X}^*_2$ and consider the expected distortion. From a similar justification starting from \eqref{orthog-1st-step} and leading to \eqref{orthog-step-next}, we can write the distortion as follows
\begin{IEEEeqnarray}{rCl}
\mathbbm{E}[\|X_2-\hat{X}_2^*\|^2]&=&\mathbbm{E}[\|X_2-\tilde{X}_2\|^2]+\mathbbm{E}[\|\tilde{X}_2-\hat{X}_2^*\|^2].\label{distortion-optimal-low}
\end{IEEEeqnarray}
Now, we study the expected term $\mathbbm{E}[\|\tilde{X}_2-\hat{X}_2^*\|^2]$ as follows
\begin{IEEEeqnarray}{rCl}
\mathbbm{E}[\|\tilde{X}_2-\hat{X}_2^*\|^2]&=& \sum_{x_1}P_{\hat{X}_1^*}(x_1)\mathbbm{E}[\|\tilde{X}_2-\hat{X}_2^*\|^2|\hat{X}^*_1=x_1].\label{D-analysis}
\end{IEEEeqnarray}
In order to analyze the above expression, we first approximate the MMSE reconstruction $\tilde{X}_2$ as follows
\begin{IEEEeqnarray}{rCl}
\tilde{X}_2&=&\mathbbm{E}_{P^{\text{noisy}}}[X_2|M_1,M_2,K]\\
&=& (1-\mu)\mathbbm{E}_P[X_2]+\mu \mathbbm{E}_{Q^{\text{noisy}}}[X_2|M_1,M_2,K]\label{low-rate-MMSE-app}\\
&=& \mathbbm{E}[X_2]+O(\mu),\label{MMSE-approximation}
\end{IEEEeqnarray}
where \eqref{low-rate-MMSE-app} follows from~\eqref{noisy-representation2}. Moreover, notice that~\eqref{MMSE-approximation} implies that
\begin{IEEEeqnarray}{rCl}
\mathbbm{E}[\|X_2-\tilde{X}_2\|^2]&=&\mathbbm{E}[\|X_2-\mathbbm{E}[X_2]+\mu (\mathbbm{E}_{Q^{\text{noisy}}}[X_2|M_1,M_2,K]-\mathbbm{E}[X_2])\|^2]\\
&=& \mathbbm{E}[\|X_2-\mathbbm{E}[X_2]\|^2]+O(\mu).\label{MMSE-distortion-final-expression}
\end{IEEEeqnarray}
Next, consider the expected term in~\eqref{D-analysis} as follows
\begin{IEEEeqnarray}{rCl}
\sum_{x_1}P_{\hat{X}_1^*}(x_1)\mathbbm{E}[\|\tilde{X}_2-\hat{X}_2^*\|^2|\hat{X}^*_1=x_1] &=& \sum_{x_1}P_{\hat{X}_1^*}(x_1)\mathbbm{E}[\|\mathbbm{E}[X_2]-\hat{X}_2^*\|^2|\hat{X}^*_1=x_1]+O(\mu)\nonumber\\\label{D2-step0-low}\\
&=& \sum_{x_1}P_{\hat{X}_1^*}(x_1)\mathbbm{E}[\|\mathbbm{E}[X_2]-X_2\|^2|X_1=x_1]+O(\mu)\nonumber\\\label{D2-step1-low}\\
&=& \sum_{x_1}P_{X_1}(x_1)\mathbbm{E}[\|\mathbbm{E}[X_2]-X_2\|^2|X_1=x_1]+O(\mu)\nonumber\\\label{D2-step2-low}\\
&=& \mathbb{E}[\|\mathbbm{E}[X_2]-X_2\|^2]+O(\mu)\\
&=& \mathbb{E}[\|\tilde{X}_2-X_2\|^2]+O(\mu),\label{D2-step3-low}
\end{IEEEeqnarray}
where 
\begin{itemize}
\item \eqref{D2-step0-low} follows from~\eqref{MMSE-approximation};
\item \eqref{D2-step1-low} follows because the $0$-PLF-JD implies that $P_{\hat{X}^*_2|\hat{X}^*_1}=P_{X_2|X_1}$ and $\mathbbm{E}[X_2]$ is just a constant;
\item \eqref{D2-step2-low} follows from $0$-PLF-JD where $P_{\hat{X}^*_1}=P_{X_1}$;
\item \eqref{D2-step3-low} follows from \eqref{MMSE-distortion-final-expression}.
\end{itemize}
Considering \eqref{distortion-optimal-low} and \eqref{D2-step3-low}, we get
\begin{IEEEeqnarray}{rCl}
\mathbbm{E}[\|X_2-\hat{X}_2^*\|^2]&=&2\mathbbm{E}[\|X_2-\tilde{X}_2\|^2]+O(\mu).
\end{IEEEeqnarray}

The proof for the third frame follows similar steps.
\end{IEEEproof}

\section{Operational RDP Region}\label{one-shot-app}
Recall the definition of iRDP region $\mathcal{C}_{\mathsf{RDP}}$ for first-order Markov sources (Definition~\ref{iRDP-region}) as follows. It is the set of all tuples $(\mathsf{R},\mathsf{D},\mathsf{P})$ satisfying
\begin{IEEEeqnarray}{rCl}
R_1 &\geq & I(X_1;X_{r,1}),\label{MaIsh-rate1}\\
R_2 &\geq & I(X_2;X_{r,2}|X_{r,1}),\label{MaIsh-rate2}\\
R_3 &\geq & I(X_3;X_{r,3}|X_{r,1},X_{r,2}),\label{MaIsh-rate3}\\
D_j &\geq & \mathbbm{E}[\|X_j-\hat{X}_j\|^2], \qquad\qquad\;\;\; j=1,2,3,\label{MaIsh-distortion}\\
P_j &\geq & \phi_j(P_{X_1\ldots X_j},P_{\hat{X}_1\ldots\hat{X}_j}), \;\;\;\;\;\; j=1,2,3,\label{MaIsh-perception}
\end{IEEEeqnarray}
for auxiliary random variables $(X_{r,1},X_{r,2},X_{r,3})$ and $(\hat{X}_1,\hat{X}_2,\hat{X}_3)$ such that
\begin{IEEEeqnarray}{rCl}
\hat{X}_1&=& \eta_1(X_{r,1}), \;\;\hat{X}_2=\eta_2(X_{r,1},X_{r,2}),\;\;\hat{X}_3=X_{r,3},\label{Markov1-new}\\
X_{r,1}&\to& X_1\to (X_2,X_3),\label{Markov2-new}\\X_{r,2}&\to& (X_2,X_{r,1})\to (X_1,X_3),\label{Markov3-new}\\X_{r,3}&\to& (X_3,X_{r,1},X_{r,2})\to (X_1,X_2),\label{Markov4-new}
\end{IEEEeqnarray}
for some deterministic functions $\eta_1(.)$ and $\eta_2(.,.)$.

\begin{theorem} For first-order Markov sources, a given $(\mathsf{D},\mathsf{P})$ and $\mathsf{R}\in \mathcal{R}(\mathsf{D},\mathsf{P})$, we have
\begin{IEEEeqnarray}{rCl}
\mathsf{R}+\log(\mathsf{R}+1)+5\in \mathcal{R}^{o}(\mathsf{D},\mathsf{P}).\label{inner-bound-1-shot}
\end{IEEEeqnarray}
Moreover, the following holds:
\begin{IEEEeqnarray}{rCl}
\mathcal{R}^{o}(\mathsf{D},\mathsf{P})\subseteq \mathcal{R}(\mathsf{D},\mathsf{P}).\label{outer-bound-1-shot}
\end{IEEEeqnarray}
\end{theorem} 
\begin{IEEEproof}
\begin{figure*}[t]
\centering
  \includegraphics[scale=0.3]{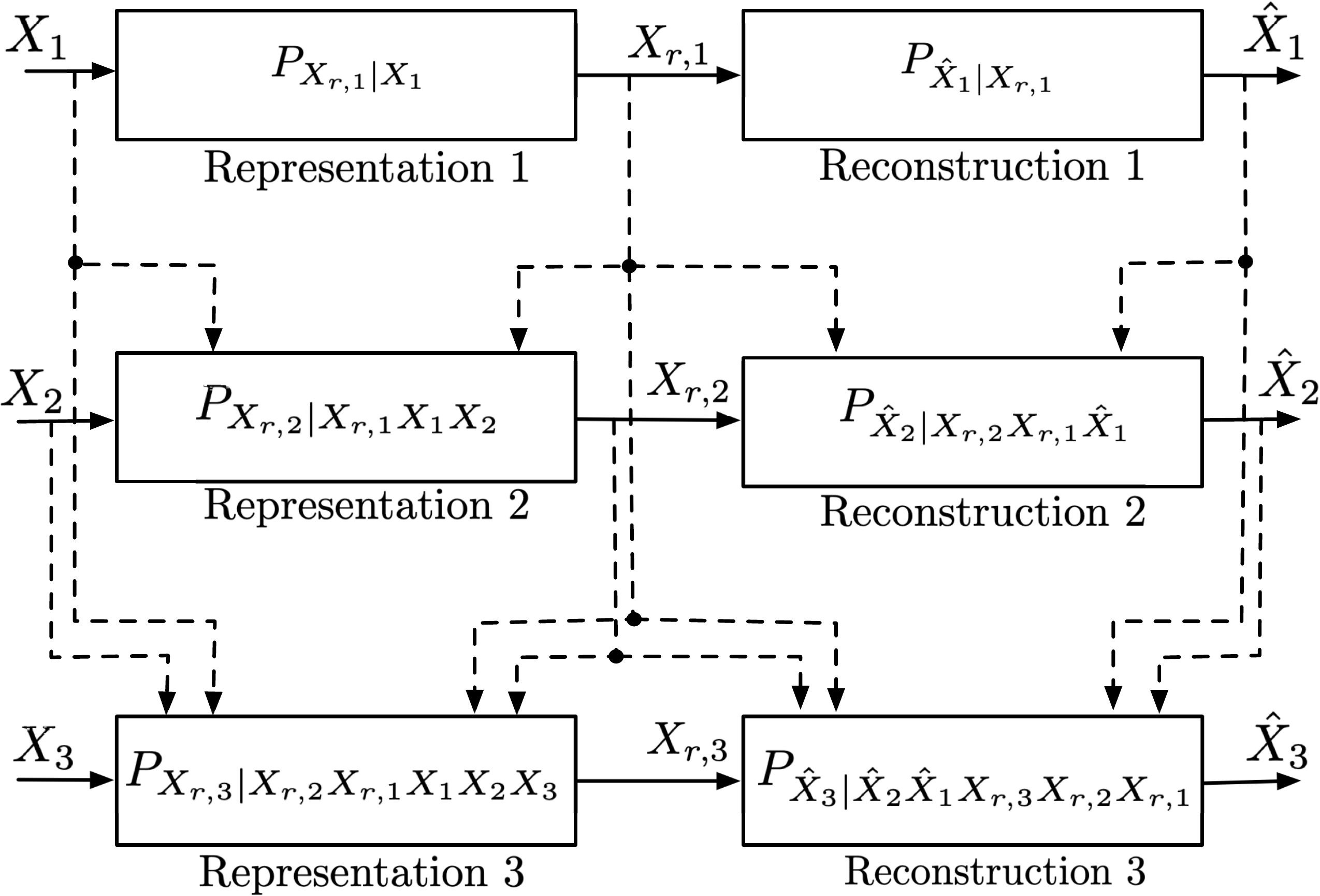}
\caption{Encoded representations and reconstructions of the iRDP region $\mathcal{C}_{\mathsf{RDP}}$.}
\label{enc-dec2}
\end{figure*}
Before stating the achievable scheme, we first discuss the strong functional representation lemma \cite{LiElGamal}. It states that for jointly distributed random variables $X$ and $Y$, there exists a random variable $U$ independent of $X$, and function $\phi$ such that $Y = \phi(X, U)$. Here, $U$ is not necessarily unique. The
strong functional representation lemma states further that there exists a $U$ which has information of $Y$ in the sense that
\begin{IEEEeqnarray}{rCl}
H(Y|U)\leq I(X;Y)+\log(I(X;Y)+1)+4.
\end{IEEEeqnarray}
Notice that the strong functional representation lemma can be applied conditionally. Given $P_{XY|W}$, we can represent $Y$ as a function of $(X,W,U)$ such that $U$ is independent of $(X,W)$ and 
\begin{IEEEeqnarray}{rCl}
H(Y|W,U)\leq I(X;Y|W)+\log(I(X;Y|W)+1)+4.
\end{IEEEeqnarray}

\underline{\textit{Proof of \eqref{inner-bound-1-shot} (Inner bound)}}:

For a given $(\mathsf{D},\mathsf{P})$ and $\mathsf{R}\in\mathcal{R}(\mathsf{D},\mathsf{P})$, let $\mathsf{X}_r=(X_{r,1},X_{r,2},X_{r,3})$ be jointly distributed with $\mathsf{X}=(X_1,X_2,X_3)$ where the Markov chains \eqref{Markov2-new}--\eqref{Markov4-new} hold and the rate constraints in \eqref{MaIsh-rate1}--\eqref{MaIsh-rate3} are satisfied such that there exist $(\hat{X}_1,\hat{X}_2,\hat{X}_3)$ for which distortion-perception constraints \eqref{MaIsh-distortion}--\eqref{MaIsh-perception} hold. Denote the joint distribution of $(\mathsf{X},\mathsf{X}_r,\hat{\mathsf{X}})$ by $P_{\mathsf{X}\mathsf{X}_r\hat{\mathsf{X}}}$ and notice that according to the Markov chains in \eqref{Markov2-new}--\eqref{Markov4-new}, it factorizes as the following
\begin{IEEEeqnarray}{rCl}
P_{\mathsf{X}\mathsf{X}_r\hat{\mathsf{X}}}&=&P_{X_1X_2X_3}\cdot P_{X_{r,1}|X_1}\cdot P_{X_{r,2}|X_{r,1}X_2}\cdot P_{X_{r,3}|X_{r,2}X_{r,1}X_3}\nonumber\\&&\hspace{0.5cm} \cdot\mathbbm{1}\{\hat{X}_1=g_1(X_{r,1})\}\cdot \mathbbm{1}\{\hat{X}_2=g_2(X_{r,1},X_{r,3})\}\cdot \mathbbm{1}\{\hat{X}_3=X_{r,3}\}.\label{super-P}
\end{IEEEeqnarray}

For an illustration of encoded representations $\mathsf{X}_r$ and reconstructions $\hat{\mathsf{X}}$ in $\mathcal{R}(\mathsf{D},\mathsf{P})$ which are induced by distribution $P_{\mathsf{X}\mathsf{X}_r\hat{\mathsf{X}}}$, see Fig.~\ref{enc-dec2}.

Now, we show that $\mathsf{R}+\log(\mathsf{R}+1)+5\in\mathcal{R}(\mathsf{D},\mathsf{P})$. The achievable scheme is as follows. Fix the joint distribution $P_{\mathsf{X}_r}$ according to~\eqref{super-P} which constructs the codebook, given by
\begin{IEEEeqnarray}{rCl}
P_{\mathsf{X}_r}=P_{X_{r,1}}P_{X_{r,2}|X_{r,1}}P_{X_{r,3}|X_{r,2}X_{r,1}}.\label{codebook-construction-P}
\end{IEEEeqnarray}

\begin{figure*}[t]
\centering
  \includegraphics[scale=0.32]{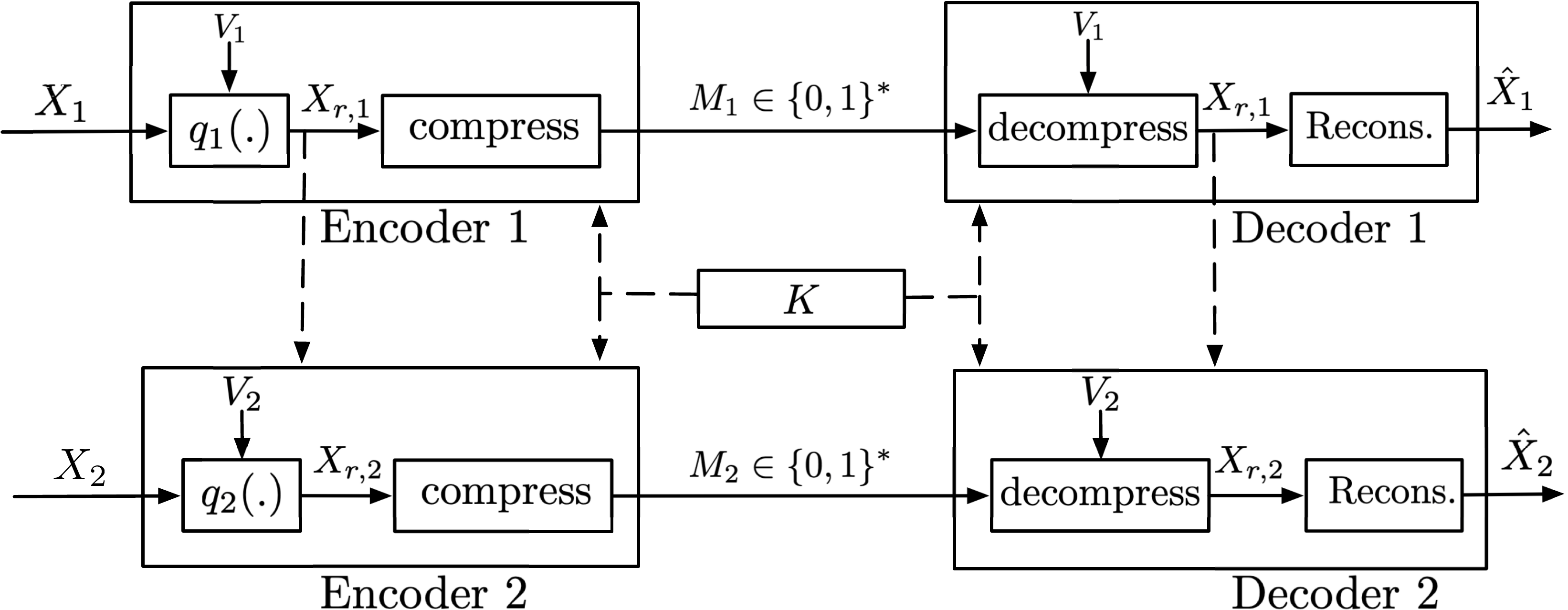}
\caption{Strong functional representation lemma for $T=2$ frames.}
\label{SFRL}
\end{figure*}

From the strong functional representation lemma \cite{LiElGamal}, we know that
\begin{itemize}
\item there exist a random variable $V_1$ independent of $X_1$ and a deterministic function $q_1$ such that $X_{r,1}=q_1(X_1,V_1)$ and
\begin{IEEEeqnarray}{rCl}
H(X_{r,1}|V_1)\leq I(X_1;X_{r,1})+\log(I(X_1;X_{r,1})+1)+4,
\end{IEEEeqnarray}
which means that the first encoder observes the source $X_1$ and applies the function $q_1$ to get $X_{r,1}$ whose  distribution needs to be preserved according to~\eqref{codebook-construction-P} (see Fig.~\ref{SFRL});
\item according to the conditional strong functional representation lemma, there exist a random variable $V_2$ independent of $(X_2,X_{r,1})$ and a deterministic function $q_2$ such that $X_{r,2}=q_2(X_{r,1},X_2,V_2)$ and
\begin{IEEEeqnarray}{rCl}
H(X_{r,2}|X_{r,1},V_2)\leq I(X_2;X_{r,2}|X_{r,1})+\log(I(X_2;X_{r,2}|X_{r,1})+1)+4.
\end{IEEEeqnarray}
At the second step, the representation $X_{r,1}$ is available at the second encoder. So, upon observing the source $X_2$, it applies the function $q_2$ to get $X_{r,2}$ whose conditional distribution given $X_{r,1}$ needs to be preserved according to~\eqref{codebook-construction-P} (see Fig.~\ref{SFRL});
\item according to the conditional strong functional representation lemma, there exist a random variable $V_3$ independent of $(X_3,X_{r,1},X_{r,2})$ and a deterministic function $q_3$ such that $X_{r,3}=q_3(X_{r,1},X_{r,2},X_3,V_3)$ and \begin{IEEEeqnarray}{rCl}H(X_{r,3}|X_{r,1},X_{r,2},V_3)\leq I(X_3;X_{r,3}|X_{r,1},X_{r,2})+\log(I(X_3;X_{r,3}|X_{r,1},X_{r,2})+1)+4.\nonumber\\\end{IEEEeqnarray}
\end{itemize}

Now, the encoding and decoding are as follows
\begin{itemize}
\item With $V_1$ available at all encoders and decoders, we can have a class of prefix-free binary codes indexed by $V_1$ with the expected codeword length not larger than $I(X_1;X_{r,1})+\log(I(X_1;X_{r,1})+1)+5$ to represent $X_{r,1}$, losslessly (see Fig.~\ref{SFRL}).
\item With $V_2$ available at the encoders and decoders, we can design a set of prefix-free binary codes indexed by $(V_2,X_{r,1})$ with expected codeword length not larger than $I(X_2;X_{r,2}|X_{r,1})+\log(I(X_2;X_{r,2}|X_{r,1})+1)+5$ to represent $X_{r,2}$, losslessly(see Fig.~\ref{SFRL}).
\item Similarly, one can represent $X_{r,3}$ losslessly with $V_3$ available at the third encoder and decoder.
\item The decoders can use functions $\hat{X}_1=\eta_1(X_{r,1})$, $\hat{X}_2=\eta_2(X_{r,1},X_{r,2})$ and $\hat{X}_3=X_{r,3}$ to get the reconstruction $\hat{\mathsf{X}}$.
\end{itemize}
This shows that $\mathsf{R}+\log(\mathsf{R}+1)+5\in \mathcal{R}^{o}(\mathsf{D},\mathsf{P})$.

\underline{\textit{Proof of \eqref{outer-bound-1-shot} (Outer Bound)}}:

For any $(\mathsf{D},\mathsf{P})$, $\mathsf{R}\in\mathcal{R}^{o}(\mathsf{D},\mathsf{P})$, shared randomness $K$, encoding functions $f_{j}\colon \mathcal{X}_1\times \ldots \times \mathcal{X}_j\times \mathcal{K}\to \mathcal{M}_{j}$ and decoding functions $g_{j} \colon \mathcal{M}_{1}\times \mathcal{M}_{2}\times \ldots \times \mathcal{M}_{j}\times \mathcal{K}\to \hat{\mathcal{X}}_j$ such that 
\begin{IEEEeqnarray}{rCl}
R_j\geq \mathbbm{E}[\ell(M_j)],\qquad j=1,2,3,
\end{IEEEeqnarray}
and
\begin{IEEEeqnarray}{rCl}
D_j &\geq & \mathbbm{E}[\|X_j-\hat{X}_j\|^2], \qquad\qquad\;\;\; j=1,2,3,\label{Dj-1-shot}\\
P_j &\geq & \phi_j(P_{X_1\ldots X_j},P_{\hat{X}_1\ldots\hat{X}_j}), \;\;\;\;\;\; j=1,2,3,\label{Pj-1-shot}
\end{IEEEeqnarray}
we lower bound the expected length of the messages. Define
\begin{IEEEeqnarray}{rCl}
X_{r,1}&:=&(M_1,K),\label{Xr1-def-1-shot}\\
X_{r,2}&:=& (M_1,M_2,K),\label{Xr2-def-1-shot}
\end{IEEEeqnarray}
and recall that according to the decoding functions, we have
\begin{IEEEeqnarray}{rCl}
\hat{X}_j=g_j(M_1,\ldots,M_j,K),\qquad j=1,2,3.
\end{IEEEeqnarray}
We can write
\begin{IEEEeqnarray}{rCl}
R_1\geq \mathbbm{E}[\ell(M_1)]&\geq & H(M_1|K)\\
&=&I(X_1;M_1|K)\\
&=&I(X_{1};M_1,K)\\
&=& I(X_{1};X_{r,1}).\label{1-shot-proof-R1}
\end{IEEEeqnarray}
Now, consider the following set of inequalities
\begin{IEEEeqnarray}{rCl}
R_2\geq \mathbbm{E}[\ell(M_2)]&\geq &H(M_2|M_1,K)\\
&=&I(X_1,X_2;M_2|M_1,K)\\
&=& I(X_1,X_{2};X_{2,r}|X_{r,1}).\label{1-shot-proof-R2}
\end{IEEEeqnarray}
Similarly, we have
\begin{IEEEeqnarray}{rCl}
R_3\geq \mathbbm{E}[\ell(M_3)]&\geq& H(M_3|M_1,M_2,K)\\
&=& I(X_1,X_2,X_3;M_3|M_1,M_2,K)\\
&\geq &I(X_1,X_2,X_{3};\hat{X}_{3}|X_{r,1},X_{r,2}).\label{1-shot-proof-R3}
\end{IEEEeqnarray}
Notice that the definitions in \eqref{Xr1-def-1-shot}--\eqref{Xr2-def-1-shot} imply the following Markov chains
\begin{IEEEeqnarray}{rCl}
X_{r,1}&\to& X_1\to (X_2,X_3),\label{1-shot-proof-Markov1}\\
X_{r,2} &\to & (X_1,X_2,X_{r,1})\to X_3.\label{1-shot-proof-Markov2}
\end{IEEEeqnarray}
On the other hand, the decoding functions of the first and second steps imply that
\begin{IEEEeqnarray}{rCl}
\hat{X}_1&=&g_1(M_1,K),\label{dec-1-proof-app}\\
\hat{X}_2&=&g_2(M_1,M_2,K),\label{dec-2-proof-app}
\end{IEEEeqnarray}
where together with definitions in~\eqref{Xr1-def-1-shot} and \eqref{Xr2-def-1-shot}, we can write
\begin{IEEEeqnarray}{rCl}
\hat{X}_1&=&g_1(M_1,K):=\eta_1(X_{r,1}),\label{1-shot-proof-g1}\\
\hat{X}_2&=&g_2(M_1,M_2,K):=\eta_2(X_{r,1},X_{r,2}),\label{1-shot-proof-g2}
\end{IEEEeqnarray}
such that $\eta_1(.)$ and $\eta_2(.,.)$ are deterministic functions.

Now, consider the fact that the set of constraints in \eqref{Dj-1-shot}--\eqref{Pj-1-shot}, \eqref{1-shot-proof-R1}, \eqref{1-shot-proof-R2}, \eqref{1-shot-proof-R3} with Markov chains in \eqref{1-shot-proof-Markov1}--\eqref{1-shot-proof-Markov2} and deterministic functions in \eqref{1-shot-proof-g1}--\eqref{1-shot-proof-g2} constitute an iRDP region, denoted by $\bar{\mathcal{C}}_{\mathsf{RDP}}$, which is the set of all tuples $(\mathsf{R},\mathsf{D},\mathsf{P})$ such that
\begin{IEEEeqnarray}{rCl}
R_1 &\geq & I(X_1;X_{r,1}),\label{rate0-ISh}\\
R_2 &\geq & I(X_1,X_2;X_{r,2}|X_{r,1}),\label{rate1-proof}\\
R_3 &\geq & I(X_1,X_2,X_3;\hat{X}_3|X_{r,1},X_{r,2}),\label{rate2-proof}\\
D_j &\geq & \mathbbm{E}[\|X_j-\hat{X}_j\|^2], \qquad\qquad\;\;\;\; j=1,2,3,\label{distortion-proof}\\
P_j &\geq & \phi_j(P_{X_1\ldots X_j},P_{\hat{X}_1\ldots\hat{X}_j}), \qquad j=1,2,3,\label{perception-proof}
\end{IEEEeqnarray}
for auxiliary random variables $(X_{r,1},X_{r,2})$ and $(\hat{X}_1,\hat{X}_2,\hat{X}_3)$ satisfying the following
\begin{IEEEeqnarray}{rCl}
\hat{X}_1&=& \eta_1(X_{r,1}), \;\;\hat{X}_2=\eta_2(X_{r,1},X_{r,2})\label{Markov1-Ish}\\
X_{r,1}&\to& X_1\to (X_2,X_3),\label{Markov2-Ish}\\
X_{r,2} &\to & (X_1,X_2,X_{r,1})\to X_3.\label{Markov3-Ish0}
\end{IEEEeqnarray}
for some deterministic functions $\eta_1(.)$ and $\eta_2(.,.)$.

 Comparing the two regions $\bar{\mathcal{C}}_{\mathsf{RDP}}$ and $\mathcal{C}_{\mathsf{RDP}}$, we identify the following differences. The Markov chain in \eqref{Markov2-new} is more restricted comparing to \eqref{Markov3-Ish0}. Moreover, the Markov chain \eqref{Markov3-new} does not exist in $\bar{\mathcal{C}}_{\mathsf{RDP}}$. The following lemma states that $\bar{\mathcal{C}}_{\mathsf{RDP}}=\mathcal{C}_{\mathsf{RDP}}$. Now, for a given $(\mathsf{D},\mathsf{P})$, let $\bar{\mathcal{R}}(\mathsf{D},\mathsf{P})$ denote the set of rate tuples $\mathsf{R}$ such $(\mathsf{R},\mathsf{D},\mathsf{P})\in \bar{\mathcal{C}}_{\mathsf{RDP}}$, then this lemma implies that $\bar{\mathcal{R}}(\mathsf{D},\mathsf{P})=\mathcal{R}(\mathsf{D},\mathsf{P})$ which completes the proof of the outer bound. Moreover, notice that the above proof only deals with the statistics of the representations and reconstructions and does not depend on the choice of the PLF. So, it holds for both PLF-FMD and PLF-JD. This concludes the proof.
 
 We conclude this section by the following lemma.

\begin{lemma}\label{ISh-thm} For first-order Markov sources, we have
\begin{IEEEeqnarray}{rCl}
\mathcal{C}_{\mathsf{RDP}}=\bar{\mathcal{C}}_{\mathsf{RDP}}.
\end{IEEEeqnarray}

\end{lemma}

\begin{IEEEproof} This result for the scenario without perception constraint has been similarly observed in \cite[Eq. (12)]{Skoglund}. The proof in this section is provided for completeness.

First, notice that the set of Markov chains in \eqref{Markov2-new}--\eqref{Markov4-new} is more restricted than the ones in \eqref{Markov2-Ish}--\eqref{Markov3-Ish0}, hence $\mathcal{C}_{\mathsf{RDP}}\subseteq \bar{\mathcal{C}}_{\mathsf{RDP}}$. Now, it remains to prove that $\bar{\mathcal{C}}_{\mathsf{RDP}}\subseteq \mathcal{C}_{\mathsf{RDP}}$. Consider the following facts
\begin{enumerate}
\item The distortion constraints in~\eqref{distortion-proof} depend only on the joint distribution of $(X_j, \hat{X}_j)$, and thus on the joint distribution of $(X_j,X_{r,1},\ldots,X_{r,j})$. So, imposing the Markov chain $X_{r,2}\to (X_2,X_{r,1})\to X_1$ does not affect the expected distortion $\mathbbm{E}[\|X_2-\hat{X}_2\|^2]$ since it does not depend on the joint distribution of $X_1$ with $(X_{r,1},X_{r,2},X_2)$. A similar argument holds for other frames;

\item The perception constraints in~\eqref{perception-proof} depend on the joint distributions  $P_{X_1\ldots X_j}$ and $P_{\hat{X}_1,\ldots,\hat{X}_j}$ (hence on $P_{X_{r,1}\ldots X_{r,j}}$). Thus, imposing $X_{r,2}\to (X_2,X_{r,1})\to X_1$ does not affect $\phi_2(P_{X_1X_2},P_{\hat{X}_1\hat{X}_2})$ since it does not depend on the joint distribution of $X_1$ with $(X_{r,1},X_{r,2},X_2)$. A similar argument holds for other frames;
\item Moreover, the rate constraints in \eqref{rate1-proof} and \eqref{rate2-proof} would be further lower bounded by 
\begin{IEEEeqnarray}{rCl}
R_2 &\geq & I(X_1,X_2;X_{r,2}|X_{r,1})\geq I(X_2;X_{r,2}|X_{r,1}),\label{lower-bound-Ish1}\\
R_3 &\geq & I(X_1,X_2,X_3;\hat{X}_3|X_{r,1},X_{r,2})\geq I(X_3;\hat{X}_3|X_{r,1},X_{r,2}).\label{lower-bound-Ish2}
\end{IEEEeqnarray}
Thus, the set of rate constraints is optimized by the set of Markov chains \eqref{Markov2-new}--\eqref{Markov4-new}.
\item The mutual information terms $I(X_1;X_{r,1})$, $I(X_2;X_{r,2}|X_{r,1})$ and $I(X_3;\hat{X}_3|X_{r,1},X_{r,2})$ depend on distributions  $P_{X_1X_{r,1}}$, $P_{X_{r,1}X_{r,2}X_2}$ and $P_{X_3\hat{X}_3X_{r,1}X_{r,2}}$, respectively. So, these distributions should be preserved by the set of Markov chains. The first two distributions are preserved by the choice of~\eqref{Markov1-new}--\eqref{Markov2-new}. 
Now, since we have first-order Markov sources (see Definition~\ref{1st-order-Markov}), preserving the joint distributions of $P_{X_{r,1}X_1}$ and $P_{X_{r,1}X_{r,2}X_2}$ is sufficient to preserve the distribution $P_{X_{r,1}X_{r,2}X_3}$. So, preserving the joint distribution of $P_{\hat{X}_3 X_{r,1}X_{r,2}}$ is sufficient to keep $I(X_3;\hat{X}_3|X_{r,1},X_{r,2})$ unchanged.
\end{enumerate}

Considering the above four facts, without loss of optimality, one can impose the following Markov chains
\begin{IEEEeqnarray}{rCl}
X_{r,1}&\to& X_1\to (X_2,X_3),\\
X_{r,2}&\to& (X_2,X_{r,1})\to (X_1,X_3),\\\hat{X}_3&\to& (X_3,X_{r,1},X_{r,2})\to (X_1,X_2).
\end{IEEEeqnarray}
This  concludes the proof for the PLF-JD. For the PLF-FMD, notice that the only difference is the second fact stated above. But, this also holds since the perception constraints depend only on $P_{X_j}$ and $P_{\hat{X}_j}$ (hence on $P_{X_{r,1}\ldots,X_{r,j}}$). 
\end{IEEEproof}

\end{IEEEproof}

\section{Gauss-Markov Source Model}\label{Gaussian-optimality-proof}
 We first remark that the Wasserstein-2 distance can also be replaced by the KL-divergence in most of the following analysis. The common properties between these two measures are convexity and the fact that they both depend on only second-order statistics when restricted to Gaussian source model.
\begin{theorem}\label{Gaussian-optimality-thm}
 For the Gauss-Markov source model, any  tuple $(\mathsf{R},\mathsf{D},\mathsf{P})\in \mathcal{C}_{\mathsf{RDP}}$ can be attained by a jointly Gaussian distribution over $(X_{r,1},X_{r,2},X_{r,3})$ and identity mappings for $\eta_j(\cdot)$ in Definition~\ref{iRDP-region}. 
\end{theorem}
\begin{IEEEproof} First, notice that a proof for the setting without perception constraint is provided in \cite{Ashishproof}. The following proof is different from \cite{Ashishproof} in some steps and also involves the perception constraint.

%For simplicity, we assume that $\sigma_1^2=\sigma_2^2=\sigma_3^2$. However, the proof can be easily extended to the case where the variances are not equal by simply scaling the choice of random variables. 
For a given tuple $(\mathsf{R},\mathsf{D},\mathsf{P})\in\mathcal{C}_{\mathsf{RDP}}$, let $X^*_{r,1}$, $X_{r,2}^*$, $\hat{X}^*_1=\eta_1(X_{r,1}^*)$, $\hat{X}^*_2=\eta_2(X_{r,1}^*,X_{r,2}^*)$ and $\hat{X}^*_3$ be random variables satisfying \eqref{Markov1-new}--\eqref{Markov3-new}. Let $P_{\hat{X}^G_1|X_1}$, $P_{\hat{X}^G_2|\hat{X}^G_1X_2}$ and $P_{\hat{X}^G_3|\hat{X}^G_1\hat{X}_2^GX_3}$ be jointly Gaussian distributions such that the following conditions are satisfied.  
\begin{IEEEeqnarray}{rCl}
\text{cov}(\hat{X}_1^G,X_1)&=&\text{cov}(\hat{X}^*_1,X_1),\label{cov-pres2}\\
\text{cov}(\hat{X}_1^G,\hat{X}^G_2,X_2)&=&\text{cov}(\hat{X}^*_1,\hat{X}^*_2,X_2),\label{cov-pres3}\\
\text{cov}(\hat{X}_1^G,\hat{X}^G_2,\hat{X}_3^G,X_3)&=&\text{cov}(\hat{X}^*_1,\hat{X}^*_2,\hat{X}^*_3,X_3),\label{cov-pres4}
\end{IEEEeqnarray}
In general, the Gaussian random variables which satisfy the constraints in \eqref{cov-pres2}--\eqref{cov-pres4} can be written in the following format
\begin{IEEEeqnarray}{rCl}
X_1&=& \nu \hat{X}_1^G+Z_1,\label{rv-relation1}\\
\hat{X}_2^G&=& \omega_1 \hat{X}_1^G+\omega_2X_2+Z_2,\label{rv-relation3}\\
\hat{X}_3^G&=& \tau_1 \hat{X}_1^G+\tau_2\hat{X}_2^G+\tau_3X_3+Z_3,\label{rv-relation-end}
\end{IEEEeqnarray}
for some real $\nu$, $\omega_1$, $\omega_2$, $\tau_1$, $\tau_2$, $\tau_3$ where $\hat{X}^G_1\sim \mathcal{N}(0,\sigma^2_{\hat{X}_1^G})$, $\hat{X}^G_2\sim \mathcal{N}(0,\sigma^2_{\hat{X}_2^G})$, $Z_1$, $Z_2$ and $Z_3$ are Gaussian random variables with zero mean and variances $\alpha_1^2, \alpha_2^2, \alpha_3^2$, independent of $\hat{X}_1^G$,  $(\hat{X}_1^G,X_2)$ and $(\hat{X}_1^G,\hat{X}_2^G,X_3)$, respectively.

We explicitly derive the coefficients $\nu$, $\omega_1$, $\omega_2$, $\tau_1$, $\tau_2$ and  $\tau_3$ in the following. Multiplying both sides of~\eqref{rv-relation1} by $\hat{X}_1^G$ and taking an expectation, we get
\begin{IEEEeqnarray}{rCl}
\mathbbm{E}[X_1\hat{X}_1^G]=\nu \sigma^2_{\hat{X}_1^G}.
\end{IEEEeqnarray}
According to~\eqref{cov-pres2}, the above equation can be written as follows
\begin{IEEEeqnarray}{rCl}
\mathbbm{E}[X_1\hat{X}_1^*]=\nu \mathbbm{E}[\hat{X}^{*2}_1].\label{equation-nu-1}
\end{IEEEeqnarray}
Multiplying both sides of~\eqref{rv-relation3} by the vector $[\hat{X}_1^G\;\; X_2]$ and taking an expectation, we have
\begin{IEEEeqnarray}{rCl}
[\mathbbm{E}[\hat{X}_1^G\hat{X}_2^G]\;\; \mathbbm{E}[X_2\hat{X}_2^G]]=[\omega_1\;\;\omega_2]\begin{pmatrix}\sigma^2_{\hat{X}_1^G} & \mathbbm{E}[X_2\hat{X}_1^G]\\ \mathbbm{E}[X_2\hat{X}_1^G] & \sigma_2^2\end{pmatrix}
\end{IEEEeqnarray}
Considering the fact that $\mathbbm{E}[X_2\hat{X}_1^G]=\rho_1\mathbbm{E}[X_1\hat{X}_1^G]$ and according to~\eqref{cov-pres3}, the above equation can be written as follows
\begin{IEEEeqnarray}{rCl}
[\mathbbm{E}[\hat{X}_1^*\hat{X}_2^*]\;\; \mathbbm{E}[X_2\hat{X}_2^*]]=[\omega_1\;\;\omega_2]\begin{pmatrix}\mathbbm{E}[\hat{X}_1^{*2}] & \rho_1\mathbbm{E}[X_1\hat{X}_1^*]\\ \rho_1\mathbbm{E}[X_1\hat{X}_1^*] & \sigma_2^2\end{pmatrix}.\label{equation-omega-2}
\end{IEEEeqnarray}
Similarly, multiplying both sides of~\eqref{rv-relation-end} by the vector $[\hat{X}_1^G\;\;\hat{X}_2^G\;\;X_3]$, taking an expectation and considering~\eqref{cov-pres4}, we get 
\begin{IEEEeqnarray}{rCl}
[\mathbbm{E}[\hat{X}_1^*\hat{X}_3^*]\;\;\mathbbm{E}[\hat{X}_2^*\hat{X}_3^*]\;\;\mathbbm{E}[X_3\hat{X}_3^*]]=[\tau_1\;\;\tau_2\;\;\tau_3]\begin{pmatrix}\mathbbm{E}[\hat{X}^{*2}_1] & \mathbbm{E}[\hat{X}^*_1\hat{X}^*_2] & \rho_1\rho_2\mathbbm{E}[X_1\hat{X}^*_1]\\\mathbbm{E}[\hat{X}_1^*\hat{X}^*_2] & \mathbbm{E}[\hat{X}^{*2}_2] & \rho_2\mathbbm{E}[X_2\hat{X}_2^*]\\\rho_1\rho_2\mathbbm{E}[X_1\hat{X}_1^*] & \rho_2\mathbbm{E}[X_2\hat{X}_2^*] & \mathbbm{E}[\hat{X}^{*2}_3]\end{pmatrix}.\nonumber\\\label{equation-tau-3}
\end{IEEEeqnarray}
Solving equations~\eqref{equation-nu-1},~\eqref{equation-omega-2} and~\eqref{equation-tau-3}, we get
\begin{IEEEeqnarray}{rCl}
\sigma^2_{\hat{X}_1^G}&=&\mathbb{E}[\hat{X}^{*2}_1],\\
\nu &=& \frac{\mathbb{E}[X_1\hat{X}^*_1]}{\mathbb{E}[\hat{X}^{*2}_1]},\\
\alpha_1^2&=& \sigma_1^2-\frac{\mathbb{E}[X_1\hat{X}^*_1]}{\mathbb{E}[\hat{X}^{*2}_1]},\\
\omega_1&=& \frac{\nu\rho_1\mathbbm{E}[\hat{X}_1^*\hat{X}_2^*]-\mathbbm{E}[X_2\hat{X}^*_2]}{\nu^2\rho_1^2\sigma^2_{\hat{X}_1^G}-\sigma_2^2},\\
\omega_2 &=& \frac{\nu\rho_1\sigma_{\hat{X}_1^G}^2\mathbbm{E}[X_2\hat{X}^*_2]-\sigma_2^2\mathbbm{E}[\hat{X}_1^*\hat{X}^*_2]}{\nu^2\rho_1^2\sigma^4_{\hat{X}_1^G}-\sigma_2^2\sigma^2_{\hat{X}_1^G}},\\
\alpha_2^2 &=& \mathbbm{E}[\hat{X}_2^{*2}]-\alpha_2^2\sigma^2_{\hat{X}_1^G}-\omega_2^2\sigma_2^2-2\omega_1\omega_2\rho_1\nu \sigma_{\hat{X}_1^G}^2.
\end{IEEEeqnarray}
For the third step, the coefficients and noise variance of \eqref{rv-relation-end} are given as follows
\begin{IEEEeqnarray}{rCl}
&&\hspace{-0.7cm}[\tau_1\;\;\tau_2\;\;\tau_3] \nonumber\\&=& [\mathbbm{E}[\hat{X}_1^*\hat{X}_3^*]\;\;\mathbbm{E}[\hat{X}_2^*\hat{X}_3^*]\;\;\mathbbm{E}[X_3\hat{X}_3^*]]\begin{pmatrix}\mathbbm{E}[\hat{X}^{*2}_1] & \mathbbm{E}[\hat{X}^*_1\hat{X}^*_2] & \rho_1\rho_2\mathbbm{E}[X_1\hat{X}^*_1]\\\mathbbm{E}[\hat{X}_1^*\hat{X}^*_2] & \mathbbm{E}[\hat{X}^{*2}_2] & \rho_2\mathbbm{E}[X_2\hat{X}_2^*]\\\rho_1\rho_2\mathbbm{E}[X_1\hat{X}_1^*] & \rho_2\mathbbm{E}[X_2\hat{X}_2^*] & \mathbbm{E}[\hat{X}^{*2}_3]\end{pmatrix}^{-1},\nonumber\\\\
\alpha_3^2 &= & \mathbbm{E}[\hat{X}^{*2}_3]-\tau_1^2\mathbbm{E}[\hat{X}^{*2}_1]-\tau_2^2\mathbbm{E}[\hat{X}^{*2}_2]-\tau_3^2\mathbbm{E}[X_3^2]\nonumber\\&&\hspace{1cm}-2\tau_1\tau_2\mathbbm{E}[\hat{X}_1^*\hat{X}_2^*]-2\tau_1\tau_3\rho_1\rho_2\mathbbm{E}[X_1\hat{X}_1^*]-2\tau_2\tau_3\rho_2\mathbbm{E}[X_2\hat{X}_2^*],
\end{IEEEeqnarray}
where $(.)^{-1}$ denotes the inverse of a matrix.

Now, we look at the rate constraints.

\underline{\textit{Rate Constraints}}:

Consider the rate constraint of the first step as follows
\begin{IEEEeqnarray}{rCl}
R_1&\geq &I(X_1;X^*_{r,1})\\
&=& H(X_1)-H(X_1|X^*_{r,1})\label{rate1-step1}\\
&\geq & H(X_1)-H(X_1|\hat{X}^*_1)\label{rate1-step3}\\
&=& H(X_1)-H(X_1-\mathbbm{E}[X_1|\hat{X}^*_1]|\hat{X}^*_1)\\
&\geq& H(X_1)-H(X_1-\mathbbm{E}[X_1|\hat{X}^*_1])\\
&\geq & H(X_1)-H(X_1-\mathbbm{E}[X_1|\hat{X}^G_1])\label{lemma2-step2}\\
&= & H(X_1)-H(X_1-\mathbbm{E}[X_1|\hat{X}^G_1]|\hat{X}^G_1)\label{rate1-step4}\\
&=& I(X_1;\hat{X}^G_1)
\end{IEEEeqnarray}
where 
\begin{itemize}
\item \eqref{rate1-step3} follows because $\hat{X}^*_1$ is a function of $X^*_{r,1}$;
    \item \eqref{lemma2-step2} follows because for a given covariance matrix in \eqref{cov-pres2}, the Gaussian distribution maximizes the differential entropy;
    \item \eqref{rate1-step4} follows because the MMSE is uncorrelated from the data and since the random variables are Gaussian, the MMSE would be independent of the data.
\end{itemize}
Next, consider the rate constraint of the second step as the following
\begin{IEEEeqnarray}{rCl}
R_2 &\geq & I(X_2;X^*_{r,2}|X^*_{r,1})\\
&=& H(X_2|X^*_{r,1})-H(X_2|X^*_{r,1},X^*_{r,2})\\
&\geq & H(X_2|X^*_{r,1})-H(X_2|\hat{X}^*_1,\hat{X}^*_2)\label{rate2-step1}\\
&\geq & H(X_2|X^*_{r,1})-H(X_2|\hat{X}^G_1,\hat{X}^G_2)\label{rate2-step4}\\
&=& H(\rho_1X_1+N_1|X^*_{r,1})-H(X_2|\hat{X}^G_1,\hat{X}^G_2)\\
&\geq & \frac{1}{2}\log \left(\rho_1^2 2^{2H(X_1|X^*_{r,1})}+2^{2H(N_1)}\right)-H(X_2|\hat{X}^G_1,\hat{X}^G_2)\label{rate2-step2}\\
&\geq  & \frac{1}{2}\log \left(\rho_1^2 2^{-2R_1}2^{2H(X_1)}+2^{2H(N_1)}\right)-H(X_2|\hat{X}^G_1,\hat{X}^G_2),\label{rate2-step3}
\label{2nd-step-lower-bound}
\end{IEEEeqnarray}
where
\begin{itemize}
    \item \eqref{rate2-step1} follows because $\hat{X}^*_1$ and $\hat{X}^*_2$ are deterministic functions of $X^*_{r,1}$ and $(X_{r,1}^*,X_{r,2}^*)$, respectively;
    \item \eqref{rate2-step4} follows because for a given covariance matrix in \eqref{cov-pres3}, the Gaussian distribution maximizes the differential entropy;
    \item \eqref{rate2-step2} follows from entropy power inequality (EPI) \cite[pp. 22]{KimElGamal};
    \item \eqref{rate2-step3} follows from \eqref{rate1-step1}.
\end{itemize}
Similarly, consider the rate constraint of the third frame as the following,
\begin{IEEEeqnarray}{rCl}
    R_3 &\geq & I(X_3;\hat{X}^*_3|X^*_{r,1},X^*_{r,2})\\
    &=& H(X_3|X^*_{r,1},X^*_{r,2})-H(X_3|X^*_{r,1},X^*_{r,2},\hat{X}^*_3)\\
    &\geq & H(X_3|X^*_{r,1},X^*_{r,2})-H(X_3|\hat{X}^*_1,\hat{X}^*_2,\hat{X}^*_3)\\
    &\geq & H(X_3|X^*_{r,1},X^*_{r,2})-H(X_3|\hat{X}^G_1,\hat{X}^G_2,\hat{X}^G_3)\\
    &=& H(\rho_2X_2+N_2|X^*_{r,1},X^*_{r,2})-H(X_3|\hat{X}^G_1,\hat{X}^G_2,\hat{X}^G_3)\\
    &\geq & \frac{1}{2}\log \left(\rho_2^2 2^{2H(X_2|X^*_{r,1},X^*_{r,2})}+2^{2H(N_2)}\right)-H(X_3|\hat{X}^G_1,\hat{X}^G_2,\hat{X}^G_3)\\
    &\geq  & \frac{1}{2}\log \left(\rho_2^2 2^{-2R_2}2^{2H(X_2|X^*_{r,1})}+2^{2H(N_2)}\right)-H(X_3|\hat{X}^G_1,\hat{X}^G_2,\hat{X}^G_3)\\
    &\geq& \frac{1}{2}\log \Big(\rho_1^2\rho_2^2 2^{-2R_1-2R_2}2^{2H(X_1)}+\rho_2^2 2^{-2R_2}2^{2H(N_1)}+2^{2H(N_2)}\Big)-H(X_3|\hat{X}_1^G,\hat{X}_2^G,\hat{X}^G_3)\nonumber\\\label{rate3-Gaussian-proof-opt}
\end{IEEEeqnarray}

Next, we look at the distortion constraint.

\underline{\textit{Distortion Constraint}}: The choices in \eqref{cov-pres2}--\eqref{cov-pres4} imply that
\begin{IEEEeqnarray}{rCl}
D_j\geq \mathbbm{E}[\|X_j-\hat{X}^*_{j}\|^2]=\mathbbm{E}[\|X_j-\hat{X}^G_{j}\|^2],\qquad j=1,2,3.\label{distortion-Gaussian-proof-opt}
\end{IEEEeqnarray}

Finally, we look at the perception constraint

\underline{\textit{Perception Constraint}}:

Define the following distribution
\begin{IEEEeqnarray}{rCl}
P_{U^*V^*}:= \arg\inf_{\substack{\tilde{P}_{UV}:\\\tilde{P}_U=P_{X_1}\\\tilde{P}_{V}=P_{\hat{X}^*_{1}}}}\mathbbm{E}_{\tilde{P}}[\|U-V\|^2].\label{step4}
\end{IEEEeqnarray}
Now, define $P_{U^GV^G}$ to be a Gaussian joint distribution with the following covariance matrix \begin{IEEEeqnarray}{rCl}\text{cov}(U^G,V^G)=\text{cov}(U^*,V^*).\label{covariance-G-match}\end{IEEEeqnarray}

Then, we have the following set of inequalities: 

\begin{IEEEeqnarray}{rCl}
P_1\geq W_2^2(P_{X_1},P_{\hat{X}^*_1})
&=& \inf_{\substack{\tilde{P}_{UV}:\\\tilde{P}_U=P_{X_1}\\\tilde{P}_{V}=P_{\hat{X}^*_{1}}}}\mathbbm{E}_{\tilde{P}}[\|U-V\|^2]\label{step5}\\
&=& \mathbbm{E}[\|U^*-V^*\|^2]\label{P-step}\\
&=& \mathbbm{E}[\|U^G-V^G\|^2]\label{step6}\\
&\geq &  W_2^2(P_{U^G},P_{V^G})\\
&=& \inf_{\substack{\hat{P}_{UV}:\\\hat{P}_U=P_{U^G}\\\hat{P}_V=P_{V^G}}}\mathbbm{E}_{\hat{P}}[\|U-V\|^2]\\
&=& \inf_{\substack{\hat{P}_{UV}:\\\hat{P}_U=P_{X_1}\\\hat{P}_V=P_{\hat{X}^G_1}}}\mathbbm{E}_{\hat{P}}[\|U-V\|^2]\label{step7}\\
&=& W_2^2(P_{X_1},P_{\hat{X}^G_1}),\label{step8}
\end{IEEEeqnarray}
where
\begin{itemize}
\item \eqref{P-step} follows from the definition in \eqref{step4};
\item \eqref{step6} follows from~\eqref{covariance-G-match} which implies that $(U^*,V^*)$ and $(U^G,V^G)$ have the same second-order statistics;
\item \eqref{step7} follows because $P_{V^G}=P_{\hat{X}^G_1}$ which is justified in the following. First, notice that both $P_{V^G}$ and $P_{\hat{X}^G_1}$ are Gaussian distributions. Denote the variance of $V^G$ by $\sigma_{V^G}^2$ and recall that the variance of $\hat{X}^G_1$ is denoted by $\sigma^2_{\hat{X}_1^G}$. According to~\eqref{covariance-G-match}, $\sigma_{V^G}^2$ is equal to the variance of $V^*$. Also, from~\eqref{step4}, we know that $P_{V^*}=P_{\hat{X}^*_1}$, hence the variances of $V^*$ and $\hat{X}^*_1$ are the same. On the other side, according to \eqref{cov-pres2}, we know that the variance of $\hat{X}^*_1$ is equal to $\sigma^2_{\hat{X}_1^G}$. Thus, we conclude that $\sigma^2_{\hat{X}_1^G}=\sigma_{V^G}^2$, which yields $P_{V^G}=P_{\hat{X}^G_1}$. A similar argument shows that $P_{U^G}=P_{X_1}$.
\end{itemize}
A similar argument holds for the perception constraint of the second and third steps for both PLFs.

 Thus, we have proved the set of Gaussian auxiliary random variables $(\hat{X}_1^G,\hat{X}_2^G,\hat{X}_3^G)$ given in \eqref{rv-relation1}--\eqref{rv-relation-end} where the coefficients are chosen according to distortion-perception constraints, provides an outer bound to $\mathcal{C}_{\mathsf{RDP}}$ which is the set of all tuples $(\mathsf{R},\mathsf{D},\mathsf{P})$ such that
\begin{IEEEeqnarray}{rCl}
R_1&\geq & I(X_1;\hat{X}_1^G),\label{R1-Gaussian-outer}\\
R_2&\geq & \frac{1}{2}\log \left(\rho_1^2 2^{-2R_1}2^{2H(X_1)}+2^{2H(N_1)}\right)-H(X_2|\hat{X}^G_1,\hat{X}^G_2),\label{R2-Gaussian-outer}\\
R_3&\geq & \frac{1}{2}\log \Big(\rho_1^2\rho_2^2 2^{-2R_1-2R_2}2^{2H(X_1)}+\rho_2^2 2^{-2R_2}2^{2H(N_1)}+2^{2H(N_2)}\Big)-H(X_3|\hat{X}_1^G,\hat{X}_2^G,\hat{X}^G_3),\nonumber\\\\
D_j&\geq &\mathbbm{E}[\|X_j-\hat{X}^G_{j}\|^2],\qquad\qquad  j=1,2,3\label{distortion-Gaussian-outer1}\\
P_j&\geq & W_2^2(P_{X_1\ldots X_j},P_{\hat{X}_1^G\ldots\hat{X}_j^G}).\label{perception-Gaussian-outer2}
\end{IEEEeqnarray}

Now, we need to show that the above RDP region is also an inner bound to $\mathcal{C}_{\mathsf{RDP}}$. This is simply verified by the following choice.  In iRDP region of \eqref{MaIsh-rate1}--\eqref{Markov4-new}, choose the following:
\begin{IEEEeqnarray}{rCl}
X_{r,j}=\hat{X}_j=\hat{X}_j^G,\qquad j=1,2,3,
\end{IEEEeqnarray}
where $(\hat{X}_1^G,\hat{X}_2^G,\hat{X}_3^G)$ satisfy \eqref{rv-relation1}--\eqref{rv-relation-end} with coefficients chosen according to distortion-perception constraints.
The lower bounds on distortion and perception constraints in \eqref{distortion-Gaussian-outer1} and \eqref{perception-Gaussian-outer2} are immediately achieved by this choice. Now, we will look at the rate constraints. The achievable rate constraint of the first step can be written as follows
\begin{IEEEeqnarray}{rCl}
R_1 &\geq  & I(X_1;\hat{X}_1^G),\label{R1-ach-Gaussian}
\end{IEEEeqnarray}
which immediately coincides with~\eqref{R1-Gaussian-outer}.
The achievable rate of the second step can be written as follows
\begin{IEEEeqnarray}{rCl}
R_2&\geq &I(X_2;\hat{X}^G_2|\hat{X}_1^G)\\
&=& H(X_2|\hat{X}_1^G)-H(X_2|\hat{X}_1^G,\hat{X}_2^G)\\
&=& H(\rho_1X_1+N_1|\hat{X}_1^G)-H(X_2|\hat{X}_1^G,\hat{X}_2^G)\\
&=&  \frac{1}{2}\log (\rho_1^22^{2H(X_1|\hat{X}_1^G)}+2^{2H(N_1)})-H(X_2|\hat{X}_1^G,\hat{X}_2^G)\label{ach-step3}\\
&\geq &\frac{1}{2}\log \left(\rho_1^2 2^{-2R_1}2^{2H(X_1)}+2^{2H(N_1)}\right)-H(X_2|\hat{X}_1^G,\hat{X}_2^G),\label{opt-ach-Gaussian-step5}
\end{IEEEeqnarray}
where 
\begin{itemize}
    \item  \eqref{ach-step3} follows because EPI holds with ``equality'' for jointly Gaussian distributions \cite[pp. 22]{KimElGamal};
    \item \eqref{opt-ach-Gaussian-step5} follows from~\eqref{R2-Gaussian-outer}.
\end{itemize}
Thus, the bound in \eqref{opt-ach-Gaussian-step5} coincides with~\eqref{rate2-step3}. A similar argument holds for the achievable rate of the third frame.

Notice that the above proof (both converse and achievability) can be extended to $T$ frames using the sequential analysis that was presented. Thus,  without loss of optimality, one can restrict to the jointly Gaussian distributions and identity functions $\eta_1(.)$ and $\eta_2(.,.)$ in iRDP region $\mathcal{C}_{\mathsf{RDP}}$.
\end{IEEEproof}
For a given rate $\mathsf{R}$, the following corollary provides the optimization programs which lead to the characterization of the DP tradeoff $\mathcal{DP}(\mathsf{R})$ for the Gauss-Markov source model.
\begin{corollary}   For a given rate tuple $\mathsf{R}$ and $T=2$ frames, the optimal reconstructions of the DP-tradeoff $\mathcal{DP}(\mathsf{R})$ can be written as follows
\begin{IEEEeqnarray}{rCl}
\hat{X}_1^G&=& \nu X_1+Z_1,\\
\hat{X}_2^G&=& \omega_1 \hat{X}_1^G+\omega_2X_2+Z_2,
\end{IEEEeqnarray}
where $Z_1$ (resp $Z_2$) is a Gaussian random variable independent of $X_1$ (resp $(\hat{X}_1^G,X_2)$) and $\hat{X}_j^G\sim\mathcal{N}(0,\hat{\sigma}^2_j)$ for $j=1,2$, and  $\nu, \omega_1, \omega_2, \hat{\sigma}_1^2, \hat{\sigma}_2^2$ are the solutions of the following optimization program for the first step,
\begin{subequations}\label{1st-step-opt-program}
\begin{IEEEeqnarray}{rCl}
&&\hspace{1cm}\min_{\nu,\hat{\sigma}_1^2}\; \sigma_1^2+\hat{\sigma}_1^2-2\nu\sigma_1^2,\label{D1-G-1st}\\
&&\text{s.t.}\qquad \nu^2\sigma_1^2\leq   \hat{\sigma}^2_1(1-2^{-2R_1}),\label{R1-G-1st}\\
&&\qquad \;\;\;\; (\sigma_1-\hat{\sigma}_1)^2\leq P_1,
\end{IEEEeqnarray}
\end{subequations}
and the following minimization problem for the second step and PLF-FMD, 
\begin{subequations}\label{PLF-FMD}
\begin{IEEEeqnarray}{rCl}
&&\hspace{1cm}\min_{\omega_1,\omega_2,\hat{\sigma}_2^2}\; \sigma_2^2+\hat{\sigma}_2^2-2\nu \omega_1\rho_1\sigma_1\sigma_2-2\omega_2\sigma_2^2,\\
&&\text{s.t.}\qquad \omega_2^2\sigma_2^2(1-2^{-2R_2}\frac{\nu^2\rho_1^2\sigma_1^2}{\hat{\sigma}_1^2})\leq (\hat{\sigma}_2^2-\omega_1^2\hat{\sigma}_1^2-2\omega_1\omega_2\nu\rho_1\sigma_1\sigma_2)(1-2^{-2R_2}),\\
&&\qquad \;\;\;\; (\sigma_2-\hat{\sigma}_2)^2\leq P_2,
\end{IEEEeqnarray}
\end{subequations}
or the following minimization problem for the second step and PLF-JD,
\begin{subequations}\label{2nd-step-opt-JD}
\begin{IEEEeqnarray}{rCl}
&&\hspace{1cm}\min_{\omega_1,\omega_2,\hat{\sigma}_2^2}\; \sigma_2^2+\hat{\sigma}_2^2-2\nu \omega_1\rho_1\sigma_1\sigma_2-2\omega_2\sigma_2^2\label{D2-G-2nd}\\
&&\text{s.t.}\qquad \omega_2^2\sigma_2^2(1-2^{-2R_2}\frac{\nu^2\rho_1^2\sigma_1^2}{\hat{\sigma}_1^2})\leq  (\hat{\sigma}_2^2-\omega_1^2\hat{\sigma}_1^2-2\omega_1\omega_2\nu\rho_1\sigma_1\sigma_2)(1-2^{-2R_2}),\label{R2-G-2nd}\\
&&\qquad \;\;\;\; \text{tr}(\Sigma_{12}+\hat{\Sigma}_{12}-2(\Sigma_{12}^{1/2}\hat{\Sigma}_{12}\Sigma_{12}^{1/2})^{1/2})
\leq P_2,\label{P2-G}
\end{IEEEeqnarray}
\end{subequations}
where $\text{tr}(.)$ denotes the trace of a matrix and
\begin{IEEEeqnarray}{rCl}
\Sigma_{12}&:=& \begin{pmatrix}\sigma_1^2 & \rho_1\sigma_1\sigma_2\\\rho_1\sigma_1\sigma_2 & \sigma_2^2\end{pmatrix},\label{Sigma12}\\
\hat{\Sigma}_{12}&:=& \begin{pmatrix} \hat{\sigma}_1^2& \omega_1\hat{\sigma}_1^2+\nu\omega_2\rho_1\sigma_1\sigma_2\\ \omega_1\hat{\sigma}_1^2+\nu\omega_2\rho_1\sigma_1\sigma_2 &\hat{\sigma}_2^2\end{pmatrix}.\label{Sigmahat12}
\end{IEEEeqnarray}
\end{corollary}
\begin{IEEEproof} We obtain the optimization programs for $T=2$ frames as follows. 

For a given rate tuple $\mathsf{R}$, the DP-tradeoff $\mathcal{DP}(\mathsf{R})$ is given by the set of all tuples $(\mathsf{D},\mathsf{P})$ such that there exists $\hat{\mathsf{X}}^G$ satisfying the following Markov chains
\begin{IEEEeqnarray}{rCl}
&&\hat{X}_1^G\to X_1\to X_2,\label{Markov1-G}\\
&&\hat{X}_2^G\to (\hat{X}_1^G,X_2)\to X_1,\label{Markov2-G}
\end{IEEEeqnarray}
and the following conditions,
\begin{IEEEeqnarray}{rCl}
R_1&\geq & I(X_1;\hat{X}_1^G),\label{R1-G}\\
R_2&\geq & I(X_2;\hat{X}_2^G|\hat{X}_1^G),\label{R2-G}
\end{IEEEeqnarray}
and 
\begin{IEEEeqnarray}{rCl}
D_j&\geq & \mathbbm{E}[\|X_j-\hat{X}_j^G\|^2],\qquad j=1,2,\label{D-G}\\
P_j&\geq & W_2^2(P_{X_1\ldots X_j},P_{\hat{X}_1^G\ldots \hat{X}_j^G}).
\end{IEEEeqnarray}
In general, the set of reconstructions that satisfy \eqref{Markov1-G}--\eqref{Markov2-G} can be written as follows
\begin{IEEEeqnarray}{rCl}
\hat{X}_1^G&=& \nu X_1+Z_1,\label{opt-recons1-G}\\
\hat{X}_2^G&=& \omega_1 \hat{X}_1^G+\omega_2X_2+Z_2.\label{opt-recons2-G}
\end{IEEEeqnarray}
Plugging the above into \eqref{R1-G} and \eqref{R2-G} yields the following rate expressions
\begin{IEEEeqnarray}{rCl}
\frac{1}{2}\log \frac{\hat{\sigma}_1^2}{\hat{\sigma}_1^2-\nu^2\sigma_1^2}&\leq& R_1,\\
\frac{1}{2}\log \frac{\hat{\sigma}_2^2-(\omega_1\hat{\sigma}_1+\frac{\omega_2\nu\rho_1\sigma_1\sigma_2}{\hat{\sigma}_1})^2}{\hat{\sigma}_2^2-\omega_1^2\hat{\sigma}_1^2-\omega_2^2\sigma_2^2-2\omega_1\omega_2\nu\rho_1\sigma_1\sigma_2}&\leq & R_2.
\end{IEEEeqnarray}
Re-arranging the terms in the above constraints yields the conditions in \eqref{R1-G-1st} and \eqref{R2-G-2nd}. Considering \eqref{D-G} with \eqref{opt-recons1-G}--\eqref{opt-recons2-G} gives the following expressions for distortions
\begin{IEEEeqnarray}{rCl}
\mathbbm{E}[\|X_1-\hat{X}_1^G\|^2]&=& \sigma_1^2+\hat{\sigma}_1^2-2\mathbbm{E}[X_1\hat{X}_1^G]=\sigma_1^2+\hat{\sigma}_1^2-2\nu\sigma_1^2,\\
\mathbbm{E}[\|X_2-\hat{X}_2^G\|^2]&=& \sigma_2^2+\hat{\sigma}_2^2-2\mathbbm{E}[X_2\hat{X}_2^G]=\sigma_2^2+\hat{\sigma}_2^2-2\omega_1\nu\rho_1\sigma_1\sigma_2-2\omega_2\sigma_2^2,
\end{IEEEeqnarray}
which are the objective functions in \eqref{D1-G-1st} and \eqref{D2-G-2nd}. Now, we evaluate the perception constraint. 
Notice that the covariance matrices of $(X_1,X_2)$ and $(\hat{X}_1^G,\hat{X}_2^G)$ are given by $\Sigma_{12}$ and $\hat{\Sigma}_{12}$ defined in~\eqref{Sigma12} and~\eqref{Sigmahat12}, respectively. The Wasserstein-2 distance between two Gaussian distributions with covariance matrices $\Sigma_{12}$ and $\hat{\Sigma}_{12}$ is given in \eqref{P2-G} as discussed in \cite[pp. 18]{Wasserstein-distance}.

Similarly, the expressions in \eqref{PLF-FMD} for the decoder based on PLF-FMD can be obtained.
\end{IEEEproof}

\section{Gauss-Markov Source Model: Extremal Rates}\label{extreme-rates-app}
In this section, we derive the achievable reconstructions for some special cases. We assume that we have only two frames, i.e., $D_3,P_3\to \infty$. Moreover, let $\sigma_1^2=\sigma_2^2:=\sigma^2$ for simplicity. In general, the reconstructions can be written as follows
\begin{IEEEeqnarray}{rCl}
\hat{X}_1^G&=& \nu X_1+Z_1,\label{X1hat-Gaussian-ach}\\
\hat{X}_2^G&=& \omega_1 \hat{X}_1^G+\omega_2X_2+Z_2,\label{X2hat-Gaussian-ach}
\end{IEEEeqnarray}
where $\hat{X}_j^G\sim\mathcal{N}(0,\hat{\sigma}^2_j)$ for $j=1,2$. Recall the optimization program of the first step in \eqref{1st-step-opt-program} as follows
\begin{subequations}
\begin{IEEEeqnarray}{rCl}
&&\hspace{1cm}\min_{\nu,\hat{\sigma}_1^2}\; \sigma^2+\hat{\sigma}_1^2-2\nu\sigma^2,\label{obj-simplified}\\
&&\text{s.t.}\qquad \nu^2\sigma^2\leq   \hat{\sigma}^2_1(1-2^{-2R_1}),\\
&&\qquad \;\;\;\; (\sigma-\hat{\sigma}_1)^2\leq P_1,
\end{IEEEeqnarray}
\end{subequations}
For a given $\hat{\sigma}_1^2$, the objective function in \eqref{obj-simplified} is a monotonically deacreasing function of $\nu$, hence one can restrict $\nu$ to be nonnegative, without loss of optimality. So, the above optimization program can be written as 
\begin{subequations}
\begin{IEEEeqnarray}{rCl}
&&\hspace{1cm}\min_{\nu,\hat{\sigma}_1^2}\; \sigma^2+\hat{\sigma}_1^2-2\nu\sigma^2,\\
&&\text{s.t.}\qquad 0\leq \nu \leq \frac{\hat{\sigma}_1}{\sigma}\sqrt{1-2^{-2R_1}},\\
&&\qquad \;\;\;\; (\sigma-\hat{\sigma}_1)^2\leq P_1,
\end{IEEEeqnarray}
\end{subequations}
Optimizing with respect to $\nu$ in the above program, we have
\begin{IEEEeqnarray}{rCl}
\nu=\frac{\hat{\sigma}_1}{\sigma}\sqrt{1-2^{-2R_1}},\label{nu-optimal}
\end{IEEEeqnarray}
where the optimization program reduces to
\begin{subequations}\label{new-1st-simplified}
\begin{IEEEeqnarray}{rCl}
&&\hspace{1cm}\min_{\hat{\sigma}_1^2}\; \sigma^2+\hat{\sigma}_1^2-2\sigma\hat{\sigma}_1\sqrt{1-2^{-2R_1}},\\
&&\text{s.t.}\qquad \;\;\;\; (\sigma-\hat{\sigma}_1)^2\leq P_1.
\end{IEEEeqnarray}
\end{subequations}
Next, recall the optimization program of the second step for PLF-FMD in \eqref{PLF-FMD} as follows
\begin{subequations}
\begin{IEEEeqnarray}{rCl}
&&\hspace{1cm}\min_{\omega_1,\omega_2,\hat{\sigma}_2^2}\; \sigma^2+\hat{\sigma}_2^2-2\nu \omega_1\rho_1\sigma^2-2\omega_2\sigma^2,\\
&&\text{s.t.}\qquad \omega_2^2\sigma^2(1-2^{-2R_2}\frac{\nu^2\rho_1^2\sigma^2}{\hat{\sigma}_1^2})\leq (\hat{\sigma}_2^2-\omega_1^2\hat{\sigma}_1^2-2\omega_1\omega_2\nu\rho_1\sigma^2)(1-2^{-2R_2}),\\\label{rate-constraint-2nd-step-original-program}
&&\qquad \;\;\;\; (\sigma-\hat{\sigma}_2)^2\leq P_2,
\end{IEEEeqnarray}
\end{subequations}
Plugging \eqref{nu-optimal} into the above program, we get
\begin{subequations}\label{2nd-opt-simplified}
\begin{IEEEeqnarray}{rCl}
&&\hspace{1cm}\min_{\omega_1,\omega_2,\hat{\sigma}_2^2}\; \sigma^2+\hat{\sigma}_2^2-2 \omega_1\rho_1\hat{\sigma}_1\sigma\sqrt{1-2^{-2R_1}}-2\omega_2\sigma^2,\label{obj-func-JD-new}\\
&&\text{s.t.}\qquad \omega_2^2\sigma^2(1-\rho_1^22^{-2R_2}(1-2^{-2R_1}))\leq (\hat{\sigma}_2^2-\omega_1^2\hat{\sigma}_1^2-2\omega_1\omega_2\rho_1\hat{\sigma}_1\sigma\sqrt{1-2^{-2R_1}})(1-2^{-2R_2}),\nonumber\\\label{R2-simplified}\\
&&\qquad \;\;\;\; (\sigma-\hat{\sigma}_2)^2\leq P_2,\label{P2-simplified}
\end{IEEEeqnarray}
\end{subequations}
The optimization program for the second step of PLF-JD is similar to the above program \eqref{2nd-opt-simplified} when \eqref{P2-simplified} is replaced by \eqref{P2-G}. In this section, we study different rate regimes and obtain the solutions of the above optimization programs. In particular, we are interested in two perception thresholds $P_2\to \infty$ and $P_2=0$ where the former corresponds to the classical rate-distortion region and the latter is the case of $0$-PLF. For the $0$-PLF-FMD, we have $\hat{\sigma}_1=\hat{\sigma}_2=\sigma$. For the $0$-PLF-JD, in addition to preserving the marginals, the correlation $\mathbbm{E}[\hat{X}_1^G\hat{X}_2^G]=\rho_1\sigma^2$ should be satisfied. For each of these cases, the optimization program in~\eqref{2nd-opt-simplified} is simplified in the following.

\underline{\textit{Optimization Program of the Second Step for $P\to\infty$}}: In this case, there is no perception constraint in the setting and the optimization program in~\eqref{2nd-opt-simplified} reduces to the following
\begin{subequations}\label{opt-program-MMSE-simplified-concl}
\begin{IEEEeqnarray}{rCl}
&&\hspace{1cm}\min_{\substack{\hat{\sigma}_2^2,\omega_1,\omega_2}}\!\! \sigma^2+\hat{\sigma}_2^2-2 \omega_1\rho_1\hat{\sigma}_1\sigma\sqrt{1-2^{-2R_1}}-2\omega_2\sigma^2,\\
&&\text{s.t.}\qquad \omega_2^2\sigma^2(1-\rho_1^22^{-2R_2}(1-2^{-2R_1}))\leq (\hat{\sigma}_2^2-\omega_1^2\hat{\sigma}_1^2-2\omega_1\omega_2\rho_1\hat{\sigma}_1\sigma\sqrt{1-2^{-2R_1}})(1-2^{-2R_2}).\nonumber\\\label{rate-constraint-MMSE-simplified}
\end{IEEEeqnarray}
\end{subequations}
This case corresponds to the classical rate-distortion tradeoff where it is shown that for a given rate, the MMSE reconstructions are indeed optimal \cite{MaIshwar,Ashishproof}. The expressions for MMSE reconstructions are given in Appendix~\ref{MMSE-representation-appendix}.

\underline{\textit{Optimization Program of the Second Step for $0$-PLF-FMD}}: In this case, we have $\hat{\sigma}_1=\hat{\sigma}_2=\sigma$. So, the optimization program in~\eqref{2nd-opt-simplified} reduces to the following
\begin{subequations}\label{opt-program-FMD-simplified-concl}
\begin{IEEEeqnarray}{rCl}
&&\hspace{1cm}\min_{\substack{\omega_1,\omega_2}}\;\; 2\sigma^2-2 \omega_1\rho_1\sigma^2\sqrt{1-2^{-2R_1}}-2\omega_2\sigma^2,\\
&&\text{s.t.}\qquad \omega_2^2(1-\rho_1^22^{-2R_2}(1-2^{-2R_1}))\leq (1-\omega_1^2-2\omega_1\omega_2\rho_1\sqrt{1-2^{-2R_1}})(1-2^{-2R_2}).\nonumber\\\label{rate-FMD-simplified}
\end{IEEEeqnarray}
\end{subequations}
Here, $\omega_1$ and $\omega_2$ only need to satisfy the rate constraint given in~\eqref{rate-FMD-simplified} which represents a larger search space than that of $0$-PLF-JD which will be discussed in the following.

\underline{\textit{Optimization Program of the Second Step for $0$-PLF-JD}}: In this case, in addition to preserving marginals $\hat{\sigma}_1=\hat{\sigma}_2=\sigma$, we need to satisfy the constraint $\mathbbm{E}[\hat{X}_1^G\hat{X}_2^G]=\rho_1\sigma^2$. Thus, the optimization program of this case has an extra condition $\omega_1+\nu\omega_2\rho_1=\rho_1$ comparing to~\eqref{opt-program-FMD-simplified-concl} and it is given as follows
\begin{subequations}\label{opt-program-JD-simplified-concl}
\begin{IEEEeqnarray}{rCl}
&&\hspace{1cm}\min_{\substack{\omega_1,\omega_2}}\;\; 2\sigma^2-2 \omega_1\rho_1\sigma^2\sqrt{1-2^{-2R_1}}-2\omega_2\sigma^2,\\
&&\text{s.t.}\qquad \omega_2^2(1-\rho_1^22^{-2R_2}(1-2^{-2R_1}))\leq (1-\omega_1^2-2\omega_1\omega_2\rho_1\sqrt{1-2^{-2R_1}})(1-2^{-2R_2}),\nonumber\\
&&\qquad\;\;\;\; \omega_1+\nu\omega_2\rho_1=\rho_1.
\end{IEEEeqnarray}
\end{subequations}
Comparing~\eqref{opt-program-JD-simplified-concl} with~\eqref{opt-program-FMD-simplified-concl}, we notice that the search space of the optimization program for $0$-PLF-JD is smaller than that of $0$-PLF-FMD. Thus, a larger distortion is expected for $0$-PLF-JD. 
 
Before studying each case of extremal rates, we introduce another constraint in the optimization program of all above three cases of perception metrics. We restrict to nonnegative $\omega_1\omega_2\rho_1$ and get an upper bound on the programs \eqref{opt-program-MMSE-simplified-concl},~\eqref{opt-program-FMD-simplified-concl} and \eqref{opt-program-JD-simplified-concl}. So, in further discussion on these programs, the constraint $\omega_1\omega_2\rho_1\geq 0$ will be also considered.

\textit{1) $R_1=R_2=\epsilon$ for small $\epsilon$}: 

In the low-rate regime, notice that we can approximate the rate term as follows 
\begin{IEEEeqnarray}{rCl}
1-2^{-2\epsilon}=2\epsilon\ln 2+O(\epsilon^2).\label{ep-app}
\end{IEEEeqnarray}
Plugging the above into \eqref{nu-optimal}, we have
\begin{IEEEeqnarray}{rCl}
\nu = \frac{\hat{\sigma}_1}{\sigma}\sqrt{2\epsilon\ln 2 +O(\epsilon^2)}.\label{nu-opt-simplified}
\end{IEEEeqnarray}
Also, inserting \eqref{ep-app} into the rate constraint of the second step \eqref{rate-constraint-2nd-step-original-program} yields the following
\begin{IEEEeqnarray}{rCl}
\omega_2^2\sigma^2(1-\rho_1^22\epsilon\ln 2 +O(\epsilon^2))&\leq& (\hat{\sigma}_2^2-\omega_1^2\hat{\sigma}_1^2-2\omega_1\omega_2\rho_1\hat{\sigma}_1\sigma\sqrt{2\epsilon\ln 2+O(\epsilon^2)})(2\epsilon\ln 2 +O(\epsilon^2)).\footnote{ The inequalities of the form $f(\epsilon)+O(\epsilon^2)\leq g(\epsilon)+O(\epsilon^2)$, where $f(\epsilon),g(\epsilon)=\Omega(\epsilon^2)$, imply that $f(\epsilon)\leq g(\epsilon)$. So, in such inequalities, we work with dominant terms ($f(\epsilon),g(\epsilon)$) and ignore the small terms $O(\epsilon^2)$. A similar argument holds if we have other orders of $\epsilon$ and the functions $f(.), g(.)$ approach zero slower than them.}\nonumber\\
\end{IEEEeqnarray}
Re-arranging the terms in the above inequality yields the following
\begin{IEEEeqnarray}{rCl}
\hat{\sigma}_2^2&\geq& \frac{\omega_2^2\sigma^2(1-\rho_1^22\epsilon\ln 2 +O(\epsilon^2))}{2\epsilon\ln 2 +O(\epsilon^2)}+\omega_1^2\hat{\sigma}_1^2+2\omega_1\omega_2\rho_1\hat{\sigma}_1\sigma\sqrt{2\epsilon\ln 2+O(\epsilon^2)}\\
&=& \omega_2^2\sigma^2\left(\frac{1}{2\epsilon\ln 2}+O\left(1\right)\right)+\omega_1^2\hat{\sigma}_1^2+2\omega_1\omega_2\rho_1\hat{\sigma}_1\sigma\sqrt{2\epsilon\ln 2+O(\epsilon^2)}\label{rate-2-simp}
\end{IEEEeqnarray}
So, in all of the optimization programs of the case $R_1=R_2=\epsilon$, the above constraint~\eqref{rate-2-simp} will replace the rate constraint of the second step.

Now, we consider different cases based on the perception measure.

\textit{a) Without a perception constraint}: In this case, using \eqref{ep-app}, the optimization program of the first step in \eqref{new-1st-simplified} simplifies to the following
\begin{IEEEeqnarray}{rCl}
D_1 &=& \min_{\hat{\sigma}_1^2}\; \sigma^2+\hat{\sigma}_1^2-2\sigma\hat{\sigma}_1\sqrt{2\epsilon\ln 2 +O(\epsilon^2)},\label{D1-simplified}
\end{IEEEeqnarray}
 which gives us the following optimal solution 
 \begin{IEEEeqnarray}{rCl}
 \hat{\sigma}_1= \sqrt{2\epsilon\ln 2+O(\epsilon^2)}\sigma=\sqrt{2\epsilon\ln 2}\sigma+O(\epsilon).
 \end{IEEEeqnarray}
 Plugging the above solution into \eqref{nu-opt-simplified} and \eqref{D1-simplified}, we get 
 \begin{IEEEeqnarray}{rCl}
 \nu = 2\epsilon\ln 2 +O(\epsilon^2),\label{nu-opt-final}
 \end{IEEEeqnarray}
 and
 \begin{IEEEeqnarray}{rCl}
 D_1= (1-2\epsilon\ln 2)\sigma^2+O(\epsilon^2).
 \end{IEEEeqnarray}
 Now, we look at the optimization program of the second step \eqref{opt-program-MMSE-simplified-concl}. For a given $\omega_1$ and $\omega_2$, the objective function is an increasing function of $\hat{\sigma}_2^2$, so optimizing over $\hat{\sigma}_2^2$ yields the following
\begin{IEEEeqnarray}{rCl}
\hat{\sigma}_2^2= \omega_2^2\sigma^2\left(\frac{1}{2\epsilon\ln 2}+O\left(1\right)\right)+\omega_1^2\hat{\sigma}_1^2+2\omega_1\omega_2\rho_1\hat{\sigma}_1\sigma\sqrt{2\epsilon\ln 2+O(\epsilon^2)}.
\end{IEEEeqnarray}
Thus, the optimization program \eqref{opt-program-MMSE-simplified-concl} is further upper bounded by the following
 \begin{IEEEeqnarray}{rCl}
\min_{\substack{\hat{\sigma}_2^2,\omega_1,\omega_2:\\\omega_1\omega_2\rho_1\geq 0}}\sigma^2+\omega_2^2\sigma^2\left(\frac{1}{2\epsilon\ln 2}+O\left(1\right)\right)+\omega_1^2\hat{\sigma}_1^2-2(1-\omega_2) \omega_1\rho_1\hat{\sigma}_1\sigma\sqrt{2\epsilon\ln 2 +O(\epsilon^2)}-2\omega_2\sigma^2.\nonumber\\
 \end{IEEEeqnarray}
The optimal solution of the above minimization is given by the following
\begin{IEEEeqnarray}{rCl}
\omega_1 &=& \rho_1 +O(\epsilon),\label{w1-opt-final}\\
\omega_2 &=& 2\epsilon\ln 2 +O(\epsilon^2).\label{w2-opt-final}
\end{IEEEeqnarray}

Thus, considering the dominant terms of \eqref{nu-opt-final}, \eqref{w1-opt-final} and \eqref{w2-opt-final}, we have
\begin{IEEEeqnarray}{rCl}
\hat{X}_1^G&=& (2\epsilon\ln 2) X_1+Z_1,\\
\hat{X}_2^G&=& \rho_1\hat{X}_1^G+(2\epsilon\ln 2) X_2+Z_2,
\end{IEEEeqnarray}
and $Z_j\sim\mathcal{N}(0,2\epsilon\sigma^2\ln 2)$ for $j=1,2$. Notice that 
\begin{IEEEeqnarray}{rCl}
D_1 &=& (1-2\epsilon\ln 2)\sigma^2,\\
D_2 &=& (1-(1+\rho_1^2)2\epsilon\ln 2)\sigma^2.
\end{IEEEeqnarray}

\textit{b) $0$-PLF-FMD}:
 In this case, we have $\hat{\sigma}_1=\hat{\sigma}_2=\sigma$. For the optimization program of the first step, \eqref{nu-optimal} reduces to the following
 \begin{IEEEeqnarray}{rCl}
 \nu = \sqrt{2\epsilon\ln 2}+O(\epsilon),\label{nu-FMD}
 \end{IEEEeqnarray}
 and $D_1$ is given in the following which is derived by \eqref{new-1st-simplified}
 \begin{IEEEeqnarray}{rCl}
 D_1 = 2(1-\sqrt{2\epsilon\ln 2})\sigma^2+O(\epsilon).
 \end{IEEEeqnarray}
 Now, we study the optimization program of the second step. The optimization program of \eqref{opt-program-FMD-simplified-concl} is further upper bounded by the following
\begin{subequations}\label{FMD-opt-eps-eps}
\begin{IEEEeqnarray}{rCl}
&&\hspace{1cm}\min_{\substack{\omega_1,\omega_2:\\\omega_1\omega_2\rho_1\geq 0}}\; 2\sigma^2-2 \omega_1\rho_1\sigma^2\sqrt{2\epsilon\ln 2 +O(\epsilon^2)}-2\omega_2\sigma^2,\\
&&\text{s.t.}\qquad 1\geq \sqrt{\omega_2^2\left(\frac{1}{2\epsilon\ln 2}+O\left(1\right)\right)+\omega_1^2+2\omega_1\omega_2\rho_1\sqrt{2\epsilon\ln 2+O(\epsilon^2)}}.\label{FMD-rate-eps-eps}
\end{IEEEeqnarray}
\end{subequations}
Now, we further simplify the inequality \eqref{FMD-rate-eps-eps} in the following. Considering the fact that $\omega_1\omega_2\rho_1\geq 0$, this inequality implies that 
\begin{IEEEeqnarray}{rCl}
&& \omega_1^2\leq 1,\\
&& \omega_2^2\leq 2\epsilon\ln 2 +O(\epsilon^2).
\end{IEEEeqnarray}
So, using the above inequalities, the RHS of \eqref{FMD-rate-eps-eps} can be upper bounded as follows
\begin{IEEEeqnarray}{rCl}
&&\hspace{-1.5cm}\sqrt{\omega_2^2\left(\frac{1}{2\epsilon\ln 2}+O\left(1\right)\right)+\omega_1^2+2\omega_1\omega_2\rho_1\sqrt{2\epsilon\ln 2+O(\epsilon^2)}}\nonumber\\
&&\leq \sqrt{\omega_2^2\left(\frac{1}{2\epsilon\ln 2}+O\left(1\right)\right)+\omega_1^2+(\omega_1^2+\omega_2^2)\rho_1\sqrt{2\epsilon\ln 2+O(\epsilon^2)}}\nonumber\\&&\leq \sqrt{\omega_2^2\left(\frac{1}{2\epsilon\ln 2}+O\left(1\right)\right)+\omega_1^2+O(\epsilon^{3/2})}.\label{reduced-constraint}
\end{IEEEeqnarray}
Now, according to \eqref{reduced-constraint}, the optimization program in \eqref{FMD-opt-eps-eps} is further upper bounded by the following
\begin{subequations}\label{mid-step-program}
\begin{IEEEeqnarray}{rCl}
&&\hspace{1cm}\min_{\substack{\omega_1,\omega_2:\\\omega_1\omega_2\rho_1\geq 0}}\; 2\sigma^2-2 \omega_1\rho_1\sigma^2\sqrt{2\epsilon\ln 2 +O(\epsilon^2)}-2\omega_2\sigma^2,\label{objective-function-new3}\\
&&\text{s.t.}\qquad 1\geq \sqrt{\omega_2^2\left(\frac{1}{2\epsilon\ln 2}+O\left(1\right)\right)+\omega_1^2+O(\epsilon^{3/2})}.
\end{IEEEeqnarray}
\end{subequations}

For a given $\omega_1$ (resp $\omega_2$), the objective function~\eqref{objective-function-new3} is a monotonically decreasing function of $\omega_2$ (resp $\omega_1$), so the optimal solution is attained on the boundary, i.e.,
\begin{IEEEeqnarray}{rCl}
1= \sqrt{\omega_2^2\left(\frac{1}{2\epsilon\ln 2}+O\left(1\right)\right)+\omega_1^2+O(\epsilon^{3/2})}\label{surface}
\end{IEEEeqnarray}
Thus, the program~\eqref{mid-step-program} further simplifies to the following
\begin{IEEEeqnarray}{rCl}
\min_{\substack{\omega_1:\\\omega_1\rho_1\geq 0}}2\sigma^2-2 \omega_1\rho_1\sigma^2\sqrt{2\epsilon\ln 2 +O(\epsilon^2)}-2\sigma^2\sqrt{(1-\omega_1^2-O(\epsilon^{3/2}))(2\epsilon\ln 2+O(\epsilon^2))}.\nonumber\\
\end{IEEEeqnarray}
The optimal solution of the above program is given by 
 \begin{IEEEeqnarray}{rCl}
\omega_1 = \frac{\rho_1}{\sqrt{1+\rho_1^2}}+O\left(\epsilon\right),\label{w1-FMD}
\end{IEEEeqnarray}
which together with~\eqref{surface} yields 
\begin{IEEEeqnarray}{rCl}
\omega_2=\sqrt{\frac{2\epsilon\ln 2}{1+\rho_1^2}}+O(\epsilon).\label{w2-FMD}
\end{IEEEeqnarray}

Thus, considering dominant terms of \eqref{nu-FMD}, \eqref{w1-FMD} and \eqref{w2-FMD}, we get
\begin{IEEEeqnarray}{rCl}
\hat{X}_1^G&=& \sqrt{2\epsilon\ln 2} X_1+Z_1,\label{cons-ach-FMD-eps1}\\
\hat{X}_2^G&=&  \frac{\rho_1}{\sqrt{1+\rho_1^2}}\hat{X}_1^G+\sqrt{\frac{2\epsilon\ln 2}{1+\rho_1^2}} X_2+Z_2,\label{cons-ach-FMD-eps2}
\end{IEEEeqnarray}
where $Z_1\sim\mathcal{N}(0,(1-2\epsilon\ln 2)\sigma^2)$ and 
\begin{IEEEeqnarray}{rCl}Z_2\sim \mathcal{N}(0,(1-\frac{\rho_1^2}{1+\rho_1^2}-\frac{1+2\rho_1^2}{1+\rho_1^2}2\epsilon\ln 2)\sigma^2).\end{IEEEeqnarray} 
Notice that
\begin{IEEEeqnarray}{rCl}
D_1 &=& 2(1-\sqrt{2\epsilon\ln 2})\sigma^2,\\
D_2 &=& 2(1-\sqrt{(1+\rho_1^2)2\epsilon\ln 2})\sigma^2.\label{D2-final-FMD}
\end{IEEEeqnarray}

For the special case of $\rho_1=1$, the expressions in~\eqref{cons-ach-FMD-eps1} and \eqref{cons-ach-FMD-eps2} simplify as follows
\begin{IEEEeqnarray}{rCl}
\hat{X}_1^G&=& \sqrt{2\epsilon\ln 2} X_1+Z_1,\\
\hat{X}_2^G&=& \sqrt{2}\sqrt{2\epsilon\ln 2}X_1+\frac{1}{\sqrt{2}}Z_1+Z_2.
\end{IEEEeqnarray}
Define $Z_{\text{FMD}}:=\frac{1}{\sqrt{2}}Z_1+Z_2$ and notice that $Z_{\text{FMD}}\sim \mathcal{N}(0,(1-4\epsilon\ln 2)\sigma^2)$. Moreover, we have 
\begin{IEEEeqnarray}{rCl}
D_1 &=& 2(1-\sqrt{2\epsilon\ln 2})\sigma^2,\\
D_2 &=& 2(1-\sqrt{4\epsilon\ln 2})\sigma^2.
\end{IEEEeqnarray}
\textit{c) $0$-PLF-JD}: In this case, the optimization program of the first step is similar to the previous case. The optimization program of the second step is given in~\eqref{opt-program-JD-simplified-concl} where the condition $\omega_1+\nu\omega_2\rho_1=\rho_1$ is introduced. According to~\eqref{nu-FMD}, $\nu=O(\sqrt{\epsilon})$ which suggests the following form for $\omega_1$,
\begin{IEEEeqnarray}{rCl}
\omega_1=\rho_1-\delta_{\epsilon},\label{JD-first-step-O-eps}
\end{IEEEeqnarray}
for some small $\delta_{\epsilon}$ that goes to zero as $\epsilon\to 0$. The parameter $\delta_{\epsilon}$ will be determined later. Plugging $\omega_1=\rho_1-\delta_{\epsilon}$ into~\eqref{surface}, we find out that only the constant term of $\omega_1$ contributes to a dominant term for $\omega_2$ which yields the following
\begin{IEEEeqnarray}{rCl}\omega_2=\sqrt{2\epsilon\ln 2(1-\rho_1^2)}+O(\epsilon).\label{JD-1st-step}\end{IEEEeqnarray}
Thus, we have 
\begin{IEEEeqnarray}{rCl}
\hat{X}_1^G&=& \sqrt{2\epsilon\ln 2} X_1+Z_1,\\
\hat{X}_2^G&=& (\rho_1-\delta_{\epsilon}) \hat{X}_1^G+\sqrt{(1-\rho_1^2)2\epsilon\ln 2} X_2+Z_2,
\end{IEEEeqnarray}
Now, applying the constraint $\mathbbm{E}[\hat{X}_1^G\hat{X}_2^G]=\rho_1\sigma^2$, we get
\begin{IEEEeqnarray}{rCl}
\delta_{\epsilon}=\rho_1\sqrt{1-\rho_1^2}(2\epsilon\ln 2).\label{delta-epsilon-justification}
\end{IEEEeqnarray}
However, notice that since $\delta_{\epsilon}=O(\epsilon)$, it does not contribute to dominant terms of distortion. So, we can simply represent $\hat{X}_1^G$ and $\hat{X}_2^G$ as follows
\begin{IEEEeqnarray}{rCl}
\hat{X}_1^G&=& \sqrt{2\epsilon\ln 2} X_1+Z_1,\\
\hat{X}_2^G&=& \rho_1 \hat{X}_1^G+\sqrt{(1-\rho_1^2)2\epsilon\ln 2} X_2+Z_2,\label{JD-last-step-O-eps}
\end{IEEEeqnarray}
where $Z_1\sim\mathcal{N}(0,(1-2\epsilon\ln 2)\sigma^2)$ and $Z_2\sim \mathcal{N}(0,(1-\rho_1^2-(1-\rho_1^2+2\rho_1^2\sqrt{1-\rho_1^2})2\epsilon\ln2)\sigma^2)$.
The following distortions are also achievable
\begin{IEEEeqnarray}{rCl}
D_1 &=& 2(1-\sqrt{2\epsilon\ln 2})\sigma^2,\\
D_2 &=& 2(1-(\rho_1^2+\sqrt{1-\rho_1^2})\sqrt{2\epsilon\ln 2})\sigma^2.\label{D2-JD-last}
\end{IEEEeqnarray}
For the special case of $\rho=1$, according to~\eqref{JD-last-step-O-eps} and~\eqref{D2-JD-last}, we have $\hat{X}_2^G=\hat{X}_1^G$ and $D_2=D_1$.

\textit{2) $R_1\to \infty$, $R_2=\epsilon$ for small $\epsilon$}: In this case, since $R_1\to \infty$, we have $\hat{X}_1^G=X_1$, $D_1=0$, and we only need to solve the optimization program of the second step. Also, we have the following approximation
\begin{IEEEeqnarray}{rCl}
1-2^{-2R_2}=1-2^{-2\epsilon}=2\epsilon\ln 2 +O(\epsilon^2).
\end{IEEEeqnarray}

We consider three different cases based on the perception constraint.

\textit{a) Without a perception constraint}: In this case, consider the optimization program~\eqref{opt-program-MMSE-simplified-concl}. For a given $\omega_1$ and $\omega_2$, the objective function is an increasing function of $\hat{\sigma}_2^2$, hence optimizing over $\hat{\sigma}_2^2$, we get
\begin{IEEEeqnarray}{rCl}
\hat{\sigma}_2^2=\frac{\omega_2^2\sigma^2(1-\rho_1^2 +O(\epsilon))}{2\epsilon\ln 2 + O(\epsilon^2)}+\omega_1^2\sigma^2+2\omega_1\omega_2\rho_1\sigma^2.
\end{IEEEeqnarray}
The program in \eqref{opt-program-MMSE-simplified-concl} is further upper bounded by the following
\begin{IEEEeqnarray}{rCl}
&&\hspace{1cm}\min_{\substack{\omega_1,\omega_2:\\\omega_1\omega_2\rho_1\geq 0}}\; \sigma^2+\frac{\omega_2^2\sigma^2(1-\rho_1^2 +O(\epsilon))}{2\epsilon\ln 2 + O(\epsilon^2)}+\omega_1^2\sigma^2+2\omega_1\omega_2\rho_1\sigma^2-2 \omega_1\rho_1\sigma^2-2\omega_2\sigma^2,\nonumber\\
\end{IEEEeqnarray}
The solution of the above optimization program is given by the following
\begin{IEEEeqnarray}{rCl}
\omega_1 &=& \rho_1-\rho_1(2\epsilon\ln2),\\
\omega_2 &=& 2\epsilon\ln 2.
\end{IEEEeqnarray}
Thus, we have
\begin{IEEEeqnarray}{rCl}
\hat{X}_1^G &=& X_1,\\
\hat{X}_2^G &=& (\rho_1-\rho_1(2\epsilon\ln2)) X_1+(2\epsilon\ln 2) X_2+Z_2,
\end{IEEEeqnarray}
 where $Z_2\sim\mathcal{N}(0,(1-\rho_1^2)\sigma^22\epsilon\ln 2)$.
So, the reconstruction of the second frame closely resembles the first frame. The distortions of the first and second frames are zero and $(1-\rho_1^2-(1-\rho_1^2)2\epsilon\ln 2)\sigma^2$, respectively.

\textit{b) $0$-PLF-FMD}: In this case, $\hat{\sigma}_1=\hat{\sigma}_2=\sigma$. Thus, the optimization program in \eqref{opt-program-FMD-simplified-concl} is further upper bounded by the following
\begin{subequations}\label{FMD-R1-inf-program}
\begin{IEEEeqnarray}{rCl}
&&\hspace{1cm}\min_{\substack{\omega_1,\omega_2:\\\omega_1\omega_2\rho_1\geq 0}}\; 2\sigma^2-2 \omega_1\rho_1\sigma^2-2\omega_2\sigma^2,\label{FMD-R1-inf-objective-function}\\
&&\text{s.t.}\qquad \omega_2^2(1-\rho_1^2 +O(\epsilon))\leq (1-\omega_1^2-2\omega_1\omega_2\rho_1)(2\epsilon\ln 2 + O(\epsilon^2)).\label{FMD-R1-inf}
\end{IEEEeqnarray}
\end{subequations}
For a given $\omega_1$ (resp $\omega_2$), the objective function~\eqref{FMD-R1-inf-objective-function} is a monotonically decreasing function of $\omega_2$ (resp $\omega_1$). So, the optimal solution is attained on the boundary, i.e., \eqref{FMD-R1-inf} is satisfied with equality given as follows 
\begin{IEEEeqnarray}{rCl}
\omega_2^2(1-\rho_1^2 +O(\epsilon))= (1-\omega_1^2-2\omega_1\omega_2\rho_1)(2\epsilon\ln 2 + O(\epsilon^2)).\label{R2-approx-marg}
\end{IEEEeqnarray}
It can be easily verified that the first-order terms of $\omega_1$ and $\omega_2$ which optimize the program are $1$ and $0$, respectively. So, we write $\omega_1$ and $\omega_2$ in the following form
\begin{IEEEeqnarray}{rCl}
\omega_1 &=& 1+(2\epsilon\ln 2)\delta_1+O(\epsilon^2),\label{w1-R1-inf}\\
\omega_2 &=& (2\epsilon\ln 2)\delta_2+O(\epsilon^2),\label{w2-R1-inf}
\end{IEEEeqnarray}
for some real $\delta_1$ and $\delta_2$. Plugging the above \eqref{w1-R1-inf} and \eqref{w2-R1-inf} into \eqref{R2-approx-marg} and considering the dominant terms, we get
\begin{IEEEeqnarray}{rCl}
\delta_2^2(1-\rho_1^2)=-2\delta_1-2\rho_1\delta_2.
\end{IEEEeqnarray}
On the other side, we can write the objective function in \eqref{FMD-R1-inf-program} as follows
\begin{IEEEeqnarray}{rCl}
&&\hspace{-1cm}2\sigma^2-2 \omega_1\rho_1\sigma^2-2\omega_2\sigma^2\nonumber\\&=& 2\sigma^2-2\rho_1\omega_1\sigma^2-2\omega_2\sigma^2+O(\epsilon^2)\\
&=& 2\sigma^2-2\rho_1\sigma^2-2(\rho_1\delta_1\sigma^2+\delta_2\sigma^2)(2\epsilon\ln 2)+O(\epsilon^2)\\
&=&2\sigma^2-2\rho_1\sigma^2-(-2\rho^2_1\delta_2\sigma^2-\rho_1(1-\rho_1^2)\delta_2^2+2\delta_2\sigma^2)(2\epsilon\ln 2)+O(\epsilon^2).
\end{IEEEeqnarray}
Differentiating the above expression with respect to $\delta_2$ and letting it be zero, we have:
\begin{IEEEeqnarray}{rCl}
\delta_2=\frac{1}{\rho_1},\qquad \delta_1=-\frac{1+\rho_1^2}{2\rho_1^2}.
\end{IEEEeqnarray}
Thus, we have
\begin{IEEEeqnarray}{rCl}
\hat{X}_1^G&=& X_1,\\
\hat{X}_2^G&=& (1-\frac{(1+\rho_1^2)2\epsilon\ln 2}{2\rho_1^2}) \hat{X}_1^G+\frac{2\epsilon \ln 2}{\rho_1}X_2+Z_2,
\end{IEEEeqnarray}
where $Z_2\sim \mathcal{N}(0,(\frac{1-\rho_1^2}{\rho_1^2})2\epsilon\ln 2 )$. Again, the reconstruction of the second frame is almost similar to the first frame and the distortion is $2(1-\rho_1-(\frac{1-\rho_1^2}{2\rho_1})2\epsilon\ln 2)\sigma^2$.

\textit{c) $0$-PLF-JD}: First consider the case where $\rho_1\neq 1$. The optimization program is given in~\eqref{opt-program-JD-simplified-concl} where the constraint $\omega_1+\nu\rho_1\omega_2=\rho_1$ is introduced. Notice that $\omega_1$ can be written in the following form
\begin{IEEEeqnarray}{rCl}
\omega_1&=&\rho_1+\delta_{\epsilon},
\end{IEEEeqnarray}
for some $\delta_{\epsilon}$ that goes to zero as $\epsilon\to 0$. The parameter $\delta_{\epsilon}$ will be determined later. Plugging $\omega_1=\rho_1+\delta_{\epsilon}$  into \eqref{R2-approx-marg} yields the following
\begin{IEEEeqnarray}{rCl}
\omega_2= \sqrt{2\epsilon\ln 2}+O(\epsilon),
\end{IEEEeqnarray}
which is derived only through the first-order term of $\omega_1$ which is $\rho_1$. Now, considering the fact that $\mathbbm{E}[\hat{X}_1^G\hat{X}_2^G]=\rho_1\sigma^2$, we obtain
\begin{IEEEeqnarray}{rCl}
\delta_{\epsilon}=-\rho_1\sqrt{2\epsilon\ln 2}.
\end{IEEEeqnarray}
Thus,  we have
\begin{IEEEeqnarray}{rCl}
\hat{X}_1^G&=& X_1,\label{R1inf-JD-X1}\\
\hat{X}_2^G&=& (\rho_1-\rho_1\sqrt{2\epsilon\ln 2}) \hat{X}_1^G+\sqrt{2\epsilon\ln 2 }X_2+Z_2,\label{R1inf-JD-X2}
\end{IEEEeqnarray}
where  $Z_2\sim \mathcal{N}(0,(1-\rho_1^2)\sigma^2)$. Here, the reconstruction of the second frame closely resembles the first frame. The distortion of the second frame is $2(1-\rho_1^2-(1-\rho_1^2)\sqrt{2\epsilon\ln 2 })\sigma^2$.

If $\rho_1=1$, we simply have $\hat{X}_2^G=\hat{X}_1^G=X_1=X_2$ which can be derived from \eqref{R1inf-JD-X1}--\eqref{R1inf-JD-X2} by letting $X_1=X_2$.

The analysis for the case of $R_1=\epsilon$ and $R_2\to \infty$ is similar and is omitted for brevity. The results of this section are summarized in Table~\ref{table-ach-recons}.

\begin{table*}[t]
\caption{Achievable reconstructions  for extremal rates and different PLFs (The first, second and third rows represent reconstructions corresponding to the MMSE, $0$-PLF-FMD and $0$-PLF-JD, respectively). }
\label{table-ach-recons}
\vskip 0.15in
\begin{center}
\begin{tiny}
\begin{sc}
\begin{minipage}{\linewidth}
\begin{tabular}{llll}
\toprule
& $R_1=R_2=\epsilon$  &  $R_1\to \infty,$  $R_2=\epsilon$ & $R_1=\epsilon, R_2=\infty$ \\
\midrule
  & \hspace{-0.3cm}$\hat{X}_1^G{=}(2\epsilon\ln 2)X_1+Z_1$  & \hspace{-0.5cm}$\hat{X}_1^G{=}X_1$ & $\hat{X}_1^G{=}(2\epsilon\ln 2)X_1+Z_1$\\[0.5ex]
&  \hspace{-0.3cm}$\hat{X}_2^G{=}\rho_1\hat{X}_1^G+ (2\epsilon\ln 2)X_2+Z_2$ & \hspace{-0.5cm}$\hat{X}_2^G{=}(\rho_1-\rho_12\epsilon\ln 2)\hat{X}^G_1+(2\epsilon\ln 2)X_2+Z_2$ & $\hat{X}_2^G{=}X_2$\\[0.5ex] 
\begin{turn}{90}MMSE\end{turn} & \hspace{-0.3cm}$Z_j{\sim} \mathcal{N}(0,2\epsilon\sigma^2\ln 2)$ & \hspace{-0.5cm}$Z_2{\sim} \mathcal{N}(0,(1-\rho_1^2)2\epsilon\sigma^2\ln 2)$ & $Z_1{\sim} \mathcal{N}(0,2\epsilon\sigma^2\ln 2)$\\[0.5ex]
&  \hspace{-0.3cm}$D_1{=}(1-2\epsilon\ln 2)\sigma^2$ & \hspace{-0.5cm}$D_1{=}0$ & $D_1{=}(1-2\epsilon\ln 2)\sigma^2$\\[0.5ex] & \hspace{-0.3cm}$D_2{=}(1-(1+\rho_1^2)2\epsilon\ln 2)\sigma^2$ & \hspace{-0.5cm}$D_2{=}(1-\rho_1^2-(1-\rho_1^2)2\epsilon\ln 2)\sigma^2$ & $D_2{=}0$\\[0.5ex]
 \hline\\
&  \hspace{-0.3cm}$\hat{X}_1^G{=}\sqrt{2\epsilon\ln 2}X_1+Z_1$ & \hspace{-0.5cm}$\hat{X}_1^G{=}X_1$ & $\hat{X}_1^G{=}\sqrt{2\epsilon\ln 2}X_1+Z_1$\\
 &  \hspace{-0.3cm}$\hat{X}_2^G{=}\frac{\rho_1}{\sqrt{1+\rho_1^2}}\hat{X}_1^G+\sqrt{\frac{2\epsilon\ln 2}{1+\rho_1^2}} X_2+Z_2$&\hspace{-0.5cm} $\hat{X}_2^G{=}(1-\frac{(1+\rho_1^2)2\epsilon\ln 2}{2\rho_1^2}) \hat{X}_1^G+\frac{2\epsilon \ln 2}{\rho_1}X_2+Z_2$ & $\hat{X}_2^G{=}X_2$\\
\begin{turn}{90}$0$-PLF-FMD\end{turn}&  \hspace{-0.3cm}$Z_1{\sim} \mathcal{N}(0,(1-2\epsilon\ln 2)\sigma^2)$ & \hspace{-0.5cm}$Z_2{\sim} \mathcal{N}(0,(\frac{1-\rho_1^2}{\rho_1^2})2\epsilon\ln 2 )$ & $Z_1{\sim} \mathcal{N}(0,(1-2\epsilon\ln 2)\sigma^2)$\\[0.5ex]
 & \hspace{-0.3cm}$Z_2{\sim} \mathcal{N}(0,(1\!-\!\frac{\rho_1^2}{1+\rho_1^2}\!-\!\frac{1+2\rho_1^2}{1+\rho_1^2}2\epsilon\ln 2)\sigma^2)$& \\[0.5ex]
&  \hspace{-0.3cm}$D_1{=}2(1-\sqrt{2\epsilon\ln 2})\sigma^2$& \hspace{-0.5cm}$D_1{=}0$ & $D_1{=}2(1-\sqrt{2\epsilon\ln 2})\sigma^2$\\
&  \hspace{-0.3cm}$D_2{=}2(1-\sqrt{(1+\rho_1^2)2\epsilon\ln 2})\sigma^2$ & \hspace{-0.5cm}$D_2{=}2(1-\rho_1-(\frac{1-\rho_1^2}{2\rho_1})2\epsilon\ln 2)\sigma^2$ &  $D_2{=}0$\\[0.5ex]
 \hline\\
 &  \hspace{-0.3cm}$\hat{X}_1^G{=}\sqrt{2\epsilon\ln 2}X_1+Z_1$ & \hspace{-0.4cm}$\hat{X}_1^G{=}X_1$ &  $\hat{X}_1^G{=}\sqrt{2\epsilon\ln 2}X_1+Z_1$\\[0.5ex]
& \hspace{-0.3cm}$\hat{X}_2^G{=}\rho_1\hat{X}_1^G+\sqrt{(1-\rho_1^2)2\epsilon\ln 2}X_2+Z_2$\footnote{\tiny{As justified in \eqref{JD-first-step-O-eps}--\eqref{JD-last-step-O-eps}, the coefficient $\omega_1$ (the coefficient of $\hat{X}_1^G$ in $\hat{X}_2^G$) has some correction terms of $O(\epsilon)$ which are ignored in the presentation of $\hat{X}_2^G$ since they do not contribute to dominant terms of distortion.}}& \hspace{-0.4cm}$\hat{X}_2^G{=}(\rho_1-\rho_1\sqrt{2\epsilon\ln 2})\hat{X}_1^G+\sqrt{2\epsilon\ln 2 }X_2+Z_2$ & $\hat{X}_2^G{=}\rho_1\hat{X}_1^G+\sqrt{1-\rho_1^2}X_2$\\[0.5ex]
\begin{turn}{90}$0$-PLF-JD\end{turn} & \hspace{-0.3cm}$Z_1{\sim} \mathcal{N}(0,(1-2\epsilon\ln2)\sigma^2)$& \hspace{-0.5cm}$Z_2{\sim} \mathcal{N}(0,(1-\rho_1^2)\sigma^2)$ & $Z_1{\sim} \mathcal{N}(0,(1-2\epsilon\ln2)\sigma^2)$\\[0.5ex]
 & \hspace{-0.3cm}$Z_2{\sim} \mathcal{N}(0,(1-\rho_1^2-(1-\rho_1^2)2\epsilon\ln2)\sigma^2)$& & \\[0.5ex]
&  \hspace{-0.3cm}$D_1{=}2(1-\sqrt{2\epsilon\ln 2})\sigma^2$& \hspace{-0.5cm}$D_1{=}0$ & $D_1{=}2\sigma^2$\\[0.5ex] & \hspace{-0.3cm}$D_2{=}2(1-(\rho^2_1+\sqrt{1-\rho_1^2})\sqrt{2\epsilon\ln 2})\sigma^2$ & \hspace{-0.5cm}$D_2{=}2(1-\rho_1^2-(1-\rho_1^2)\sqrt{2\epsilon\ln 2 })\sigma^2$ & \hspace{0cm}$D_2{=}2(1-\sqrt{1-\rho_1^2}$\\[0.5ex]
 & & & \hspace{1.1cm}$-\rho_1^2\sqrt{2\epsilon \ln 2})\sigma^2$\\
\hline
\end{tabular}
\end{minipage}
\end{sc}
\end{tiny}
\end{center}
\vskip -0.1in
\end{table*}

\section{Comparison of PLFs in Low-Rate Regime}\label{comparison-app}

\begin{theorem}  For sufficiently small $\epsilon$, let $R_j=\epsilon$ and suppose that  $\rho_j=\rho$ and $\sigma_j=\sigma$, for $j=1,\ldots, T$ . The achievable distortions
$D_{\text{FMD},j}$  (for $0$-PLF-FMD), and $D_{\text{JD},j}$ (for $0$-PLF-JD) are: 
\begin{IEEEeqnarray}{rCl}
%D_{\text{FMD},1}&=& 2(1-\sqrt{2\epsilon\ln 2})\sigma^2,\\
D_{\text{FMD},j}=2(1-\Delta_{\text{FMD},j}\sqrt{2\epsilon\ln 2})\sigma^2,\quad  D_{\text{JD},j}=2(1-\Delta_{\text{JD},j}\sqrt{2\epsilon\ln 2})\sigma^2,
\end{IEEEeqnarray}
where  $\Delta_{\text{FMD},j}:=\sqrt{1+\rho^2\frac{(2\rho^2)^{j-1}-1}{2\rho^2-1}}$ and $\Delta_{\text{JD},j}:=\rho^{2(j-1)}+\mathbbm{1}\{j\geq 2\}\cdot\sqrt{1-\rho^2}(\sum_{i=0}^{j-2}\rho^{2i})$.
\end{theorem}

\begin{IEEEproof} We extend the proof in the previous section for the low-rate regime to $T$ frames.

\underline{\textit{Distortion Analysis for $0$-PLF-FMD}}:

We follow similar steps to \eqref{nu-FMD}--\eqref{D2-final-FMD} for optimization problems of the third and fourth frames and then use induction to derive expressions for $T$ frames. For simplicity, we assume that $\rho_j=\rho$ for all $j$. Notice that in the following proof, $(\hat{X}_1^G,\hat{X}_2^G)$ are as in \eqref{X1hat-Gaussian-ach}--\eqref{X2hat-Gaussian-ach} where $\nu$, $\omega_1$ and $\omega_2$ are already derived in \eqref{nu-FMD}--\eqref{D2-final-FMD}.

Now, consider the reconstruction of the third frame as follows
\begin{IEEEeqnarray}{rCl}
\hat{X}_3^G=\tau_1\hat{X}_1^G+\tau_2\hat{X}^G_2+\tau_3X_3 + Z_3,
\end{IEEEeqnarray}
for some $\tau_1,\tau_2,\tau_3$, where $\hat{X}_3^G\sim \mathcal{N}(0,\sigma^2)$ and $Z_3$ is a Gaussian random variable independent of $(\hat{X}_1^G,\hat{X}_2^G,X_3)$. The rate constraint of the third step is given by 
\begin{IEEEeqnarray}{rCl}
R_3\geq I(X_3;\hat{X}_3^G|\hat{X}_1^G,\hat{X}_2^G).
\end{IEEEeqnarray}
Evaluating the above constraint with the choice of random variables $(\hat{X}_1^G,\hat{X}_2^G,\hat{X}_3^G)$ and re-arranging the terms, we get
\begin{IEEEeqnarray}{rCl}
&&\hspace{-0.2cm}\tau_3^2\sigma^2 (1-2^{-2R_3}(\rho^42^{-2R_1-2R_2}+\rho^2(1-\rho^2)2^{-2R_2}-\rho^2))\leq\nonumber\\&& (1-2^{-2R_3})(1-\tau_1^2-\tau_2^2-2\tau_1\tau_2\omega_1\nu-2\tau_1\tau_2\omega_2\nu\rho-2\tau_2\tau_3\omega_1\nu\rho^2-2\tau_2\tau_3\omega_2\rho-2\tau_1\tau_3\nu\rho^2)\sigma^2.\nonumber\\
\end{IEEEeqnarray}
Similar to \eqref{surface}, considering the dominant terms of the above rate constraint and the fact that the solution of the optimization problem is attained when the above inequality is satisfied with ``equality'', we get 
\begin{IEEEeqnarray}{rCl}
(1-\tau_1^2-\tau_2^2+O(\epsilon^{3/2}))(2\epsilon\ln 2 +O(\epsilon^2))=\tau_3^2(1+O(\epsilon)).
\end{IEEEeqnarray}
The distortion can be written as follows
\begin{IEEEeqnarray}{rCl}
\mathbbm{E}[\|X_3-\hat{X}_3^G\|^2] &=& 2\sigma^2-2\tau_3\sigma^2-2\tau_2\omega_2\rho\sigma^2-2\tau_2\omega_1\nu\rho^2\sigma^2-2\tau_1\nu\rho^2\sigma^2.
\end{IEEEeqnarray}
So, the goal is to solve the following optimization problem for the third step
\begin{IEEEeqnarray}{rCl}
&&\min_{\tau_1,\tau_2,\tau_3}\;\; 2\sigma^2-2\tau_3\sigma^2-2\tau_2\omega_2\rho\sigma^2-2\tau_2\omega_1\nu\rho^2\sigma^2-2\tau_1\nu\rho^2\sigma^2\\
&&\text{s.t.}:\qquad\qquad (1-\tau_1^2-\tau_2^2+O(\epsilon^{3/2}))(2\epsilon\ln 2 +O(\epsilon^2))=\tau_3^2(1+O(\epsilon)).
\end{IEEEeqnarray}
We restrict the search space to  $\tau_1,\tau_2,\tau_3\geq 0$ and get an upper bound to the above optimization program as follows
\begin{IEEEeqnarray}{rCl}
&&\min_{\tau_1,\tau_2,\tau_3\geq 0}\;\; 2\sigma^2-2\tau_3\sigma^2-2\tau_2\omega_2\rho\sigma^2-2\tau_2\omega_1\nu\rho^2\sigma^2-2\tau_1\nu\rho^2\sigma^2\\
&&\text{s.t.}:\qquad\qquad (1-\tau_1^2-\tau_2^2+O(\epsilon^{3/2}))(2\epsilon\ln 2 +O(\epsilon^2))=\tau_3^2(1+O(\epsilon)).
\end{IEEEeqnarray}
The above optimization problem is equivalent to the following
\begin{IEEEeqnarray}{rCl}
&&\min_{\tau_1,\tau_2\geq 0}\Bigg(2\sigma^2-2\sqrt{\frac{(2\epsilon\ln 2+O(\epsilon^2))(1-\tau_1^2-\tau_2^2+O(\epsilon^{3/2}))}{1+O(\epsilon)}}\sigma^2\nonumber\\&&\hspace{4cm}-2\tau_2\omega_2\rho\sigma^2-2\tau_2\omega_1\nu\rho^2\sigma^2-2\tau_1\nu\rho^2\sigma^2\Bigg).\label{T-frame-FMD}
\end{IEEEeqnarray}
We proceed with solving the above optimization program.
Taking the derivative of the objective function with respect to $\eta_1$ and $\eta_2$ yields the following:
\begin{IEEEeqnarray}{rCl}
\frac{\eta_2}{\sqrt{1-\eta_1^2-\eta_2^2}}&=&\rho\sqrt{1+\rho^2}+O(\epsilon),\\
\frac{\eta_1}{\sqrt{1-\eta_1^2-\eta_2^2}}&=& \rho^2+O(\epsilon).
\end{IEEEeqnarray}
Solving the above set of equations, we get
\begin{IEEEeqnarray}{rCl}
\eta_1 &=& \frac{\rho^2}{\sqrt{1+\rho^2+2\rho^4}}+O(\epsilon),\\
\eta_2 &=& \frac{\rho\sqrt{1+\rho^2}}{\sqrt{1+\rho^2+2\rho^4}}+O(\epsilon).
\end{IEEEeqnarray}
Thus, considering the dominant terms, we get the following reconstruction for the third frame
\begin{IEEEeqnarray}{rCl}
\hat{X}_3^G&=&\frac{\rho^2}{\sqrt{1+\rho^2+2\rho^4}}\hat{X}_1^G+\frac{\rho\sqrt{1+\rho^2}}{\sqrt{1+\rho^2+2\rho^4}}\hat{X}_2^G+\frac{\sqrt{2\epsilon\ln 2}}{\sqrt{1+\rho^2+2\rho^4}}X_3+Z_3.
\end{IEEEeqnarray}
The above reconstruction yields the following distortion for the third frame
\begin{IEEEeqnarray}{rCl}
\mathbbm{E}[\|X_3-\hat{X}_3^G\|^2]=2(1-\sqrt{2\epsilon\ln 2 (1+\rho^2+2\rho^4)})\sigma^2.
\end{IEEEeqnarray}
Finally, consider the reconstruction of the fourth frame as follows
\begin{IEEEeqnarray}{rCl}
\hat{X}^G_4 = \lambda_1\hat{X}^G_1+\lambda_2\hat{X}^G_2+\lambda_3\hat{X}^G_3+\lambda_4X_4+Z_4,
\end{IEEEeqnarray}
where $\hat{X}_4^G\sim \mathcal{N}(0,\sigma^2)$. The rate constraint of the fourth step implies that 
\begin{IEEEeqnarray}{rCl}
(1-\lambda_1^2-\lambda_2^2-\lambda_3^2+O(\epsilon))(2\epsilon\ln 2 +O(\epsilon))=\lambda_4^2(1+O(\epsilon)).
\end{IEEEeqnarray}
The distortion can be written as follows
\begin{IEEEeqnarray}{rCl}
\mathbbm{E}[\|X_4-\hat{X}_4^G\|^2] &=& 2\sigma^2-2\lambda_4\sigma^2-2\lambda_3\rho\tau_3\sigma^2-2\lambda_3\rho^2\tau_2\omega_2\sigma^2-2\lambda_3\rho^3\tau_2\omega_1\nu\sigma^2\nonumber\\&&\hspace{0.2cm}-2\lambda_3\rho^3\tau_1\nu\sigma^2-2\lambda_2\rho^3\omega_1\nu\sigma^2-2\lambda_2\rho^2\omega_2\sigma^2-2\lambda_1\rho^3\nu\\
&=& 2\sigma^2-2\sqrt{(2\epsilon\ln 2) (1-\lambda_1^2-\lambda_2^2-\lambda_3^2)}\sigma^2-2\lambda_3\rho\tau_3\sigma^2\nonumber\\&&\hspace{0.2cm}-2\lambda_3\rho^2\tau_2\omega_2\sigma^2-2\lambda_3\rho^3\tau_2\omega_1\nu\sigma^2-2\lambda_3\rho^3\tau_1\nu\sigma^2\nonumber\\&&\hspace{0.2cm}-2\lambda_2\rho^3\omega_1\nu\sigma^2-2\lambda_2\rho^2\omega_2\sigma^2-2\lambda_1\rho^3\nu+O(\epsilon).
\end{IEEEeqnarray}
We take the derivative of the above expression with respect to $\lambda_1$, $\lambda_2$ and $\lambda_3$ and we get
\begin{IEEEeqnarray}{rCl}
\frac{\lambda_1}{\sqrt{1-\lambda_1^2-\lambda_2^2-\lambda_3^2}}&=& \rho^3+O(\epsilon),\\
\frac{\lambda_2}{\sqrt{1-\lambda_1^2-\lambda_2^2-\lambda_3^2}}&=& \rho^2\sqrt{1+\rho^2}+O(\epsilon),\\
\frac{\lambda_3}{\sqrt{1-\lambda_1^2-\lambda_2^2-\lambda_3^2}}&=& \rho\sqrt{1+\rho^2+2\rho^4}+O(\epsilon).
\end{IEEEeqnarray}
Solving the above set of equations yields the following
\begin{IEEEeqnarray}{rCl}
\lambda_1 &=& \frac{\rho^3}{\sqrt{1+\rho^2+2\rho^4+4\rho^6}}+O(\epsilon),\\
\lambda_2 &=& \frac{\rho^2\sqrt{1+\rho^2}}{\sqrt{1+\rho^2+2\rho^4+4\rho^6}}+O(\epsilon),\\
\lambda_3 &=& \frac{\rho\sqrt{1+\rho^2+2\rho^4}}{\sqrt{1+\rho^2+2\rho^4+4\rho^6}}+O(\epsilon).
\end{IEEEeqnarray}
Thus, considering the dominant terms, we can write 
\begin{IEEEeqnarray}{rCl}
\hat{X}_4^G &=& \frac{\rho^3}{\sqrt{1+\rho^2+2\rho^4+4\rho^6}}\hat{X}_1^G+ \frac{\rho^2\sqrt{1+\rho^2}}{\sqrt{1+\rho^2+2\rho^4+4\rho^6}}\hat{X}_2^G\nonumber\\&&\hspace{0.5cm}
+\frac{\rho\sqrt{1+\rho^2+2\rho^4}}{\sqrt{1+\rho^2+2\rho^4+4\rho^6}}\hat{X}_3^G
+\frac{\sqrt{2\epsilon\ln 2}}{\sqrt{1+\rho^2+2\rho^4+4\rho^6}}X_4+Z_4.
\end{IEEEeqnarray}
The distortion term then becomes:
\begin{IEEEeqnarray}{rCl}
&&\mathbbm{E}[\|X_4-\hat{X}_4^G\|^2]=2(1-\sqrt{2\epsilon\ln 2(1+\rho^2+2\rho^4+4\rho^6}))\sigma^2.
\end{IEEEeqnarray}
Now, we use induction to derive the terms for $T$ frames.  Define
\begin{IEEEeqnarray}{rCl}
\Delta_{\text{FMD},j}&:=& \sqrt{1+\sum_{i=1}^{j-1}2^{j-1-i}\rho^{2(j-i)}},\qquad j=2,\ldots,T\\
&=& \sqrt{1+\rho^2\frac{(2\rho^2)^{j-1}-1}{2\rho^2-1}}.
\end{IEEEeqnarray}
Thus, we have
\begin{IEEEeqnarray}{rCl}
\hat{X}_j^G=\sum_{i=1}^{j-1}\frac{\Delta_{\text{FMD},i}\rho^{j-i}}{\Delta_{\text{FMD},j}}\hat{X}_i^G+\frac{\sqrt{2\epsilon\ln 2}}{\Delta_{\text{FMD},j}}X_j+Z_j,\qquad j=2,\ldots,T,
\end{IEEEeqnarray}
where $Z_j$ is a Gaussian random variable independent of $(\hat{X}_1^G,\ldots,\hat{X}_{j-1}^G,X_j)$ and its variance is such that $\mathbbm{E}[(\hat{X}_j^G)^2]=\sigma^2$. The distortion is given by the following expression
\begin{IEEEeqnarray}{rCl}
D_{\text{FMD},j}=\mathbbm{E}[\|X_j-\hat{X}_j\|^2]=2(1-\Delta_{\text{FMD},j}\sqrt{2\epsilon\ln 2})\sigma^2,\qquad j=2,\ldots,T.
\end{IEEEeqnarray}
For the special case where $\rho=1$, then the distortion simplifies to the following
\begin{IEEEeqnarray}{rCl}
\mathbbm{E}[\|X_j-\hat{X}_j\|^2]=2(1-2^{\frac{j-1}{2}}\sqrt{2\epsilon\ln 2})\sigma^2,\qquad j=2,\ldots,T,
\end{IEEEeqnarray}
which shows an exponential decrease at each step.

\underline{\textit{Distortion Analysis for $0$-PLF-JD}}:

In this case, the proof for $T$ frames is similar to \eqref{JD-1st-step}--\eqref{D2-JD-last}. Thus, we have
\begin{IEEEeqnarray}{rCl}
\hat{X}_j^G&=&\rho\hat{X}_{j-1}^G+\sqrt{(1-\rho^2)2\epsilon\ln 2}X_j+Z_j, j=2,\ldots,T,\label{JD-T-optimal-recon}
\end{IEEEeqnarray}
where $Z_j$ is a Gaussian random variable independent of $(\hat{X}_{j-1}^G,X_j)$ and its variance is such that $\mathbbm{E}[(\hat{X}_T^G)^2]=\sigma^2$. It should be mentioned that preserving the correlation coefficients, e.g., $\mathbbm{E}[\hat{X}_j^G\hat{X}_{j-1}^G]=\rho$, needs some correction terms of $O(\epsilon)$ as discussed in \eqref{delta-epsilon-justification}. However, as shown in \eqref{D2-JD-last}, these correction terms do not contribute to dominant terms of distortion and hence, they can be ignored in the presentation of~\eqref{JD-T-optimal-recon}. Now, define 
\begin{IEEEeqnarray}{rCl}
\Delta_{\text{JD},j}:=\rho^{2(j-1)}+\sqrt{1-\rho^2}(\sum_{i=0}^{j-2}\rho^{2i}),\qquad j=2,\ldots,T,
\end{IEEEeqnarray}
and notice that
\begin{IEEEeqnarray}{rCl}
D_{\text{JD},j}&:=&\mathbbm{E}[\|X_j-\hat{X}_j\|^2]\\
&=& 2\sigma^2-2\mathbbm{E}[X_j\hat{X}_j]\\
&=& 2\sigma^2-2\mathbbm{E}[X_j(\rho\hat{X}^G_{j-1}+\sqrt{(1-\rho^2)2\epsilon \ln 2}X_j)]\\
&=& 2\sigma^2-2\mathbbm{E}[X_j(\rho^{j-1}X_1+\sqrt{1-\rho^2}(\rho^{j-2}X_2+\ldots +X_j))]\sqrt{2\epsilon\ln 2}\sigma^2\\
&=&2(1-\Delta_{\text{JD},j}\sqrt{2\epsilon \ln 2})\sigma^2.
\end{IEEEeqnarray}
For the special case of $\rho=1$, we get $\Delta_{\text{JD},j}=1$ which remains a constant across different steps.
\end{IEEEproof}

\section{Universality Statement for Gauss-Markov Source Model}
\subsection{MMSE Representations for a Given Rate}\label{MMSE-representation-appendix}
For a given rate tuple $\mathsf{R}$, the minimum distortions achievable by MMSE representations are derived in \cite{Ashishproof, MaIshwar} and are given by  
\begin{IEEEeqnarray}{rCl}
D_1^{\min} &= & \sigma_1^22^{-2R_1},\\
D_2^{\min} &= & (\rho_1^2\frac{\sigma_2^2}{\sigma_1^2}D_1^{\min}+\sigma_{N_1}^2)2^{-2R_2},\\
D_3^{\min} &= & (\rho_2^2\frac{\sigma_3^2}{\sigma_2^2}D_2^{\min}+\sigma_{N_2}^2)2^{-2R_3},
\end{IEEEeqnarray}
where 
\begin{IEEEeqnarray}{rCl}
\sigma_{N_1}^2&:=& (1-\rho_1^2)\sigma_2^2,\\
\sigma_{N_2}^2&:=& (1-\rho_2^2)\sigma_3^2.
\end{IEEEeqnarray}
The above distortions are achieved by the following optimal reconstructions $\hat{\mathsf{X}}_r$ given in \cite{MaIshwar}. Notice that the MMSE representation is $\mathsf{X}_r^{\text{RD}}=\hat{\mathsf{X}}_r$, i.e., the functions $\eta_1(.)$ and $\eta_2(.,.)$ of iRDP region $\mathcal{C}_{\mathsf{RDP}}$ (Definition~\ref{iRDP-region}) are identity functions (this statement follows from Theorem~\ref{Gaussian-optimality-thm}). Now, we choose the reconstruction $\hat{\mathsf{X}}_r$ in the following.

The reconstruction $\hat{X}_{r,1}$ is chosen such that $\hat{X}_{r,1}\to X_1\to (X_2,X_3)$ holds a Markov chain and
\begin{IEEEeqnarray}{rCl}
X_1=\hat{X}_{r,1}+Z_1,
\end{IEEEeqnarray}
where $\hat{X}_{r,1}\sim \mathcal{N}(0,\sigma_1^2-D_{1}^{\min}) $ and $Z_1\sim \mathcal{N}(0,D_1^{\min})$ are independent random variables. Then, the reconstruction $\hat{X}_{r,2}$ is chosen as follows. Let 
\begin{IEEEeqnarray}{rCl}
W_2:=\rho_1\frac{\sigma_2}{\sigma_1}Z_1+N_1,
\end{IEEEeqnarray}
which is the innovation from $\hat{X}_{r,1}$ to $X_2$. Now, we find the random variables $\hat{W}_2$ and $Z_2$ such that 
\begin{IEEEeqnarray}{rCl}
W_2=\hat{W}_2+Z_2,
\end{IEEEeqnarray}
where $\hat{W}_2\sim \mathcal{N}(0,\rho_1^2\frac{\sigma_2^2}{\sigma_1^2}D_1^{\min}+\sigma_{N_1}^2-D_2^{\min})$ and $Z_2\sim \mathcal{N}(0,D_2^{\min})$ are independent from each other, and the Markov chain $\hat{W}_2\to (X_2,\hat{X}_{r,1})\to (X_1,X_3)$ holds. Now, define 
\begin{IEEEeqnarray}{rCl}
\hat{X}_{r,2}:=\rho_1\frac{\sigma_2}{\sigma_1}\hat{X}_{r,1}+\hat{W}_2.
\end{IEEEeqnarray}
Finally, we choose the reconstruction $\hat{X}_{r,3}$ as follows. Let 
\begin{IEEEeqnarray}{rCl}
W_3:=\rho_2\frac{\sigma_3}{\sigma_2}Z_2+N_2,
\end{IEEEeqnarray}
which is the innovation from $\hat{X}_{r,2}$ to $X_3$. Now, we find random variables $\hat{W}_3$ and $Z_3$ such that 
\begin{IEEEeqnarray}{rCl}
W_3=\hat{W}_3+Z_3,
\end{IEEEeqnarray}
where $\hat{W}_3\sim \mathcal{N}(0,\rho_2^2\frac{\sigma_3^2}{\sigma_2^2}D_2^{\min}+\sigma_{N_2}^2-D_3^{\min})$ and $Z_2\sim \mathcal{N}(0,D_3^{\min})$ are independent from each other, and the Markov chain $\hat{W}_3\to (X_3,\hat{X}_{r,1},\hat{X}_{r,2})\to (X_1,X_2)$ holds. Now, define 
\begin{IEEEeqnarray}{rCl}
\hat{X}_{r,3}:=\rho_1\frac{\sigma_3}{\sigma_2}\hat{X}_{r,2}+\hat{W}_3.
\end{IEEEeqnarray}
Thus, the optimal reconstruction $\hat{\mathsf{X}}_r$ is chosen and it satisfies the rate constraint $\mathsf{R}$.

\subsection{Universality Statement}\label{Gaussian-universality-app}

\begin{theorem}  For a given rate tuple $\mathsf{R}$ with strictly positive components, let the MMSE representation  be denoted as $\mathsf{X}_r^{\text{RD}}= (X^{\text{RD}}_{r,1},X^{\text{RD}}_{r,2},X^{\text{RD}}_{r,3})$. Let $(\mathsf{D},\mathsf{P})\in\mathcal{DP}(\mathsf{R})$ and let $\hat{\mathsf{X}}=(\hat{X}_1,\hat{X}_2,\hat{X}_3)$ be the corresponding reconstruction achieving it. Then  there exist $\kappa_1$, $\theta_1$, $\theta_{2}$, $\psi_{1}$, $\psi_{2}$ and $\psi_{3}$ and noise variables $(Z_1$, $Z_2$, $Z_3)$ independent of $(X^{\text{RD}}_{r,1},X^{\text{RD}}_{r,2},X^{\text{RD}}_{r,3})$,  which satisfy the following
\begin{IEEEeqnarray}{rCl}
\hat{X}_{1} = \kappa_{1}X^{\text{RD}}_{r,1}+Z_1,\quad
\hat{X}_{2} = \theta_{1}X^{\text{RD}}_{r,1}+\theta_2X^{\text{RD}}_{r,2}+Z_2,\quad
\hat{X}_{3} = \psi_{1}X^{\text{RD}}_{r,1}+\psi_{2}X^{\text{RD}}_{r,2}+\psi_{3}\hat{X}^{\text{RD}}_{r,3}+Z_3. \notag 
\end{IEEEeqnarray}

For a given positive rate tuple $\mathsf{R}$, let the MMSE representation  $\mathsf{X}_r^{\text{RD}}$ be in the set $\mathcal{P}^{\text{RD}}(\mathsf{R})$. Also,  let $(\mathsf{D},\mathsf{P})\in\mathcal{DP}(\mathsf{R})$ and $\mathsf{X}_r$, $\hat{\mathsf{X}}$ be the corresponding representation and reconstruction achieving it.
\end{theorem}
\begin{IEEEproof} First, notice that according to the proof of Theorem~\ref{Gaussian-optimality-thm} for the Gauss-Markov source model, one can set $\hat{\mathsf{X}}=\mathsf{X}_r$ in iRDP region of $\mathcal{C}_{\mathsf{RDP}}$, without loss of optimality. So, in the following proof, the  reconstruction $\mathsf{X}_r$ and representation $\hat{\mathsf{X}}$ are used interchangeably, in some places.

We show the following statement. If 
\begin{IEEEeqnarray}{rCl}
R_1&\geq & I(X_1;X_{r,1}),\label{R1-univ-Gaus-app}\\
R_2&\geq & I(X_2;X_{r,2}|X_{r,1}),\label{R2-univ-Gaus-app}\\
R_3&\geq & I(X_3;X_{r,3}|X_{r,1},X_{r,2}),\label{R3-univ-Gaus-app}
\end{IEEEeqnarray}
then, there exist $\kappa_1$, $\theta_1$, $\theta_{2}$, $\psi_{1}$, $\psi_{2}$ and $\psi_{3}$ and noise variables $Z_1$, $Z_2$, $Z_3$ independent of $X^{\text{RD}}_{r,1}$, $(X^{\text{RD}}_{r,1},X^{\text{RD}}_{r,2})$, $(X^{\text{RD}}_{r,1},X^{\text{RD}}_{r,2},X^{\text{RD}}_{r,3})$, respectively, which satisfy the following
\begin{IEEEeqnarray}{rCl}
\hat{X}_{1} &=& \kappa_{1}X^{\text{RD}}_{r,1}+Z_1,\label{transform1-app}\\
\hat{X}_{2} &=& \theta_{1}X^{\text{RD}}_{r,1}+\theta_2X^{\text{RD}}_{r,2}+Z_2,\label{transform2-app}\\
\hat{X}_{3} &=& \psi_{1}X^{\text{RD}}_{r,1}+\psi_{2}X^{\text{RD}}_{r,2}+\psi_{3}\hat{X}^{\text{RD}}_{r,3}+Z_3.\label{transform3-app}
\end{IEEEeqnarray}
If \eqref{R1-univ-Gaus-app}--\eqref{R3-univ-Gaus-app} are satisfied with equality, then the noise random variables in \eqref{transform1-app}--\eqref{transform3-app} do not exist and a linear combination is sufficient for converting $(X_{r,1}^{\text{RD}},X_{r,2}^{\text{RD}},X_{r,3}^{\text{RD}})$ to $(\hat{X}_1,\hat{X}_2,\hat{X}_3)$.

First, we prove the statement when all of inequalities in \eqref{R1-univ-Gaus-app}--\eqref{R3-univ-Gaus-app} hold with ``equality''. 
We provide the proof for $T=2$ frames. The extension to arbitrary number of frames is straightforward. 
To that end, we first prove the following two lemmas.
\begin{lemma}\label{step1-lemma} Without loss of optimality, the reconstruction of the first step $\hat{X}_1$ satisfies the following
\begin{IEEEeqnarray}{rCl}
\gamma_{1}\hat{X}_{1}=W_1,
\end{IEEEeqnarray}
where
\begin{IEEEeqnarray}{rCl}
\gamma_{1}:=\frac{\mathbbm{E}[X_1\hat{X}_{1}]}{\sigma^2_{\hat{X}_{1}}},\label{gam1-def}
\end{IEEEeqnarray}
and $W_1$ is a Gaussian random variable that its statistics do not depend on the pair $(D_1,P_1)$.
\end{lemma}
\begin{IEEEproof} 
According to Theorem~\ref{Gaussian-optimality-thm}, we know that $(X_1,\hat{X}_{1})$ are jointly Gaussian. So,  we can write $X_1$ as follows
\begin{IEEEeqnarray}{rCl}
X_1 = \gamma_{1} \hat{X}_{1}+T_{1},\label{X1-diff-noise22}
\end{IEEEeqnarray}
where $T_{1}$ is a Gaussian random variable independent of $\hat{X}_{1}$ with a constant variance $\sigma_1^22^{-2R_1}$. Notice that~\eqref{X1-diff-noise22} can be written as follows
\begin{IEEEeqnarray}{rCl}
\hat{X}_{1}=\alpha_{1} (X_1+Q),\label{alpha-Xhat}
\end{IEEEeqnarray}
where $Q$ is a Gaussian random variable independent of $X_1$ with a zero-mean and variance $\frac{\sigma_1^22^{-2R_1}}{1-2^{-2R_1}}$ and 
\begin{IEEEeqnarray}{rCl}
\alpha_{1} := \frac{1}{\gamma_{1}}(1-2^{-2R_1}).
\end{IEEEeqnarray}
From \eqref{alpha-Xhat}, we get
\begin{IEEEeqnarray}{rCl}
\gamma_{1}\hat{X}_{1}=(1-2^{-2R_1})(X_1+Q).
\end{IEEEeqnarray}
Now, defining $W_1:=(1-2^{-2R_1})(X_1+Q)$ yields the desired result.
\end{IEEEproof}
\begin{lemma}\label{step2-lemma} Without loss of optimality, the reconstructions of the first and second steps $(\hat{X}_1,\hat{X}_2)$ satisfy the following
\begin{IEEEeqnarray}{rCl}
\lambda_{1} \hat{X}_{1}+\lambda_{2} \hat{X}_{2}=W_2,
\end{IEEEeqnarray}
where 
\begin{IEEEeqnarray}{rCl}
\lambda_{1}&:=& \frac{\rho_1\mathbbm{E}[X_1\hat{X}_{1}]\hat{\sigma}^2_{X_{2}}-\mathbbm{E}[\hat{X}_{1}\hat{X}_{2}]\mathbbm{E}[X_2\hat{X}_{2}]}{\hat{\sigma}^2_{X_{1}}\hat{\sigma}^2_{X_{2}}-\mathbbm{E}^2[\hat{X}_{1}\hat{X}_{2}]},\label{lam1-def}\\
\lambda_{2}&:=& \frac{\rho_1\mathbbm{E}[X_1\hat{X}_{1}]\mathbbm{E}[\hat{X}_{1}\hat{X}_{2}]-\hat{\sigma}^2_{X_{1}}\mathbbm{E}[X_2\hat{X}_{2}]}{\hat{\sigma}^2_{X_{1}}\hat{\sigma}^2_{X_{2}}-\mathbbm{E}^2[\hat{X}_{1}\hat{X}_{2}]},\label{lam2-def}
\end{IEEEeqnarray}
and $W_2$ is a Gaussian random variable that its statistics do not depend on the pairs $(D_1,P_1)$ and $(D_2,P_2)$. 
\end{lemma}
\begin{IEEEproof} According to Theorem~\ref{Gaussian-optimality-thm}, we know that $(X_1, X_2,\hat{X}_{1},\hat{X}_2)$ are jointly Gaussian. So,  we can write $X_2$ as follows
\begin{IEEEeqnarray}{rCl}
X_2 = \lambda_{1} \hat{X}_{1}+\lambda_2\hat{X}_2+T_{2},\label{X2-diff-noise2}
\end{IEEEeqnarray}
where $T_{2}$ is a Gaussian random variable independent of $(\hat{X}_{1},\hat{X}_2)$ with a constant variance of $\sigma_{X_2|\hat{X}_1}^22^{-2R_2}$ where
\begin{IEEEeqnarray}{rCl}
\sigma_{X_2|\hat{X}_1}^2:=\frac{1}{2}\log\left(\rho_1^2\sigma_1^22^{-2R_1}+2^{2H(N_1)}\right).
\end{IEEEeqnarray}
Notice that~\eqref{X2-diff-noise2} can be written as follows
\begin{IEEEeqnarray}{rCl}
\lambda_{1} \hat{X}_{1}+\lambda_2\hat{X}_2=(1-2^{-2R_2})(X_2+Q'),
\end{IEEEeqnarray}
where $Q'$ is a Gaussian random variable independent of $X_2$ with a zero-mean and variance $\frac{\sigma_{X_2|\hat{X}_1}^22^{-2R_2}}{1-2^{-2R_2}}$. Defining $W_2:=(1-2^{-2R_2})(X_2+Q')$ yields the desired result.
\end{IEEEproof}
Now, we proceed with the proof of the theorem.  According to Lemma~\ref{step1-lemma}, there exist real $\gamma_{1}$ and $\gamma'_{1}$ such that 
\begin{IEEEeqnarray}{rCl}
\gamma_{1}\hat{X}_{1}=\gamma'_{1}X^{\text{RD}}_{r,1}.\label{first-frame-reconstruction}
\end{IEEEeqnarray}
Define \begin{IEEEeqnarray}{rCl}
\kappa_{1}:=\frac{\gamma'_{1}}{\gamma_{1}}.\label{kap}\end{IEEEeqnarray} Then, according to Lemma~\ref{step2-lemma}, there exist $\lambda_{1}$, $\lambda_{2}$, $\lambda'_{1}$ and $\lambda'_{2}$ such that
\begin{IEEEeqnarray}{rCl}
&&\lambda_{1}\hat{X}_{1}+\lambda_{2}\hat{X}_{2}=\lambda'_{1}X^{\text{RD}}_{r,1}+\lambda'_{2}X^{\text{RD}}_{r,2}.
\end{IEEEeqnarray}
The above equation can be written as 
\begin{IEEEeqnarray}{rCl}
\hat{X}_{2}&=& \frac{\lambda'_{1}-\lambda_{1}\kappa_{1}}{\lambda_{2}}X^{\text{RD}}_{r,1}+\frac{\lambda'_{2}}{\lambda_{2}}X^{\text{RD}}_{r,2}\\
&:=&\theta_{1}X^{\text{RD}}_{r,1}+\theta_{2}X^{\text{RD}}_{r,2}. \label{second-frame-reconstruction}
\end{IEEEeqnarray}
A similar justification holds for the third frame.

Next, we prove the statement when at least one of the rate constraints in \eqref{R1-univ-Gaus-app}--\eqref{R3-univ-Gaus-app} hold with strict inequality. In the following, we construct new reconstructions $(\hat{X}'_1,\hat{X}'_2)$ based on $(\hat{X}_1,\hat{X}_2)$ such that they satisfy the rate constraints $(R_1,R_2)$ with equality. Then, we will be able to apply the two lemmas we proved to show that $(\hat{X}_1,\hat{X}_2)$ are linearly related to MMSE reconstructions $(X_{r,1}^{\text{RD}},X_{r,2}^{\text{RD}})$. 

\underline{\textit{Construction of $\hat{X}'_1$}}:

Now, let
\begin{IEEEeqnarray}{rCl}
\hat{R}_1:=I(X_1;\hat{X}_1), 
\end{IEEEeqnarray}
where $\hat{R}_1\leq R_1$. Also, recall that 
\begin{IEEEeqnarray}{rCl}
R_1=I(X_1;X^{\text{RD}}_{r,1}).
\end{IEEEeqnarray}
Now, let $\hat{X}'_1$ such that $\hat{X}'_1\to X^{\text{RD}}_{r,1}\to X_1$ holds and 
\begin{IEEEeqnarray}{rCl}
\hat{X}'_1=X^{\text{RD}}_{r,1}+W_1,\label{X1-deg}
\end{IEEEeqnarray}
where $W_1\sim \mathcal{N}(0,\nu_1^2)$ independent of $\hat{X}_1$ and $\nu_1^2$ will be determined in the following. Notice that $I(X_1;\hat{X}'_1)$ is a monotonically decreasing function of $\nu_1^2$. So, one choose $\nu_1^2$ such that 
\begin{IEEEeqnarray}{rCl}
I(\hat{X}'_1;X_1)=I(X_1;\hat{X}_1)=\hat{R}_1.
\end{IEEEeqnarray}
Now, according to Lemma~\ref{step1-lemma}, since $\hat{X}'_1$ and $\hat{X}_1$ have the same rates, there exists a coefficient $\kappa'_1$ such that 
\begin{IEEEeqnarray}{rCl}
\hat{X}_1&=&\kappa'_1\hat{X}'_1\\
&=& \kappa'_1X_{r,1}^{\text{RD}}+\kappa'_1W_1.
\end{IEEEeqnarray}
Now, define $Z_1:=\kappa'_1W_1$ and notice that
\begin{IEEEeqnarray}{rCl}
\hat{X}_1=\kappa'_1X_{r,1}^{\text{RD}}+Z_1.\label{X1-new-deg}
\end{IEEEeqnarray}

\underline{\textit{Construction of $\hat{X}'_2$}}:

Next, consider the second step. Define 
\begin{IEEEeqnarray}{rCl}
\hat{R}_2:=I(X_2;\hat{X}_2|\hat{X}_1),
\end{IEEEeqnarray}
where $\hat{R}_2\leq R_2$. Also, recall that 
\begin{IEEEeqnarray}{rCl}
R_2=I(X_2;X_{r,2}^{\text{RD}}|X_{r,1}^{\text{RD}}).
\end{IEEEeqnarray}
Define $\tilde{X}_2:=\mathbbm{E}[X_2|X_{r,1}^{\text{RD}},X_{r,2}^{\text{RD}}]$ to be the MMSE reconstruction and consider that
\begin{IEEEeqnarray}{rCl}
R_2 &=& I(X_2;X_{r,2}^{\text{RD}}|X_{r,1}^{\text{RD}})\\
&=& I(X_2;\tilde{X}_2|X_{r,1}^{\text{RD}}),
\end{IEEEeqnarray}
where the last equality follows because both Markov chains $X_2\to (X_{r,1}^{\text{RD}},X_{r,2}^{\text{RD}})\to \tilde{X}_2$ and $X_2\to \tilde{X}_2\to (X_{r,1}^{\text{RD}},X_{r,2}^{\text{RD}})$ hold where the latter one is satisfied for Gaussian random variables for which we can write $X_2=\mathbbm{E}[X_2|X_{r,1},X_{r,2}]+W'$ such that $W'$ is independent of $(X_{r,1}^{\text{RD}},X_{r,2}^{\text{RD}})$.

Now, we show that $I(X_2;\tilde{X}_2|X_{r,1}^{\text{RD}})\leq I(X_2;\tilde{X}_2|\hat{X}'_1)$. This is justified in the following
\begin{IEEEeqnarray}{rCl}
I(X_2;\tilde{X}_2|\hat{X}'_1) &=& I(X_2;\tilde{X}_2|X^{\text{RD}}_{r,1}+W_1)\\
&=&H(X_2|X^{\text{RD}}_{r,1}+W_1)-H(X_2|\tilde{X}_2,X^{\text{RD}}_{r,1}+W_1)\\
&\geq & H(X_2|X^{\text{RD}}_{r,1}+W_1,W_1)-H(X_2|\tilde{X}_2,X^{\text{RD}}_{r,1}+W_1)\\
&=& H(X_2|X^{\text{RD}}_{r,1},W_1)-H(X_2|\tilde{X}_2,X^{\text{RD}}_{r,1}+W_1)\\
&\geq & H(X_2|X^{\text{RD}}_{r,1},W_1)-H(X_2|\tilde{X}_2)\\
&=& H(X_2|X^{\text{RD}}_{r,1})-H(X_2|\tilde{X}_2)\label{gaus-univ-step1}\\
&=& H(X_2|X^{\text{RD}}_{r,1})-H(X_2|\tilde{X}_2,X_{r,1}^{\text{RD}})\label{gaus-univ-step2}\\
&=& I(X_2;\tilde{X}_2|X_{r,1}^{\text{RD}}),
\end{IEEEeqnarray}
where \eqref{gaus-univ-step1} follows because $W_1$ is independent of $(X_2,X_{r,1}^{\text{RD}})$ and \eqref{gaus-univ-step2} follows from the Markov chain $X_2\to \tilde{X}_2\to X_{r,1}^{\text{RD}}$.

Define
\begin{IEEEeqnarray}{rCl}
R'_2:=I(X_2;\tilde{X}_2|\hat{X}'_1),
\end{IEEEeqnarray}
and consider the fact that $R'_2\geq R_2$. Now, we introduce $\hat{X}'_2$ such that $\hat{X}'_2\to (\tilde{X}_2,\hat{X}'_1)\to X_2$ forms a Markov chain and 
\begin{IEEEeqnarray}{rCl}
\hat{X}'_2=\tilde{X}_2+\hat{X}'_1+W_2,\label{X2-deg}
\end{IEEEeqnarray}
where $W_2\sim \mathcal{N}(0,\nu_2^2)$ independent of $(\tilde{X}_2,\hat{X}_1)$ and $\nu_2^2$ will be determined in the following. Since $I(X_2;\hat{X}'_2|\hat{X}'_1)$ is a monotonically decreasing function of $\nu_2^2$, we can choose $\nu_2^2$ such that 
\begin{IEEEeqnarray}{rCl}
I(X_2;\hat{X}'_2|\hat{X}'_1)= I(X_2;\hat{X}_2|\hat{X}_1)=\hat{R}_2.
\end{IEEEeqnarray}
Then, according to Lemma~\ref{step2-lemma}, there exist $\lambda_1'$, $\lambda'_2$, $\hat{\lambda}_1$ and $\hat{\lambda}_2$ such that
\begin{IEEEeqnarray}{rCl}
\lambda'_1\hat{X}'_1+\lambda'_2\hat{X}'_2=\hat{\lambda}_1\hat{X}_1+\hat{\lambda}_2\hat{X}_2.
\end{IEEEeqnarray}
Plugging \eqref{X1-deg}, \eqref{X1-new-deg} and \eqref{X2-deg} into the above expression and letting $\tilde{X}_2=\alpha X^{\text{RD}}_{r,1}+\beta X^{\text{RD}}_{r,2}$ for some $\alpha,\beta$, we get
\begin{IEEEeqnarray}{rCl}
(\lambda'_1+(1+\alpha)\lambda'_2-\hat{\lambda}_1\kappa')X_{r,1}^{\text{RD}}+\lambda'_2\beta X_{r,2}^{\text{RD}}+(\lambda'_1+\lambda'_2)W_1+\lambda'_2W_2-\hat{\lambda}_1Z_1 = \hat{\lambda}_2\hat{X}_2.
\end{IEEEeqnarray}
Now define
\begin{IEEEeqnarray}{rCl}
\theta_1&:=&\frac{\lambda'_1+(1+\alpha)\lambda'_2-\hat{\lambda}_1\kappa'}{\hat{\lambda}_2},\\
\theta_2&:=& \frac{\lambda'_2\beta}{\hat{\lambda}_2},\\
Z_2&:=& \frac{(\lambda'_1+\lambda'_2)}{\hat{\lambda}_2}W_1+\frac{\lambda'_2}{\hat{\lambda}_2}W_2-\frac{\hat{\lambda}_1}{\hat{\lambda}_2}Z_1.
\end{IEEEeqnarray}
Thus, we have 
\begin{IEEEeqnarray}{rCl}
\hat{X}_{2} &=& \theta_{1}X^{\text{RD}}_{r,1}+\theta_2X^{\text{RD}}_{r,2}+Z_2.
\end{IEEEeqnarray}
Notice that the above proof only uses the information about reconstructions of the operating points in DP-tradeoff and it does not depend on the choice of PLF. So, it holds for both PLF-JD and PLF-FMD. This concludes the proof.
\end{IEEEproof}

\subsection{Gaussian Example}\label{Gaussian-example-app}
 Assume that the sources are symmetric in the sense that $\sigma_1^2=\sigma_2^2=\sigma_3^2=1$, $\rho_1=\rho_2=\rho_3:=\rho$ for some $0<\rho \leq 1$. Also, suppose that the perception thresholds are symmetric, i.e., $P_1=P_2=P_3:=P$ for some $0<P\leq 1$. We choose the rate tuple $\mathsf{R}$ such that the minimum distortions $D_j^{\min}=D$ for $j\in\{1,2,3\}$. According to Appendix~\ref{MMSE-representation-appendix}, such rates are given by
 \begin{IEEEeqnarray}{rCl}
 R_1&=& \frac{1}{2}\log \frac{1}{D},\\
 R_2&=& \frac{1}{2}\log \frac{\rho^2D+(1-\rho)}{D},\\
 R_3&=& \frac{1}{2}\log \frac{\rho^2D+(1-\rho^2)}{D}.
 \end{IEEEeqnarray}
 
The covariance matrix of the MMSE representations $\text{cov}(X_{r,1}^{\text{RD}},X_{r,2}^{\text{RD}},X_{r,3}^{\text{RD}})$ is given by $(1-D)\Sigma$ where 
\begin{IEEEeqnarray}{rCl}
\Sigma:=\begin{pmatrix}1 & \rho& \rho^2\\\rho & 1 & \rho \\ \rho^2& \rho & 1\end{pmatrix}.
\end{IEEEeqnarray}
If we introduce the $0$-PLF while keeping the rates as those of MMSE reconstructions, it can be shown that the optimal distortions are all equal to $D_1=D_2=D_3=2-2\sqrt{1-D}$.
Denote the reconstructions by $(\hat{X}^0_{D_1},\hat{X}^0_{D_2},\hat{X}^0_{D_3})$ and  notice that the covariance matrix of the reconstructions is equal to that of the sources and is given by $\Sigma$.
Thus, the covariance matrix of $(X_{r,1}^{\text{RD}},X_{r,2}^{\text{RD}},X_{r,3}^{\text{RD}})$ is $(1-D)$ times the covariance matrix of $(\hat{X}^0_{D_1},\hat{X}^0_{D_2},\hat{X}^0_{D_3})$. So, the reconstructions $(X_{r,1}^{\text{RD}},X_{r,2}^{\text{RD}},X_{r,3}^{\text{RD}})$ and $(\hat{X}^0_{D_1},\hat{X}^0_{D_2},\hat{X}^0_{D_3})$ can be transformed to each other by the scaling factor $\frac{1}{\sqrt{1-D}}$. This inspires the idea that reconstructions corresponding to different tuples $(\mathsf{D},\mathsf{P})$ are linearly related to those of MMSE representations which is the essence of the following Theorem~\ref{universal-thm}. Moreover, both PLFs either based on FMD or JD perform similarly in this example since individually scaling the reconstruction of each frame  finally ends up in matching the covariance matrix of all frames.

\section{Justification of low-rate regime for Moving MNSIT}
\label{sec:just_mnist}
In the MovingMNIST dataset, the digit in I-frame is generated uniformly across the $32 {\times} 32$ center region in a $64{\times}64$ image, meaning that $\log(32{\times}32){=}10$ bits are required to localize the digits and any lower rate would result in much less correlated reconstructions. As such, one can consider $R_1{=}12$ bits (2 extra bits for content and style) as a low rate. For P-frames, the movement is uniformly constrained within a $10{\times}10$ region so any rate $R_2{\leq} \log_2(10{\times}10){=}6.6$ bits (excluding residual compensation) can be considered a low rate.

\section{Experiment Details}\label{sec:extra-experiment}

\subsection{Training Setup and Overview}
\label{exp-setup}
Our compression architecture is built on the scale-space flow model \cite{agustsson2020scale}, which allows end-to-end training without relying on pre-trained optical flow estimators. For better compression efficiency, we replace the residual compression module with the conditioning one \cite{li2021deep}. In the following, we will interchangeably refer $X_1$ as the I-frame and subsequent ones as P-frames. The annotation for the encoder, decoder, and critic (discriminator) will be referred to as $f,g$, and $h$ respectively and their specific functionality (e.g motion compression, joint perception critic) will be described within context through a subscript/superscript.

\begin{figure}[t]
    \centering
    \includegraphics[width=1.0\textwidth]{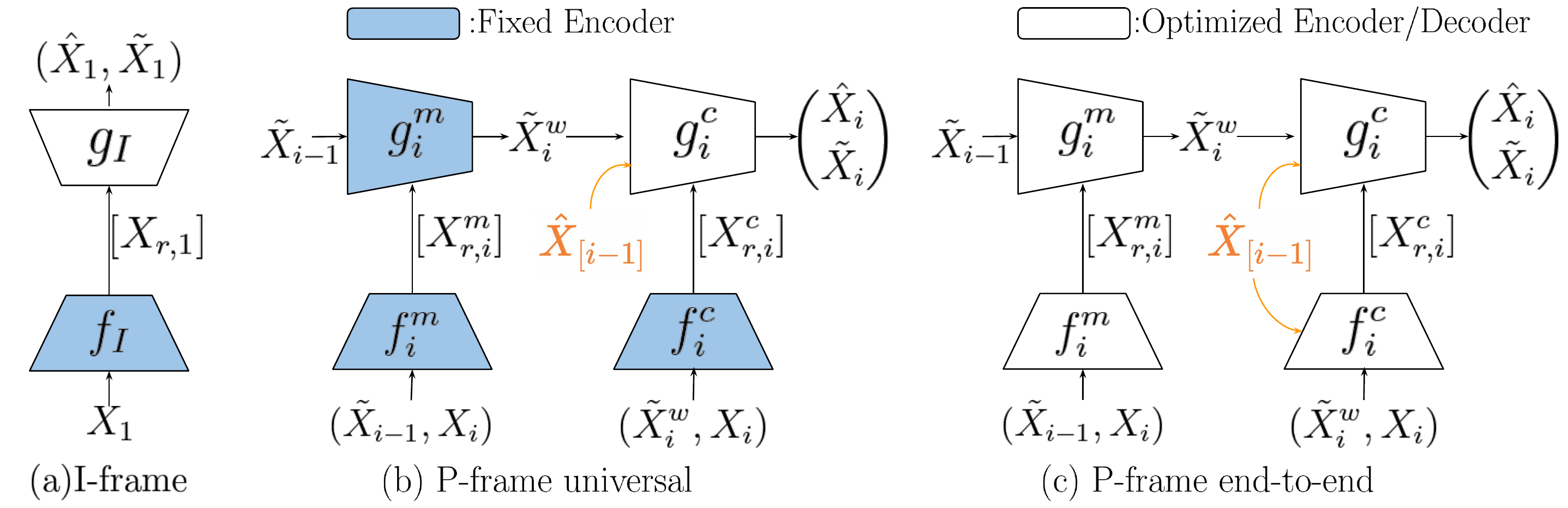}
    \caption{\textcolor{black}{Compression diagram  for (a) I-frame (b) P-frame with universal representation and (c) P-frame with optimized representation. For simplicity, we do not show the shared randomness $K$. }}
    \label{fig:figure_architecture}
\end{figure}

\noindent \emph{Distortion and Perception Measurement: }  We follow the setup in prior works \cite{blau2018perception,Jun-Ashish2021} for distortion and perception measurement. Specifically, we use MSE loss $\mathbbm{E}[||X{-}\hat{X}||^2]$ as a distortion metric and Wasserstein-1 distance as a perception metric, which can be estimated through the WGAN critics (following the Kanotorovich-Rubinstein duality). For the marginal perception metric, we optimize our critics $h_m$ to classify between original image $X$ and synthetic ones $\hat{X}$. This will then allow us to measure $W_1(P_X, P_{\hat{X}})$ since:
\begin{equation}
    W_1(P_X, P_{\hat{X}}) = \sup_{h_m \in \mathcal{F}} \mathbbm{E}[h_m(X)] - \mathbbm{E}[h_m(\hat{X})]
\end{equation}
where $\mathcal{F}$ is a set of all bounded 1-Lipschitz functions. Similarly, the joint perception metric is realized through $W_1(P_{X_1\ldots X_j}, P_{\hat{X}_1\ldots \hat{X}_j})$  by training a critic $h_j$ that classifies between synthetic and authentic sequences:
\begin{equation}
    W_1(P_{X_1\ldots X_j}, P_{\hat{X}_1\ldots \hat{X}_j}) = \sup_{h_j \in \mathcal{F}} \mathbbm{E}[h_j(X_1,...,X_i)] - \mathbbm{E}[h_j(\hat{X}_1,...,\hat{X}_i)]
\end{equation}
In practice, the set of 1-Lipschitz functions is limited by the neural network architecture. Also, although our analysis employs the Wasserstein-2 distance as a perception metric, it is worth noting that the ideal reconstructions (0-PLF) for this metric and the one used in our study should be identical. 
%$$W_1(p_X, p_{\hat{X}} ) = \mathbbm{E}$$

\noindent \emph{I-frame Compressor:} We compress I-frames in a similar fashion as previous works \cite{blau2018perception,Jun-Ashish2021}. Our encoder $f_I$ and decoder $g_I$ in Figure \ref{fig:figure_architecture}a contain a series of convolution operations and we control the rate $R_1$ by varying the dimension and quantization level in the bottleneck. The model utilizes common randomness through the dithered quantization operation. For a given rate $R_1$, we vary the amount of DP tradeoff by controlling the hyper-parameter $\lambda_i^{\text{marginal}}$ in the following minimization objective $\mathcal{L}_1$:   
\begin{equation}
    \mathcal{L}_1 = \mathbbm{E}[||X_1 - \hat{X}_1||^2] + \lambda_i^{\text{marginal}} W_1(P_{X_1}, P_{\hat{X}_1} )
    \label{marginal_objective}
\end{equation}
Following the results from Zhang et al.\cite{Jun-Ashish2021}, we fix the encoder after optimizing the encoder-decoder pair for MSE representations. We then fix the encoder and train another decoder to obtain the  optimal reconstruction with perfect perception, i.e, $W_1(P_X, P_{\hat{X}}) \approx 0$. We will leverage these universal representation results to compress P-frames (both end-to-end and universal).

\noindent \emph{P-frame Compressor: } We describe the loss functions before explaining our architectures. Given previous reconstructions $\hat{X}_{[i-1]}{:=}\{\hat{X}_1,\hat{X}_2,...,\hat{X}_{i-1}\}$, one can adjust the distortion-joint perception tradeoff by controlling the hyper-parameter $\lambda_{i}^{\text{joint}}$ in the following objective  $\mathcal{L}_i$. 
\begin{equation}
    \mathcal{L}_i^{\text{joint}} = \mathbbm{E}[||X_i - \hat{X}_i||^2] + \lambda_{i}^{\text{joint}} W_1(P_{X_{[i]}}, P_{\hat{X}_{[i]}} )
\end{equation}
Note that in order to achieve 0-PLF-JD, previous reconstructions $\hat{X}_{[i-1]}$ must also achieve 0-PLF-JD, since it is impossible to reconstruct such $\hat{X}_i$ if the previous $\hat{X}_{[i-1]}$ are not temporally consistent\footnote{ This follows from the inequality: $W^2_2(P_{X_1,X_2}, P_{\hat{X}_1,\hat{X}_2}) {\geq} W^2_2(P_{X_1},P_{\hat{X}_1}) {+} W^2_2(P_{X_2},P_{\hat{X}_2}) $}. For the FMD metric, we use the loss function in (\ref{marginal_objective}). 

In the \textit{universal model} in Figure \ref{fig:figure_architecture}b, the motion encoder $f^m_i$ compresses and sends the quantized flow fields $[X^m_{r,i}]$ between the MMSE reconstruction $\tilde{X}_{i-1}$ and $X_i$.  Given $[X^m_{r,i}]$, the flow decoder and warping module $g^m_i$ will transform $\tilde{X}_{i-1}$ into $\tilde{X}_{i}^w$ (predicted frame).  We use $f^c_i$ to compress the residual information $[X^c_{r,i}]$ between $X_i$ and $\tilde{X}_{i}^w$ \footnote{Here, we use conditioning \cite{li2021deep} instead of sending $X_i - \tilde{X}_{i-1}^w$ as in the original work \cite{agustsson2020scale}}, which will be decoded  by $g^c_i$. We note that for MMSE representation,  $g^c_i$ only requires $\tilde{X}^w_i$ as a conditional input  while an additional conditioning input $\hat{X}_{[i-1]}$ is required when perceptual optimization is involved. Together, $f^m_i, g^m_i$, $f^c_i$, and $g^c_i$ are optimized for MMSE reconstructions. To train for different DP tradeoffs, we fix $f^m_i, g^m_i,f^c_i$ and adapt the new decoder $\hat{g}^c_i$ (conditioning on $\tilde{X}^w_i, \hat{X}_{[i-1]}$). We note that fixing  $g^m_i$ for universal representation is essential since $[X^c_{r,i}]$ is dependent on the outputs $\tilde{X}_{i}^w$ of $g^m_i$. 

In the \textit{end-to-end model} in Figure \ref{fig:figure_architecture}c, we use an MMSE representation to estimate the motion vector, as in the case of the universal model. The only difference is that the encoder $f^c_i$ also uses previous $\hat{X}_i$ and the encoders will be jointly trained with the decoders.

\subsection{Networks Architecture}

In this section, we describe the network architecture for universal and end-to-end P-frame compressor models. \footnote{For the I-frame compressor, we follow the DCGAN implementation by Denton et al\cite{SVG}, adding the dithered quantization layer in the encoder's last layer( \url{https://github.com/edenton/svg/blob/master/models/dcgan_64.py})}. In the architecture layout, we denote BN2D and SN for the \emph{Batchnorm2D} and \emph{Spectral Normalization} layers.  Convolutional and transposed convolutional layer are denoted as ``conv'' and ``upconv'' respectively, which is accompanied by number of filters, kernel size, stride, and padding. %Finally, we refer ``channels'' as the number of image channels (1 for MNIST and 3 for KTH), and $d_{f^m_i}, d_{f^c_i}$ are respectively the output dimension of the motion encoder and residual encoder, which together with the number of quantization levels,  will specify the rate. 

\emph{Motion Encoder and Decoder.} The universal and optimized end-to-end model shares the same architecture for the motion encoder and decoder. ($f^m_i$ and $g^m_i$ respectively). We follow the original implementations \cite{agustsson2020scale} and present the convolutional architecture in Table \ref{motion_compressor_table}. Different from the original implementation, however, we replace the last layer with dithered quantization layer (as in \cite{Jun-Ashish2021}) in our implementation. The output dimension of the motion encoder is denoted as $d_m$.

\begin{table}[h]
    % \small
    \centering
    \caption{Motion Encoder $f^m_i$ and Decoder $g^m_i$.} \label{motion_compressor_table}
    \begin{tabular}[t]{ |c| }
    \multicolumn{1}{c}{\textbf{(a)} Encoder $f^m_i$} \\
	\hline 
        Input-$64{\times}64{\times}(2{\times} \text{channels})$ \\ 
        \hline 
        conv (64:5:2:0), BN2D, l-ReLU \\
        \hline 
        conv (64:5:2:0), BN2D, l-ReLU \\
        \hline 
        conv (64:5:2:0), BN2D, l-ReLU \\
        \hline 
        conv (64:5:2:0), BN2D, l-ReLU \\
        \hline 
        conv ($d_m$:4:2:0), BN2D \\
        \hline
        Quantizer \\
    \hline
    \end{tabular}
    \quad
    \begin{tabular}[t]{ |c| }
    \multicolumn{1}{c}{\textbf{(b)} Decoder $g^m_i$} \\
	\hline 
        Input-$d_{m}$  \\ 
        \hline 
        upconv (64:4:1:0), BN2D, l-ReLU \\
        \hline 
        upconv (64:5:2:0), BN2D, l-ReLU \\
        \hline
        upconv (64:5:2:0), BN2D, l-ReLU \\
        \hline 
        upconv (64:5:2:0), BN2D, l-ReLU \\
        \hline 
        upconv (3:5:2:0), BN2D
        \\
    \hline
    \end{tabular}
    
\end{table}

\emph{Residual Encoder and Decoder.} The architecture of the conditional residual encoder is shown in Table \ref{deep compressor}a, where we stack multiple frames along their channel dimension as an input. As described previously, in the residual encoder, the universal model requires only $X_i,\tilde{X}^w_i$  while the end-to-end model will receive $X_i,\tilde{X}^w_i$ and $\hat{X}_{[i-1]}$. We denote the output dimension of this residual encoder as $d_r$. In the decoding part, the decoder will first condition all the previous reconstructions $\hat{X}_[i-1]$ by projecting them into an embedding vector of size $192$ (conditioning module in Table \ref{deep compressor}b). Then we concatenate this vector with the output of $f^r_i$. The concatenated vector will be fed into the decoder (Table \ref{deep compressor}c) to produce the reconstruction $\hat{X}_i$.

\begin{table}[h]
    % \small
    \centering
    \caption{Residual Encoder, Conditional Module, and Residual Decoder.} 
    \begin{tabular}[t]{ |c| }
    \multicolumn{1}{c}{\textbf{(a)}Encoder $f^c_i$} \\
	\hline 
        Input  \\ 
        \hline 
        conv (64:5:2:0), BN2D, l-ReLU \\
        \hline 
        conv (64:5:2:0), BN2D, l-ReLU \\
        \hline 
        conv (64:5:2:0), BN2D, l-ReLU \\
        \hline 
        conv (64:5:2:0), BN2D, l-ReLU \\
        \hline 
        conv ($d_r$:4:1:0), BN2D \\
        \hline
        Quantizer \\
    \hline
    \end{tabular}
    \quad
    \begin{tabular}[t]{ |c| }
    \multicolumn{1}{c}{\textbf{(b)}Conditional Module} \\
	\hline 
        Input\\ 
        \hline 
        conv (64:5:2:0), BN2D, l-ReLU \\
        \hline 
        conv (64:5:2:0), BN2D, l-ReLU \\
        \hline 
        conv (64:5:2:0), BN2D, l-ReLU \\
        \hline 
        conv (64:5:2:0), BN2D, l-ReLU \\
        \hline 
        conv (192:4:1:0), BN2D \\
    \hline
    \end{tabular}
    \quad
    \begin{tabular}[t]{ |c| }
    \multicolumn{1}{c}{\textbf{(c)}Decoder} \\
	\hline 
        Input-($d_r {+} 192$)  \\ 
        \hline 
        upconv (64:4:1:0) uc4s1, BN2D, l-ReLU \\
        \hline 
        upconv (64:5:2:0), BN2D, l-ReLU \\
        \hline
        upconv (64:5:2:0), BN2D, l-ReLU \\
        \hline 
        upconv (64:5:2:0), BN2D, l-ReLU \\
        \hline 
       upconv (\text{channels}:5:2:0), BN2D
        \\
    \hline
    \end{tabular}
    
    \label{deep compressor}
\end{table}

\emph{FMD and JD Critics. } For the video critics, our PLF-JD critic architecture is inspired by the work of Kwon and Park\cite{kwon2019predicting}, where we concatenate frames sequentially along their channel dimensions. For both PLF-FMD and PLF-JD critics, we add spectral normalization layers for better convergence. Their architecture is shown in Table \ref{critic_table}.

\begin{table}[h]
    % \small
    \centering
    \caption{PLF-FMD and PLF-JD critic for frame $i$.} \label{critic_table}
    \begin{tabular}[t]{ |c| }
    \multicolumn{1}{c}{\textbf{(a)} PLF-FMD Critic} \\
	\hline 
        Input--$64{\times}64{\times}\text{channels}$  \\ 
        \hline 
        SN, conv (64:4:2:1), l-ReLU \\
        \hline 
        SN, conv (128:4:2:1), l-ReLU \\
        \hline
        SN, conv (256:4:2:1), l-ReLU \\
        \hline
        conv (512:4:2:1), l-ReLU \\
        \hline
        Linear  \\  
    \hline
    \end{tabular}
    \quad
    \begin{tabular}[t]{ |c| }
    \multicolumn{1}{c}{\textbf{(b)} PLF-JD Critic} \\
	\hline 
        Input--$64{\times}64{\times}(i{\times} \text{channels})$  \\ 
         \hline 
        SN, conv (64:4:2:1), l-ReLU \\
        \hline 
        SN, conv (128:4:2:1), l-ReLU \\
        \hline
        SN, conv (256:4:2:1), l-ReLU \\
        \hline
        conv (512:4:2:1), l-ReLU \\
        \hline
        Linear  \\ 
    \hline
    \end{tabular}
    \vspace{-0.5cm}
\end{table}

\noindent \emph{Rate and output dimension} The rate $R$ is computed by $\log_2(d_{enc}{\times} L)$, where $L$ is the number of quantization levels and $d_{enc} {=} d_r {+} d_m$. Table \ref{tab:rate_param} provides configurations of the rate, $d_{m},d_r$, and $L$  in the experiment.

\noindent \emph{Training Details:} We use a batch size of 64, RMSProp optimizer with a learning rate of $5{\times}10^{-5}$, and train each model with $360$ epochs, where the training set contains 60000 images. To accelerate training, we  pre-train each model for 60 epochs with the MSE objective only. Under WGAN-GP framework\cite{gulrajani2017improved}, we use the gradient penalty of 10 and update the encoders/decoders for every 5 iterations. The parameters $\lambda$ controlling the tradeoff are in Table.\ref{tab:comp_parameters}. Training takes 2 days per model on a single NVIDIA P100 GPU. For the MovingMNIST factor of two bound and permanence of error experiments, we repeat the training 3 times.
\begin{table}[h]
  \vspace{-0.5cm}
  \centering
  \caption{Rate, embedding dimension $d_{m}, d_r$ and quantization level $L$.}
  \label{tab:rate_param}
  
  \begin{subtable}[b]{0.4\textwidth}
    \centering
    \caption{P-frame encoder, $R_1=\infty$.}
    \begin{tabular}{|c|c|c|c|}
      \hline
      $R_2$ & $d_{m}$ & $d_{r}$ & $L$ \\
      \hline
      1 bit & 1 & 0 & 2 \\
      \hline
      2 bits & 1 & 1 & 2 \\
      \hline
      3.17 bits & 1 & 1 & 3 \\
      \hline
    \end{tabular}
    
    \label{tab:rateb}
  \end{subtable}
  \quad
  \begin{subtable}[b]{0.4\textwidth}
    \centering
    \caption{P-frame encoder, $R_1=\epsilon$ (12 bits).}
    \begin{tabular}{|c|c|c|c|}
      \hline
      $R_2$ & $d_{m}$  & $d_{r}$& $L$ \\
      \hline
      4 bit & 2 & 2 & 2 \\
      \hline
      8 bits & 4 & 2 & 2\\
      \hline
      12 bits & 6 & 2 & 2 \\
      \hline
    \end{tabular}
     
    \label{tab:ratec}
  \end{subtable}
\vspace{-0.5cm}
\end{table}

\begin{table}[h]
	\centering
 \caption{Perception loss and their associated $\lambda$}
 
	\small
	\begin{tabular}[t]{|c|l|}
		\hline
		$\text{Perception Loss}$ & \multicolumn{1}{|c|}{$\lambda \times 10^{-3}$}\\
		\hline \hline
		$\text{Joint Distance (JD)}$ & \makecell{$0.0,0.7, 1.0, 1.15,1.2,1.25,1.3,1.5,1.7$ \\ $2.0,3.0,5.0,8.0,10.0,40.0,80.0$} \\
		\hline
		$\text{Frame Marginal Distance (FMD)}$  & $0.0,0.4,0.7,1.0,1.5,2.0,2.5,3.0,3.5,4.0,7.0,10.0,40.0$ \\
  \hline
	\end{tabular}
 \label{tab:comp_parameters}
\end{table}

\subsection{Permanence of Error on KTH Datasets}\label{Additional Experiments}\label{extra-experiments}
The KTH dataset is a widely-used benchmark dataset in computer vision research, consisting of video sequences of human actions performed in various scenarios. We show more examples supporting our argument for the permanence of error on this realistic dataset. We use 16 bits for each frame. In general, the 0-PLF-JD decoder consistently outputs correlated but incorrect reconstructions due to the error induced by the first reconstructions, i.e., the P-frames will follow the wrong direction induced from the I-frame reconstruction. Besides the moving direction, we also notice that the type of actions (i.e. walking, jogging, and running) is also affected. On the other hand, while losing some temporal cohesion, MMSE and 0-PLF FMD decoders manage to fix the movement error.

\begin{figure}[t]
  \includegraphics[width=\textwidth,trim={0cm 0cm 0.0cm 0cm},clip]{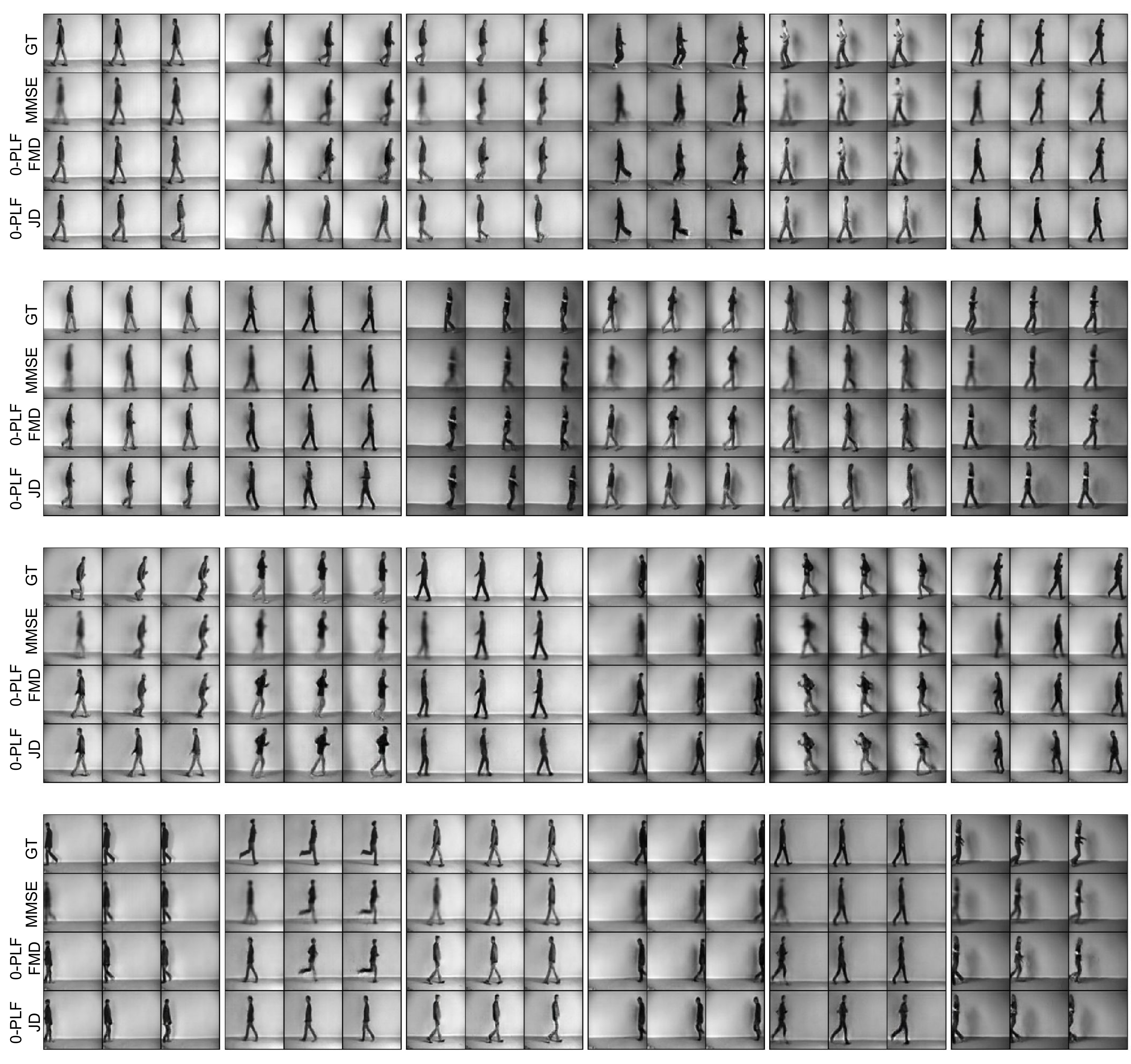}
  \caption{Additional Experimental Results for the Permanence of Error Phenomenon on KTH Dataset.}
  \label{fig:KTH_more examples}
\end{figure}

\subsection{RDP Tradeoff for 3 frames}
\begin{figure*}[ht]
\centering
\vspace{0cm}
\hspace{0cm}
\begin{subfigure}[t]{0.485\textwidth}
  \includegraphics[width=\textwidth,trim={0cm 0.2cm 0cm 0cm},clip]{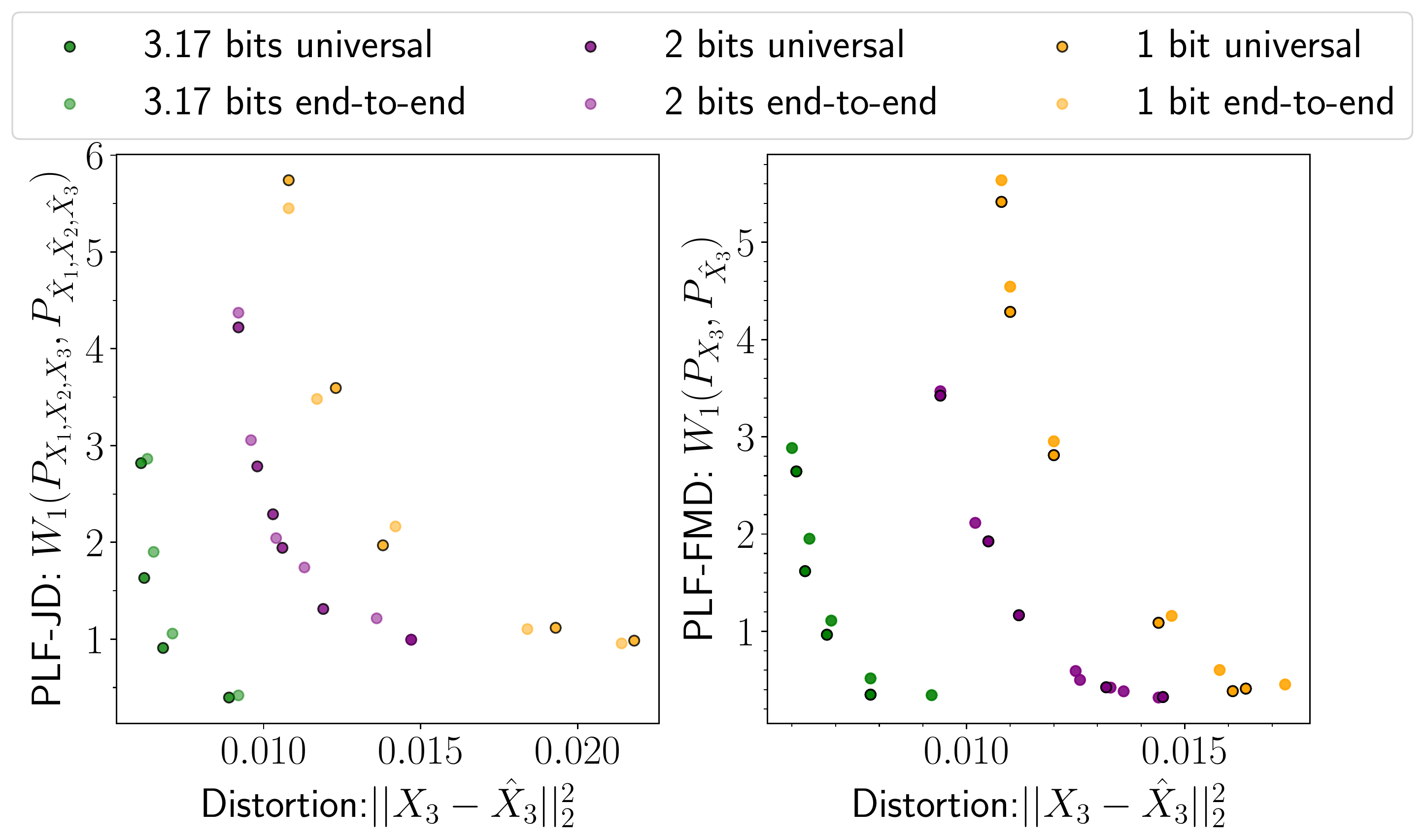}
  \caption{Case 1: $R_1{=}\infty$, $R_2{=}R_3{=}\{1, 2, 3.17\}$ bits}
  \label{fig:RDP_tradeoff_3framesa}
\end{subfigure}\hspace{0.2cm}
\begin{subfigure}[t]{0.48\textwidth}
  \includegraphics[width=\textwidth,trim={0cm 0.2cm 0cm 0cm},clip]{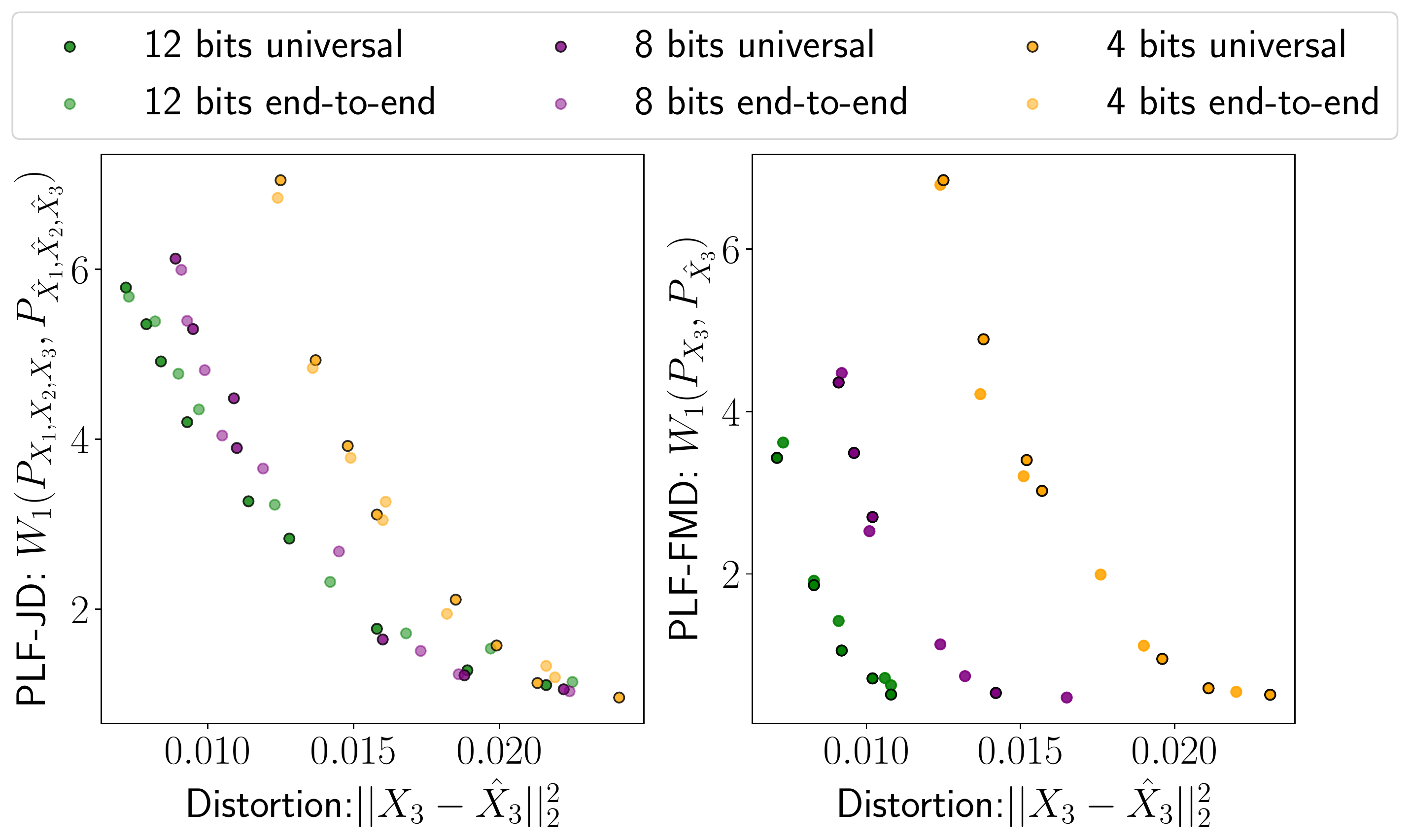}
  \caption{Case 2: $\!R_1{=}12$, $R_2{=}R_3{=}\{4, 8, 12\}$ bits}
  \label{fig:RDP_tradeoff_3framesb}
\end{subfigure}%\vspace{-0.15cm}
\caption{RDP tradeoff curves for end-to-end and universal models. We plot the tradeoff for the two regimes: $R_1{=}\infty$ and $R_1{=}\epsilon$ in (a) and (b) respectively. The universal and optimal curves are close to each other.}
\label{fig:RDP_tradeoff_3frames}
\end{figure*}
We extend our experimental results for the RDP-tradeoff and the principal of universality to the case of GOP size 3. As mentioned in the main paper, while the universal model only requires MMSE representations, the optimal end-to-end model also requires the MMSE reconstructions from previous frames to provide best estimates for motion flow vectors. Practically, this is challenging for our employed architecture since only previous $\hat{X}_1, \hat{X}_2$ are available. As a result, to compare the RDP tradeoff between universal and end-to-end model, we also provide the end-to-end model with the MMSE estimate from previous frames while noting that this is  unfeasible in practice. Interestingly, we show in Figure \ref{fig:RDP_tradeoff_3frames} the RDP tradeoff curves for the third frame $X_3$ and its reconstruction $\hat{X}_3$, observing that the universal and optimized curves are still relatively close to each other. When $(R_1,R_2,R_3){=}(\infty, \epsilon, \epsilon)$, we note that the distortion for $X_3$ is larger than $X_2$ since the allocated rate is not enough to correct the motion. Finally, for the case $(R_1,R_2,R_3){=}(\epsilon, \epsilon, \epsilon)$, we note that the curves again converge as in the case of $(R_1,R_2){=}(\epsilon, \epsilon)$ due to the incorrect reconstruction in the I-frame. 
%is required to provide the best estimates the motion flow vectors

\subsection{Diversity and Correlation}
\label{sec:diversity}
\begin{figure}[h]
\centering
  \includegraphics[width=0.6\textwidth,trim={0cm 0cm 0.0cm 0cm},clip]{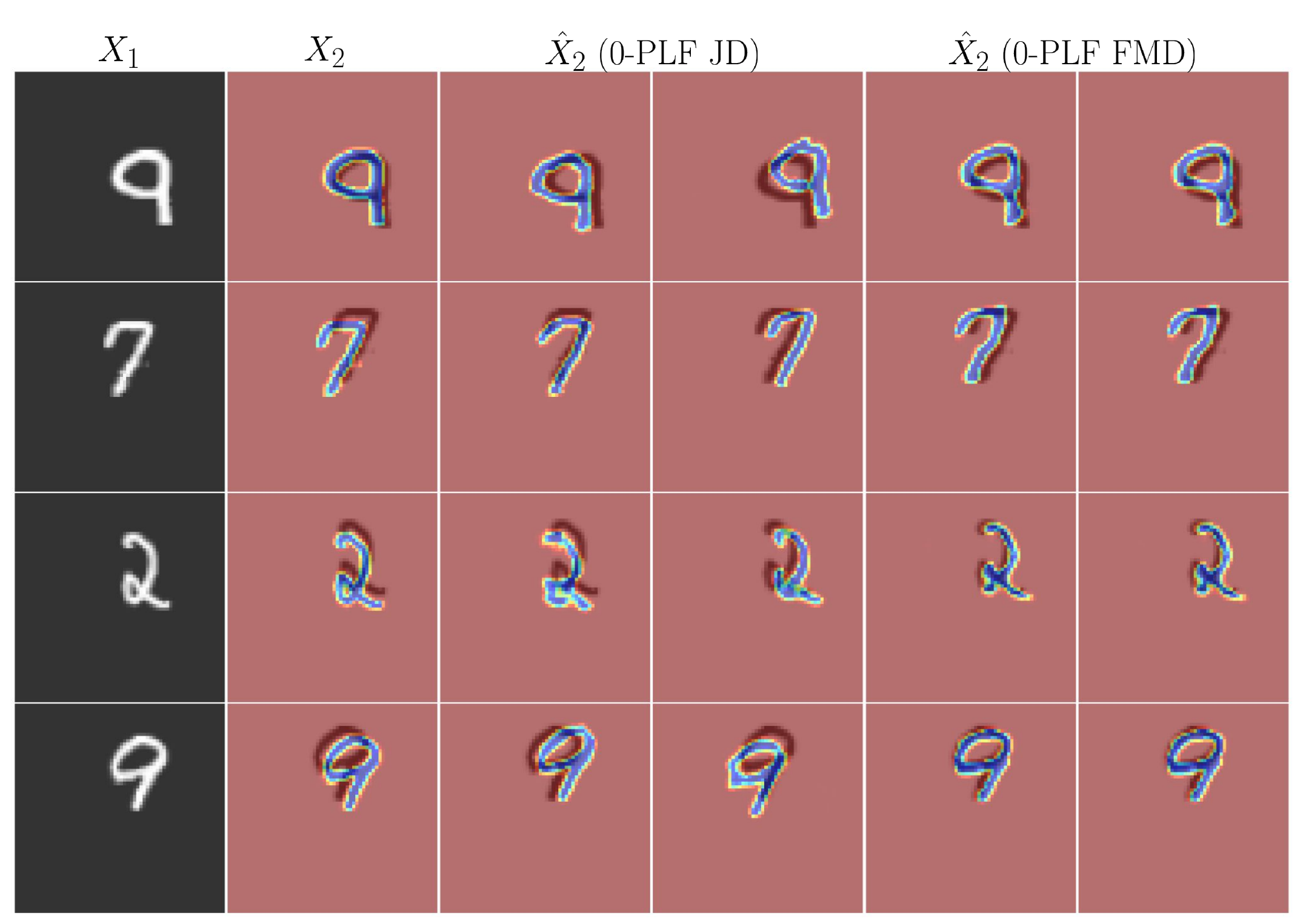}
  \caption{Diversity in reconstruction $\hat{X}_2$ for 0-PLF-JD and correlation with previous frames $\hat{X}_1$ for 0-PLF-JMD. We show $X_1$ in the first column. From the second column, the light-dark region represents $X_1$ and the color digit represents $X_2,\hat{X}_2$. For each perception metric, we show two samples.}
  \label{fig:diversity_visual}
\end{figure}

When $(R_1,R_2){=}(\infty,\epsilon)$, our theoretical analysis predicted that the decoder optimized for JD is capable of producing diverse reconstructions. On the other hand, an optimized decoder for FMD will tend to produce  reconstructions that are highly correlated with the previous reconstruction $\hat{X}_1$\footnote{$X_1=\hat{X}_1$ in this regime.}. In our experiment, we also observe such behavior, summarized in Table \ref{tab:diverse} and show several examples for $R_2=2$ bits in Figure \ref{fig:diversity_visual}.  We observe that reconstructions from the joint metric deviate more randomly from $X_1$ than the marginal reconstructions. The marginal reconstructions, on the other hand, stay much closer to their original reconstruction $\hat{X}_1$. 
%$\mathrm{E[(\hat{X}_2 - X_1)^2]}$. 

We measure the diversity in $\hat{X}_2$ reconstruction using $\mathrm{E}[\text{Var}(\hat{X}_2|X_1, X_2)]$ and the correlation with $\hat{X_1}$ by $\mathrm{E[\text{sim}(\hat{X}_2, X_1)]}$, where $\text{sim}(u,v)$ is the cosine distance between $u,v$. Table \ref{tab:diverse}a shows that as we increase the number of bits in $R_2$, the diversity decreases as the decoder can reconstruct the frame more precisely. In Table \ref{tab:diverse}b, we see that the joint metric keeps the correlation relatively constant, showing that it actually preserves the temporal consistency. On the other hand, as the rate becomes larger, 0-PLF-FMD reconstruction tends to be less correlated with the previous frame $X_1$. Finally, we note that our architecture innately utilizes common randomness to produce diverse reconstructions and does not suffer from mode-collapse behavior in general conditional GAN settings \cite{hong2019diversity}.

\begin{table}[t]
  %\centering
  \caption{Diversity (a) between $\hat{X}_2$ and Correlation Measures (b) between $\hat{X}_2$ and $X_1$.}
  \label{tab:diverse}
  \begin{subtable}[b]{0.5\textwidth}
    \centering
    \caption{Diversity Measures $\uparrow$.}
    \label{tab:diversea}
    \begin{tabular}{|c|c|c|}
      \hline
      $R_2$ & Joint & Marginal \\
      \hline
      1 bit & 0.0096 & 0.0004 \\
      \hline
      2 bits & 0.0082 & 0.0029 \\
      \hline
      3.17 bits & 0.0042 & 0.0022 \\
      \hline
    \end{tabular}
    
  \end{subtable}
  \quad
  \begin{subtable}[b]{0.4\textwidth}
  \caption{Correlation Measures. $\uparrow$}
    \label{tab:diverseb}
    \centering
    \begin{tabular}{|c|c|c|}
      \hline
      $R_2$ & Joint & Marginal \\
      \hline
      1 bit & 0.5218 & 0.6202 \\
      \hline
      2 bits & 0.5190 & 0.5969 \\
      \hline
      3.17 bits & 0.5205 & 0.5508 \\
      \hline
    \end{tabular}
    
  \end{subtable}
  
\end{table}

\section{Error Permanence on UVG Dataset}\label{appendix-UVG-dataset}
\begin{figure}[h]
  \includegraphics[width=\textwidth,trim={0cm 0.5cm 0cm 0.15cm},clip]{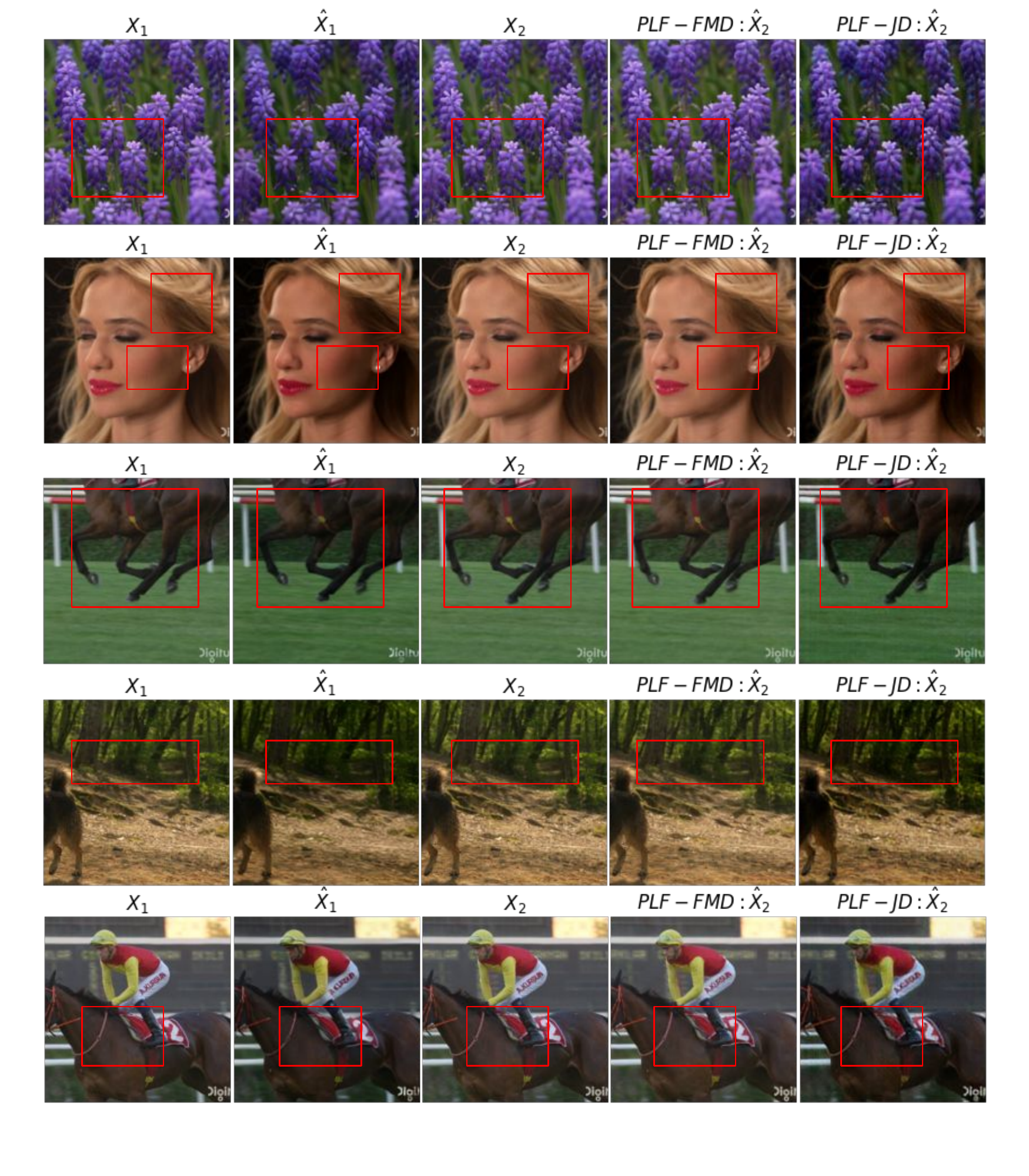}
  \vspace{-0.95cm}
  \caption{Visualization of error permanence. The PLF-JD reconstructions propagate the flaws in the color tone from
the previous I-frame reconstruction while the PLF-FMD is able to fix such error. Compression rate for I-frame and P-frame are ~0.144bpp (low rate) and 4.632bpp (high rate) respectively.}
  \label{fig:UVG_more examples}
\end{figure}

We demonstrate the phenomenon of error permanence in a large-scale scenario using the UVG dataset. Our P-frame compressor is trained on the Vimeo-90k dataset. As illustrated in Figure \ref{fig:UVG_more examples}, when reconstructing I-frames, an inaccurate color tone is introduced, which persists when employing PLF-JD. However, PLF-FMD effectively rectifies this issue within P-frames. Numerical results are in Table.~\ref{UVG_table}.

\begin{table}
\centering
\begin{tabular}{|c|c|c|c|}
\hline
 & MMSE & PLF-FMD & PLF-JD \\
\hline
Distortion (MSE) & 0.0026 & 0.0032 & 0.0168 \\
\hline
\end{tabular}
\caption{Distortion $(X_2-\hat{X}_2)^2$ evaluated across 3900 UVG frames. The PLF-JD reconstructions exhibit notably greater distortion compared to MMSE and PLF-FMD, aligning with our theoretical findings. }
\label{UVG_table}
\end{table}

\section{Limitations}
This work  studies the effects of different perception loss functions, namely the PLF-JD and PLF-FMD, on the performance of lossy causal video compression.  Our theoretical analysis and experiment reveal the error permanence phenomenon and show the universality principle, suggesting that MMSE representation can be transformed into other points on the DP tradeoffs. 

In practice, one might want to combine these two losses, for example, perfect framewise realism (0-PLF FMD) while retaining some degree of temporal cohesion (PLF-JD small), which is not considered in this work. Furthermore, analysis for other types of video compression schemes, such as with B-frame, and scaling the universality compression architecture to high-definition videos are also desired. 

%We should suggest that the right perception metric should be some weighted average of the perception metrics of: (i) per-frame marginal of the I-frames; (ii) conditional distribution of the P-frames conditioned on the previous I-frames. (iiI) conditional distribution of the B-frames conditioned on the previous and the next I-frames. 

\end{document}